\documentclass[11pt,a4paper]{article}
\pdfoutput=1 

\usepackage{amsmath,amssymb,amsfonts,color,scalefnt}
\usepackage{epsfig,amsthm}
\usepackage{soul}
\usepackage[makeroom]{cancel}
\usepackage{mathrsfs}
\usepackage{epstopdf}
\usepackage[utf8]{inputenc}
\usepackage{hyperref}
\usepackage{slashed}
\usepackage{graphicx}
\usepackage{pict2e}
\usepackage{tikz}
\usepackage{cite}
\usepackage{float}
\usepackage[percent]{overpic}
\usepackage{multirow}
\usepackage{physics}
\usepackage{booktabs}
\usepackage{mwe}
\usepackage [autostyle, english = american]{csquotes}
\usepackage{bbm}
\MakeOuterQuote{"}

\newcommand\Tstrut{\rule{0pt}{2.6ex}}       % "top" strut
\newcommand\Bstrut{\rule[-0.9ex]{0pt}{0pt}} % "bottom" strut
\newcommand{\TBstrut}{\Tstrut\Bstrut} % top&bottom struts
\newcommand{\mtree}{\text{mtree}}
\newcommand{\bfZ}{\mathbf{Z}}
\newcommand{\beq}{\begin{eqnarray}}
\newcommand{\eeq}{\end{eqnarray}}
\newcommand{\ba}{\begin{eqnarray}}
\newcommand{\ea}{\end{eqnarray}}
\newcommand{\be}{\begin{equation}}
\newcommand{\ee}{\end{equation}}
\newcommand{\bpmatrix}{\begin{pmatrix}}
\newcommand{\epmatrix}{\end{pmatrix}}
\renewcommand{\braket}[1]{\left(#1\right)}

\newcommand{\abbrev}{\scalefont{.9}}
\newcommand{\eqn}[1]{Eq.\,\ref{#1}}
\newcommand{\fig}[1]{Fig.\,\ref{#1}}

\newcommand{\lo}{{\abbrev LO}}

\newcommand{\ocp}{{\abbrev CP}}
\newcommand{\nlo}{{\abbrev NLO}}

\renewcommand{\Re}{{\rm Re}}

\newcommand{\mc}[1]{\ensuremath{\mathcal{#1}}}
\newcommand{\mbf}[1]{\ensuremath{\mathbf{#1}}}

\newcommand{\nn}{\nonumber}

\newcommand{\comment}[1]{\ignorespaces}

\newcommand{\s}{\newline \vspace*{-3.5mm}}
\newcommand{\process}[1]{\ensuremath{H^\pm \rightarrow W^\pm h_{#1}}}

\renewcommand{\mathbbm}[1]{\text{\usefont{U}{bbm}{m}{n}#1}}

\begin{document}

\title{
	\vspace*{-3cm}
	\phantom{h} \hfill\mbox{\small KA-TP-21-2019}
\\[-1.1cm]
\phantom{h} \hfill\mbox{\small IFIRSE-TH-2019-8} 
	\\[1cm]
	%\vspace{13mm}   
	\textbf{Gauge Dependences of Higher-Order Corrections to \\ NMSSM Higgs
		Boson Masses and \\ the Charged Higgs Decay \process{i}}}

\date{}
\author{
Thi Nhung Dao$^{1\,}$\footnote{E-mail: \texttt{dtnhung@ifirse.icise.vn}},
Lukas Fritz$^{2,3\,}$\footnote{E-mail:
	\texttt{lukas.fritz@psi.ch}}, Marcel Krause$^{3\,}$\footnote{E-mail:
	\texttt{marcel.krause@kit.edu}},
Margarete M\"{u}hlleitner$^{3\,}$\footnote{E-mail:
	\texttt{margarete.muehlleitner@kit.edu}},
Shruti Patel$^{3,4\,}$\footnote{E-mail: \texttt{shruti.patel@kit.edu}}
\\[9mm]
{\small\it
$^1$Institute For Interdisciplinary Research in Science and Education, ICISE,}\\
{\small\it 590000 Quy Nhon, Vietnam.}\\[3mm]
{\small\it$^2$Paul Scherrer Institut, CH-5232 Villigen PSI, Switzerland.}\\[3mm]
{\small\it
$^3$Institute for Theoretical Physics, Karlsruhe Institute of Technology,} \\
{\small\it Wolfgang-Gaede-Str. 1, 76131 Karlsruhe, Germany.}\\[3mm]
{\small\it$^4$Institute for Nuclear Physics, Karlsruhe Institute of Technology,
76344 Karlsruhe, Germany.}\\[3mm]
}
\maketitle

%
%
%\date{\today}

\begin{abstract}
In this paper we compute the electroweak corrections to the charged
Higgs boson decay into a $W$ boson and a neutral Higgs boson in the
CP-conserving NMSSM. We calculate the process in a general $R_\xi$
gauge and investigate the dependence of the loop-corrected decay width
on the gauge parameter $\xi$. The gauge dependence arises from the
mixing of different loop orders. Phenomenology requires the inclusion of mass and mixing
corrections to the external Higgs bosons in order to match the
experimentally measured mass values. As a result, we move away from a
strict one-loop calculation and consequently mix orders in
perturbation theory. Moreover, determination of the loop-corrected
masses in an iterative procedure also results in the mixing of
different loop orders. Gauge dependence then arises from  the mismatch
with tree-level Goldstone boson couplings that are applied in the loop
calculation, and from the gauge dependence of the loop-corrected
masses themselves. We find that the gauge dependence is significant.
\end{abstract}

%\maketitle
%\tableofcontents

\thispagestyle{empty}
\vfill
\newpage

%-----------------------------
\section{Introduction}
\label{sec:intro}
%-----------------------------
The discovery of the Higgs boson by the LHC experiments ATLAS and CMS
\cite{Aad:2012tfa,Chatrchyan:2012xdj} structurally completed the
Standard Model (SM). Subsequent measurements revealed a very SM-like
behavior of the Higgs boson. Due to open questions that cannot be
answered within the SM, however, theories beyond the SM are
considered, many of which contain extended Higgs sectors. So far, no
direct signs of New Physics have been observed. This moves the Higgs
sector itself into the focus of searches for indirect manifestations
beyond the SM. Due to the very SM-like nature of the Higgs boson,
sophisticated experimental techniques together with precise
theoretical predictions are required for these investigations to be
successful. In particular, higher-order (HO) corrections to the Higgs
boson observables, their production cross sections, decay widths, and
branching ratios have to be taken into account. \s 

A clear manifestation of extended Higgs sectors would be the discovery of a charged Higgs boson that is not present in the SM. Charged Higgs bosons appear {\it
  e.g.}~in the next-to-minimal supersymmetric extension of the SM
(NMSSM)
\cite{Barbieri:1982eh,Dine:1981rt,Ellis:1988er,Drees:1988fc,Ellwanger:1993xa,Ellwanger:1995ru,Ellwanger:1996gw,Elliott:1994ht,King:1995vk,Franke:1995tc,Maniatis:2009re,Ellwanger:2009dp}
which is the model that we consider in this work. More 
specifically, we work in the framework of the scale-invariant
CP-conserving NMSSM. The main decay channels of the charged Higgs 
boson are those into fermionic final states, but also decays into a
Higgs and gauge boson final state, or into electroweakinos can become
numerically important depending on the specific parameter values. In
this paper, we compute the electroweak corrections to the decay of the
charged Higgs boson into a $W$ boson and a light CP-even Higgs
boson. We restrict ourselves to pure on-shell decays. The aim of this
paper is not only to quantify the relative importance of the
electroweak corrections, but in particular we also highlight
problems with respect to gauge dependences that occur in the
computation of the HO corrections. In a gauge theory
gauge-breaking effects do not appear when the computation is
restricted to a fixed order, here the one-loop level. This
changes, however, when different loop orders are mixed, see {\it
  e.g.}~also the discussions in
\cite{Williams:2007dc,Fowler:2009ay,Williams:2011bu,Benbrik:2012rm,Gonzalez:2012mq,Nhung:2013lpa,Goodsell:2017pdq,Domingo:2017rhb,Domingo:2018uim,Baglio:2019nlc}. We encounter
such a mixing when we include loop corrections to the mass of the
involved external Higgs boson. Since the tree-level upper bound of
the SM-like Higgs boson is below the observed $125.09$ GeV
\cite{Aad:2015zhl}, loop corrections have to be included to shift its
mass to the measured value. This introduces a mismatch between the
loop-corrected mass of the external neutral Higgs boson and its
corresponding tree-level mass, which is used in the propagators of the
internal lines and in the tree-level Higgs-Goldstone boson couplings
that occur in the computation of the one-loop amplitude\footnote{Note
  that tree-level masses and tree-level couplings have to be used in
  the one-loop diagrams in order to ensure the cancelation of
  ultraviolet (UV) divergences.}. While the latter can be cured by an appropriate
change of the Higgs-Goldstone boson coupling as we will outline below (see also 
\cite{Benbrik:2012rm,Gonzalez:2012mq,Goodsell:2017pdq,Domingo:2017rhb,
  Domingo:2018uim} for a discussion), the former cannot be cured by
introducing loop-corrected masses for the internal lines since it will break UV finiteness. Furthermore, we encounter additional gauge dependences  due to the
gauge dependence of the loop-corrected Higgs boson masses and their
loop-corrected mixing matrix. The loop-corrected Higgs boson masses
are defined through the complex poles of the propagator matrix which
are evaluated by using an iterative method. While this method gives
precise values of the complex poles, it mixes the contributions of
different orders of perturbation theory and therefore introduces a
dependence on the gauge parameter. The loop-corrected mixing matrix is
used to resum the large corrections that stem from the mixing between
different neutral Higgs bosons so that the loop corrections remain
small and the higher-order prediction are reliable. In addition, the
loop-corrected mixing matrix ensures the on-shell property of the
external Higgs boson.  This mixing matrix is, however, gauge
dependent by definition. With the loop corrections to the light Higgs
boson masses and the mixing matrix being substantial also the
gauge dependence turns out to be significant and much more
important than in the minimal supersymmetric extension of the SM
(MSSM) as discussed in this work. The purpose of the present paper is
to quantify this effect and to investigate different approximations
with respect to their gauge dependences. \s

The outline of the paper is as follows. In Sec.~\ref{sec:HiggsSec}
we introduce the Higgs sector of the NMSSM at tree level and at higher
orders, and set our notation. In Sec.~\ref{sec:one-loop-decay} we
describe our computation of the 
electroweak one-loop corrections to the charged Higgs decay into a
gauge plus Higgs boson final state. In particular we present the decay
width at strict one-loop order. We follow up with a general discussion
of gauge dependences encountered in the decay before presenting
improved effective decay widths that include higher-order-corrected
external Higgs states in different approximations. In the numerical
results of Sec.~\ref{sec:results} we analyze the gauge dependence
of the loop-corrected masses themselves and subsequently the decay
amplitudes and decay widths. We analyze the latter two with respect to
their gauge dependences by including various approximations in the
treatment of the loop-corrected external Higgs states. We also compare
the results with the the size of the gauge dependences in the MSSM
limit. We conclude with a small discussion of the relative size of the
electroweak corrections as a function of the relevant NMSSM
parameters. In Sec.~\ref{sec:concl} we summarize our results.

%%%%%%%%%%%%%%%%%%%%%%%%%%%%%%%%%%%%%%%%%%%%%%%%%%%%%%%%%%%%%
%----------------------------------------
\section{Higgs sector of the NMSSM}
\label{sec:HiggsSec}
%----------------------------------------

In this paper, we calculate within the NMSSM the one-loop
corrections to the decays of the charged Higgs boson into a $W ^\pm $
boson and a neutral CP-even Higgs boson. 
To that end, we briefly introduce the Higgs sector of the NMSSM and
set up the notation required both for the calculation of the charged
Higgs decays as well as for the discussion of the
higher-order-corrected neutral Higgs boson masses. Since we apply the
same approximations and renormalization conditions as in our previous
works on higher-order corrections to the NMSSM Higgs boson masses 
and trilinear self-couplings \cite{Ender:2011qh,Graf:2012hh,Muhlleitner:2014vsa,Dao:2019qaz,Nhung:2013lpa,Muhlleitner:2015dua}, we remain here as brief as possible and
refer, where appropriate, to the corresponding literature for more information. 
We work in the framework of an NMSSM wherein a gauge-singlet chiral
superfield $\hat{S}$ is added to the MSSM field content, and the
superpotential couplings of this singlet are constrained by a
$\mathbb{Z}_3$ symmetry. In terms of the two Higgs doublet superfields
$\hat{H}_u$ and $\hat{H}_d$ and the singlet superfield $\hat{S}$ the
NMSSM superpotential is written as
\be
\mathcal{W}_\text{NMSSM} = \mathcal{W}_\text{MSSM} +
\frac{1}{3} \kappa {\hat S}^3 -\epsilon_{ij} \lambda \hat S \hat H_d^i
\hat H_u^j\,,
\label{eq:nmssmsuperpot}
\ee
with the totally antisymmetric tensor $\epsilon_{ij}$ ($i,j=1,2$) and
$\epsilon_{12}= \epsilon^{12}=1$, where $i,j$ denote the indices of the
fundamental $SU(2)_L$ representation. Working in the framework of the
CP-conserving NMSSM, the dimensionless parameters
$\lambda$ and $\kappa$ are taken to be real.
The MSSM superpotential $\mathcal{W}_\text{MSSM}$ is expressed in
terms of the quark and lepton superfields and their charge conjugates as
denoted by the superscript $c$, \textit{i.e.}~$\hat Q, \hat{U}^c, \hat{D}^c, \hat L$ and $\hat{E}^c$, as
\be
\mathcal{W}_\text{MSSM} = \epsilon_{ij} [y_e \hat{H}^i_d \hat{L}^j
	\hat{E}^c + y_d \hat{H}_d^i \hat{Q}^j \hat{D}^c - y_u \hat{H}^i_u
	\hat{Q}^j \hat{U}^c] \,.
\label{mssmsuperpot}
\ee
For better readability, the color and generation indices have been
suppressed, and $\mu$ (\textit{i.e.}~the supersymmetric Higgs mass
parameter of the MSSM) is set to 0 due to the applied 
  $\mathbb{Z}_3$ symmetry. We neglect flavor mixing so that the 
Yukawa couplings $y_u, y_d$ and $y_e$, which in general are $3 \times
3$ matrices in flavor space, are diagonal. \s

The soft supersymmetry (SUSY) breaking NMSSM Lagrangian is given in terms of the
scalar component fields $H_{u}$, $H_d$ and $S$ by
\begin{align}
	{\cal L}_{\text{soft},\text{ NMSSM}} = & -m_{H_d}^2 H_d^\dagger H_d - m_{H_u}^2
	H_u^\dagger H_u -
	m_{\tilde{Q}}^2 \tilde{Q}^\dagger \tilde{Q} - m_{\tilde{L}}^2 \tilde{L}^\dagger \tilde{L}
	- m_{\tilde{u}_R}^2 \tilde{u}_R^*
	\tilde{u}_R - m_{\tilde{d}_R}^2 \tilde{d}_R^* \tilde{d}_R
	\nonumber      \\\nonumber
& - m_{\tilde{e}_R}^2 \tilde{e}_R^* \tilde{e}_R - (\epsilon_{ij} [y_e A_e H_d^i
		\tilde{L}^j \tilde{e}_R^* + y_d
		A_d H_d^i \tilde{Q}^j \tilde{d}_R^* - y_u A_u H_u^i \tilde{Q}^j
		\tilde{u}_R^*] + \mathrm{h.c.})      \\
& -\frac{1}{2}(M_1 \tilde{B}\tilde{B} + M_2
	\tilde{W}_j\tilde{W}_j + M_3 \tilde{G}\tilde{G} + \mathrm{h.c.}) \\ \nonumber
& - m_S^2 |S|^2 +
	(\epsilon_{ij} \lambda
	A_\lambda S H_d^i H_u^j - \frac{1}{3} \kappa
	A_\kappa S^3 + \mathrm{h.c.}) \;,
	\label{eq:Lsoft}
\end{align}
where the summation over all three quark and lepton generations is implicit.
Here, we denote by $\tilde{Q}$ and $\tilde{L}$ the complex scalar
components of the corresponding left-handed quark and lepton
superfields, so that {\it e.g.} for the first generation we have
$\tilde{Q} = (\tilde{u}_L, \tilde{d}_L)^T$ and $\tilde{L}= (\tilde{\nu}_L,\tilde{e}_L)^T$.
The $M_k$ ($k=1,2,3$) represent the gaugino mass parameters of the
bino, wino and gluino fields $\tilde{B}$, $\tilde{W}_j$ ($j=1,2,3$)
and $\tilde{G}$, the $m_X^2$ are the squared soft SUSY-breaking mass
parameters of the scalar fields $X=S, H_d, H_u, \tilde{Q},
	\tilde{u}_R, \tilde{d}_R, \tilde{L}, \tilde{e}_R$ and $A_x$
($x=\lambda,\kappa,d,u,e$) are the soft SUSY-breaking
trilinear couplings. In general, the soft SUSY-breaking trilinear
couplings and the gaugino mass parameters can be complex, but in
the limit of CP conservation all parameters are taken to be real.
%
%%%%%%%%%%%%%%%%%%%%%%%%%%%%%%%%%%%%%%%%%%%%%%%%%%%%%%%%%%%%%
%-------------------------------------------
\subsection{The Higgs Sector at Tree Level}
\label{sec:HiggsSec-Tree}
%-------------------------------------------

The Higgs potential at tree level reads
\beq
V_{H}  &=& (|\lambda S|^2 + m_{H_d}^2)H_d^\dagger H_d+ (|\lambda S|^2
+ m_{H_u}^2)H_u^\dagger H_u +m_S^2 |S|^2 \nonumber \\
&& + \frac{1}{8} (g_2^2+g_1^{2})(H_d^\dagger H_d-H_u^\dagger H_u )^2
+\frac{1}{2} g_2^2|H_d^\dagger H_u|^2 \label{eq:NMSSMHiggspotential} \\
&&   + |-\epsilon^{ij} \lambda  H_{d,i}  H_{u,j} + \kappa S^2 |^2+
\big[-\epsilon^{ij}\lambda A_\lambda S   H_{d,i}  H_{u,j}  +\frac{1}{3} \kappa
	A_{\kappa} S^3+\mathrm{h.c.} \big] \;,
\nonumber
\eeq
with $g_1$ and $g_2$ being the $U(1)_Y$ and $SU(2)_L$ gauge couplings,
respectively. The two Higgs doublets and the singlet can be expanded
around their vacuum expectation values (VEVs) $v_u, v_d$ and $v_s$ as
\beq
H_u = \mqty(h_u^+ \\ \frac{1}{\sqrt{2}}(v_u + h_u + i a_u)),\quad
H_d = \mqty(\frac{1}{\sqrt{2}}(v_d + h_d + i a_d)\\  h_d^-)
\quad \mbox{and}\quad
S = \frac{1}{\sqrt{2}} (v_s + h_s + i a_s)\,.
\label{eq:vevexpansion}
\eeq
The fields $h_u, h_d$ and $h_s$ are the CP-even parts and $a_u$, $a_d$
and $a_s$ are the CP-odd parts of the neutral components of the fields
$H_u$, $H_d$ and $S$, respectively. The electrically charged
components are denoted by $h^{+}_u$ and $h^{-}_d$. The VEVs of the two
Higgs doublets, $v_u$ and $v_d$, are related to the VEV $v \approx
246$ GeV of the SM as 
\be
v^2 = v_u^2 + v_d^2 \,,
\label{eq:vevs}
\ee
with the ratio between them being
%the VEVs 
defined as
\be
\tan \beta = \frac{v_u}{v_d}\,,
\label{eq:tanb}
\ee
such that $v_u$ and $v_d$ can be expressed in terms of $v$ and $\tan \beta$.
The potential $V_H$ is minimized by the VEVs, which implies that the first
derivatives of the potential with respect to the Higgs fields must be
zero. This leads to the definition of the tadpole parameters $t_{\phi}$,
\be
t_{\phi} \equiv \left<\frac{\partial V_H}{\partial
	\phi}\right>\,, \quad  \phi \in \qty {h_u, h_d, h_s,
	a_u, a_d, a_s}\,,
\ee
which have to vanish. Since we are working in CP-conserving NMSSM, the
tadpole parameters that we have at tree level are given by
\begin{align}
	t_{h_d} & = \frac{\lambda}{2} \left(  - \sqrt{2} v_s v_u A_\lambda + \lambda v_d(v_s^2 + v_u^2) - \kappa v_s^2 v_u \right) + \frac{1}{8} (g_1^2 + g_2^2) v_d (v_d^2 - v_u^2) + m_{H_d}^2 v_d    \label{eq:tad1}             \\
	t_{h_u} & = \frac{\lambda}{2} \left(- \sqrt{2} v_s v_d A_\lambda + \lambda v_u (v_s^2 + v_d^2) - \kappa v_s^2 v_d \right) + \frac{1}{8} (g_1^2 + g_2^2) v_u (v_u^2 - v_d^2) + m_{H_u}^2 v_u     \label{eq:tad2}             \\
	t_{h_s} & = \frac{v_s}{2} \left(\sqrt{2} v_s \kappa A_\kappa +
                  \lambda^2 (v_d^2 + v_u^2) - 2 v_d v_u \kappa \lambda
                  + 2 \kappa^2 v_s^2 \right) - \frac{1}{\sqrt{2}} v_d
                  v_u \lambda A_\lambda + m_S^2 v_s ~. \label{eq:tad3}
\end{align}

In the CP-conserving NMSSM, there is no mixing between \ocp-even
and \ocp-odd Higgs fields so that the bilinear parts of the Higgs potential read
\begin{align}
	V_H \supset \begin{pmatrix} h_d^+ & h_u^+ \end{pmatrix} \bold{M_{H^{\pm}}} \begin{pmatrix} h_d^- \\ h_u^- \end{pmatrix} + \frac{1}{2} \begin{pmatrix} h_d & h_u & h_s \end{pmatrix} \bold{M_{h}} \begin{pmatrix} h_d \\ h_u \\ h_s \end{pmatrix} + \frac{1}{2} \begin{pmatrix} a_d & a_u & a_s \end{pmatrix} \bold{M_{a}} \begin{pmatrix} a_d \\ a_u \\ a_s \end{pmatrix}\,,
	\label{eq:tadpoles}
\end{align}
with separate mass matrices $\bold{M_{H^{\pm}}}, \bold{M_{h}}$ and
$\bold{M_{a}}$ for the charged, CP-even and CP-odd Higgs fields,
respectively. The explicit expressions of these tree-level mass
matrices can be found in \cite{Ender:2011qh}. The charged, neutral
CP-even and CP-odd mass eigenstates are obtained from the interaction
states through rotations with the unitary matrices ${R^{H^{\pm}}}$, $R^h$ and $R^a$ as
\begin{align}
	\mqty(G^\pm \\ H^\pm) = R^{H^\pm} \mqty(h_d^\pm \\ h_u^\pm),\quad \mqty(h_1 \\ h_2 \\ h_3) = R^h \mqty(h_d \\ h_u \\ h_s)\quad \text{and} \quad  \mqty(
	G^0         \\ a_1 \\ a_2) = R^a \mqty(a_d \\ a_u \\ a_s) \,.
	\label{eq:tree-level-R}
\end{align}
These rotation matrices diagonalize the mass matrices such
that
\begin{align}
	\bold{M_{H^{\pm}}^{\text{diag}}} = R^{H^\pm}\bold{M_{H^{\pm}}}{R^{H^{\pm}}}^{\dagger},\,\,\,\,
	\bold{M_{h}^{\text{diag}}} = R^{h}\bold{M_{h}}{R^{h}}^{\dagger}\,\,\,\,\text{and}\,\,\,\,
	\bold{M_{a}^{\text{diag}}} = R^{a}\bold{M_{a}}{R^{a}}^{\dagger}.
	\label{eq:tree-level-masses}
\end{align}
The obtained mass eigenstates are ordered by ascending mass so that we
have three CP-even Higgs states $h_i$ ($i=1,2,3$) with masses $m_{h_1}
\leq m_{h_2} \leq m_{h_3}$, two CP-odd states $a_j$ ($j=1,2$) with
masses $m_{a_1} \leq m_{a_2}$ and a charged Higgs pair $H^{\pm}$ with
mass $m_{H^\pm}$ as
the physical Higgs bosons. The fields $G^0, G^\pm$ form 
the massless charged and neutral Goldstone modes\footnote{By adding
  the 't Hooft linear gauge-fixing Langrangian, $G^0$ has mass $\sqrt{\xi_Z} M_Z$ while
 the $G^\pm$ have mass  $\sqrt{\xi_W} M_W$, where  $\xi_Z,\xi_W$ are the
 gauge parameters, and $M_Z, M_W$ denote the $Z, W^\pm$ gauge
 boson masses, respectively.}. In general, the 
analytic forms of the mass eigenvalues are rather intricate, but
analytic expressions for expansions in special parameter regions can
be found in \cite{Miller:2003ay}. On the other hand, the squared mass
of the charged Higgs boson is at tree level given by the simple
expression 
\be
m_{H^\pm}^2 = M_W^2 + \frac{\lambda v_s}{\sin\s 2\beta} \left(
\sqrt{2} A_\lambda + \kappa v_s\right) - \frac{\lambda v^2}{2} \;.
\label{eq:chargedmass}
\ee
Note that analogous to the MSSM there is an upper
bound on the squared SM-like Higgs boson mass at tree level. In the NMSSM, it
is given by
\beq
M_Z^2 \cos^2 2\beta + \frac{\lambda^2 v^2}{2} \sin^2 2\beta \;.
\label{eq:upperbound}
\eeq
In order to comply with the
measured value of $m_H = 125.09$ GeV \cite{Aad:2015zhl}, loop
corrections therefore have to be included in the computation of the
Higgs boson mass. \s 

The Higgs potential in \eqn{eq:NMSSMHiggspotential} is parametrized by
the parameter set 
\beq 
m_{H_d}^2 , m_{H_u}^2, m_S^2, g_1, g_2, v_d, v_u, v_s,
\lambda, \kappa, A_\lambda, A_\kappa \;. 
\eeq 
For a physical interpretation, it is convenient to substitute some of these
parameters with more intuitive ones, such as {\it e.g.}~the masses of gauge
bosons which are measurable quantities, or the tadpole parameters\footnote{Whether the tadpole parameters can be called physical
  quantities is debatable but certainly their introduction is
  motivated by physical interpretation.}. 
We can use Eqs.~(\ref{eq:vevs}) and (\ref{eq:tanb}) to eliminate $v_u$
and $v_d$ in favour of $v$ and $\tan \beta$, and
Eqs.~(\ref{eq:tad1})-(\ref{eq:tad3}) to replace the soft SUSY-breaking parameters $m_{H_d}^2, m_{H_u}^2$ 
and $m_S^2$ in $V_H$ with $t_{h_d}$, $t_{h_u}$ and $t_{h_s}$. Furthermore,
$A_\lambda$ can be replaced by $m_{H^\pm}^2$ using
Eq.~(\ref{eq:chargedmass}). Finally, $g_1, g_2$ and $v$ are substituted by
the squared masses $M_W^2$ and $M_Z^2$ of the $W^\pm$ and $Z$ bosons and the
electric charge $e$ via
\be
M_W^2 = \frac{v^2 g_2^2}{4},\,\,\,\, M_Z^2 = \frac{v^2 (g_1^2 +
	g_2^2)}{4},\,\,\,\,\text{and}\,\,\,\, e = \frac{g_1 g_2}{\sqrt{g_1^2
		+ g_2^2}}\,.
\ee

In summary, our set of free parameters in the Higgs sector is given by
\be
t_{h_d}, t_{h_u}, t_{h_s}, m_{H^\pm}^2, M_W^2, M_Z^2, e,  \tan \beta,
v_s, \lambda, \kappa,  A_\kappa ~.
\label{eq:freepara}
\ee
Finally, the MSSM limit of the NMSSM Higgs sector can be obtained by setting
\be
\lambda \rightarrow 0, \, \kappa \rightarrow 0, \,\,\,\,
\kappa/\lambda \equiv \text{constant},
\ee
and keeping all other parameters, including
\beq
\mu_\text{eff} \equiv \frac{\lambda v_s }{\sqrt{2}}
\eeq
and $A_{\kappa}$, fixed. In this limit the mixing between singlet and
doublet Higgs fields vanishes.

%%%%%%%%%%%%%%%%%%%%%%%%%%%%%%%%%%%%%%%%%%%%%%%%%%%%%%%%%%%%%
%------------------------------------------
\subsection{The Loop-Corrected Higgs Sector}
\label{sec:HiggsSec-Loop}
%------------------------------------------
For the determination of the loop-corrected Higgs boson masses, the
UV-divergent self-energies have to be calculated. The divergent
integrals are regularized by the SUSY-conserving dimensional
reduction scheme \cite{Siegel:1979wq,Stockinger:2005gx}. Evaluating
the self-energies in $D=4-2\epsilon$ dimensions, the divergences can
be parametrized by the regulator $\epsilon$, leading to poles
$1/\epsilon$ in the limit of $\epsilon \rightarrow 0$, \textit{i.e.}
in physical $D=4$ space-time dimensions. Also in the 
one-loop corrections to the process $H^{\pm} \to W^\pm h_i$ we
encounter UV divergences. The UV divergences are cancelled by the
renormalization of the Higgs fields and the parameters relevant for
the calculation\footnote{Note that we do not
renormalize the rotation matrices  $ R^{H^\pm}, R^{h}, R^{a}$. For
more details, {\it cf.}~\cite{Ender:2011qh}.}.  In order to do so, the
bare parameters $p_0$ of the 
Lagrangian are replaced by the renormalized ones, $p$, and their corresponding
counterterms, $\delta p$,
\begin{align}
	p_0 = p + \delta p ~.
\end{align}
Analogously, the bare fields $\phi_0$ in the Lagrangian are expressed
via the renormalized fields $\phi$ and the wave-function
renormalization constants (WFRCs) $Z_\phi$ as
\begin{align}
	\phi_0 = \sqrt{Z_\phi} \phi = \left(1 + \frac{\delta\s Z_\phi}{2}\right) \phi\,.
\end{align}
In accordance with our previous works on  higher-order corrections to
the NMSSM Higgs boson masses 
\cite{Ender:2011qh,Graf:2012hh,Muhlleitner:2014vsa,Dao:2019qaz},
we apply a mixed on-shell (OS) and $\overline{\text{DR}}$
renormalization scheme to fix the parameter and WFRCs. The free
parameters of Eq.~(\ref{eq:freepara}) are defined
to be either OS or $\overline{\text{DR}}$ parameters as follows
\be
\underbrace{t_{h_d}, t_{h_u}, t_{h_s}, m_{H^\pm}^2, m_W^2, m_Z^2, e}_{\text{OS}},
\underbrace{ \tan \beta,   v_s,\lambda,\kappa,A_{\kappa}}_{\overline{\text{DR}}} \;.
\label{eq:defparset}
\ee
The renormalization scheme for the parameters is chosen such that the
quantities which can be related to physical observables are defined
on-shell, whereas the rest of the parameters are defined as
$\overline{\text{DR}}$ parameters\footnote{The tadpoles will be
  required to minimize the potential also at higher orders and in this
  sense are called OS parameters. The electric charge is fixed
through the OS $e^+e^-\gamma$ vertex such that this vertex does not
receive any corrections at the one-loop level in the Thomson
limit. For more details, we refer {\it e.g.}
to~\cite{Graf:2012hh}. }. In addition, we introduce the 
WFRCs for the doublet and singlet fields before
rotation into the mass eigenstates as
\begin{align}
	H_d & \rightarrow \sqrt{Z_{H_d}} H_d =\left( 1+ \frac{\delta Z_{H_d}}{2}\right) H_d \\
	H_u & \rightarrow \sqrt{Z_{H_u}} H_u =\left( 1+ \frac{\delta Z_{H_u}}{2} \right)H_u \\
	S   & \rightarrow \sqrt{Z_{S}} S = \left( 1+ \frac{\delta Z_{S}}{2} \right)S.
\end{align}
We apply $\overline{\text{DR}}$ conditions for the
WFRCs of the Higgs fields. We introduce a WFRC for
the $W$ boson field, needed in the computation of the loop
corrections to the charged Higgs boson decay, as 
\be
W^\pm \to \sqrt{Z_W} W^\pm = \left( 1 + \frac{\delta Z_W}{2}\right)W^\pm \,.
\ee
The WFRC $\delta Z_W$ is defined through the OS condition
\beq
\delta Z_W = -\left.\frac{\partial \Sigma^T_{WW}}{\partial
p^2}\right|_{p^2 = M_W^2} \;,
\eeq
where $\Sigma^T_{WW}$ denotes the transverse part of the $W$ boson
self-energy. \s

%It is well-known that 
The Higgs boson masses and hence the mixing matrices receive large
radiative corrections. Therefore it is necessary to include these
corrections at the highest order possible to improve the theoretical
predictions. Recently, we completed the two-loop
order ${\cal O}(\alpha_t^2)$ corrections to the neutral Higgs boson
masses in the CP-violating NMSSM \cite{Dao:2019qaz}, thus improving
our previous results, which were available 
to two-loop order ${\cal O}(\alpha_t \alpha_s)$ \cite{Muhlleitner:2014vsa}.
The Higgs boson masses corrected up to two-loop order
${\cal O}(\alpha_t \alpha_s + \alpha_t^2)$ require also the
renormalization of the top/stop sector at one-loop order. The
computation of the two-loop corrections together with the
renormalization of the parameters in the above defined mixed
OS-$\overline{\mbox{DR}}$ scheme has been described in great detail in
~\cite{Muhlleitner:2014vsa,Dao:2019qaz}\footnote{The one- and/or
  two-loop of corrections to NMSSM Higgs boson masses were also
  studied in\cite{Ellwanger:1993hn,Elliott:1993ex,Elliott:1993uc,Elliott:1993bs,Pandita:1993tg,Ham:2001kf,Ham:2001wt,Ham:2003jf,Funakubo:2004ka,Ellwanger:2005fh,Ham:2007mt,Degrassi:2009yq,Cheung:2010ba,Staub:2010ty, Ender:2011qh,Graf:2012hh,Goodsell:2014pla,Staub:2015aea,Drechsel:2016jdg,Drechsel:2016htw,Goodsell:2016udb,Domingo:2017rhb}.}, hence we do not repeat 
it here and instead refer to these references for details. The CP-conserving
limit of these results given in the CP-violating NMSSM is
straightforward, further information can also be found in
\cite{Ender:2011qh} where the one-loop calculation is presented for
the real NMSSM. We review here, however, important points and highlight differences
for the purpose of discussing the parameter
dependence. In the following, we focus on the CP-even Higgs
bosons. Their loop-corrected masses are
defined as the real parts of the poles of the propagator matrix. These
complex poles are the zeros of the determinant of the renormalized
two-point correlation function $\hat{\Gamma}(p^2)$, where $p^2$
denotes the external squared four-momentum. The renormalized two-point
correlation function is expressed as\footnote{Here and in the
  following, the hat denotes the renormalized quantity.} 
\ba
\hat{\Gamma}  (p^2) &=& i \left( p^2 \mathbbm{1} - \hat
M^2_S(p^2,\xi) \right) \;,
\ea
with
\ba
\hat M^2(p^2,\xi) &=& \mqty(m_{h_1}^2 - \hat\Sigma_{h_1
	h_1}(p^2,\xi) &- \hat\Sigma_{h_1 h_2}(p^2,\xi)&- \hat\Sigma_{h_1
	h_3}(p^2,\xi) \\
-\hat\Sigma_{h_2 h_1}(p^2,\xi) & m_{h_2}^2 - \hat\Sigma_{h_2
	h_2}(p^2,\xi)& -\hat\Sigma_{h_2 h_3}(p^2,\xi) \\
-\hat\Sigma_{h_3 h_1}(p^2,\xi) & -\hat\Sigma_{h_3 h_2}(p^2,\xi)&
m_{h_3}^2 -\hat\Sigma_{h_3 h_3}(p^2,\xi)) \;,
\label{eq:gamma-loop}
\ea
where the renormalized self-energy $\hat{\Sigma}_{h_ih_j}(p^2,\xi)$ of
the transition $h_i\to h_j$ ($i,j=1,2,3$) is given by
\be
\hat{\Sigma}_{h_ih_j}(p^2,\xi) = \hat{\Sigma}^{1l}_{h_ih_j}(p^2,
\xi) + \hat{\Sigma}^{\alpha_t\alpha_s}_{h_ih_j}(0) +
\hat{\Sigma}^{\alpha_t^2}_{h_ih_j}(0) \;.
\ee
Here, $\hat{\Sigma}^{1l}(p^2, \xi)$ denotes the full one-loop
self-energy with full momentum-dependent contributions computed in
general $R_\xi$ gauge, where $\xi$ stands for the gauge parameters $\xi_W, \xi_Z$\footnote{We do not consider the gauge parameter $\xi _A$ of the photon which is set to unity,
        \textit{i.e.} $\xi _A = 1$. This choice 
	does not affect the results of our investigation and prevents the
	appearance of high-rank tensor loop integrals with too many
        vanishing arguments that are infrared (IR)-divergent and hence, they
        numerically blow up.}. The 
last two terms are the two-loop corrections of order ${\cal O}(\alpha_t\alpha_s)$
\cite{Muhlleitner:2014vsa} and ${\cal O}(\alpha_t^2)$
\cite{Dao:2019qaz}, respectively, which are evaluated in the
approximation of vanishing external momentum. These contributions do not introduce
additional gauge-dependent terms in the renormalized self-energies as they
are evaluated in the gaugeless limit. We want to point out that the full
one-loop renormalized self-energies $\hat{\Sigma}^{1l}(p^2, \xi)$
in general $R_\xi$ gauge are newly computed by us and implemented
in {\tt NMSSMCALC}. We computed them both in the standard tadpole scheme 
and in the Fleischer-Jegerlehner scheme\footnote{In the standard
  tadpole scheme, the tadpoles are renormalized OS while in the
  Fleischer-Jegerlehner scheme tadpoles are not renormalized
  \cite{Fleischer:1980ub,Denner:2016etu,Krause:2016oke}.} and the
results are identical.  We apply the iterative procedure described and 
applied in~\cite{Ender:2011qh} to extract the zeros of the
determinant. In the first iterative step for the calculation of the
$n^\text{th}$ CP-even Higgs boson mass, the squared external momentum in the
renormalized self-energies is chosen to be at its squared tree-level mass
value, \textit{i.e.}~$p^2 = m _{h_n} ^2$. The mass matrix
$ \hat M^2(p^2,\xi)$ is then diagonalized,
yielding the $n^\text{th}$ eigenvalue as a first approximation to the
squared mass of the $n^\text{th}$ CP-even Higgs boson. 
In the next step of the iteration, this value is taken as the new OS value for
$p^2$, and the mass matrix  is again
diagonalized. This procedure is repeated until the $n^\text{th}$
eigenvalue changes by less than $10^{-9}$ GeV$^2$
between two consecutive steps of the iteration. The resulting complex
pole is denoted by $\tilde M_{H_{n}}^2$ and the loop-corrected Higgs mass by $M_{H_n} =
\sqrt{\Re(\tilde M_{H_{n}}^2)}$. The loop-corrected CP-even Higgs
masses are sorted in ascending order $M_{H_1}<M_{H_2}<M_{H_3}$.
Note that we denote the loop-corrected masses and Higgs mass
eigenstates by capital letters $M$ and $H_i$, respectively, whereas the corresponding
tree-level values and eigenstates are denoted by lowercase letters,
{\it i.e.}~$m$ and $h_i$. 
The iterative procedure automatically mixes different orders of perturbation theory
and therefore introduces intricate gauge dependences\footnote{The
    gauge dependence of the electroweakino masses determined by the
    iterative method has been discussed in \cite{McKay:2017rjs}.}.  
This will be investigated in more detail in the
numerical section. \s

Besides the computation of the loop-corrected masses, the code {\tt
  NMSSMCALC} allows us to compute the loop-corrected mixing matrices
in several approximations which are discussed also in
\cite{Frank:2006yh}. First, we introduce the matrix $R^0$ for 
the rotation of the mass matrix in the approximation of vanishing external momentum,
\be
\text{diag}(M_{0,H_1}^2,M_{0,H_2}^2,M_{0,H_3}^2)=R^0 \hat M^2(0, \xi) (R^0)^T \;.
\label{eq:R0}
\ee
The corresponding loop-corrected mass eigenvalues are
  denoted by an index 0, hence $M_{0,H_i}^2$ ($i=1,2,3$).
In this approximation the mixing matrix $R^0$ is unitary, but does not
capture the proper OS properties of the external loop-corrected
states as momentum-dependent effects are neglected. \s

A different approach leads to the rotation matrix $R^{\mtree}$. Here we diagonalize the
mass matrix with the elements evaluated at fixed momentum squared 
which is given by the arithmetic average of the squared masses,
\beq
p_{\text{mtree}}^2 = \frac{m_{h_i}^2+m_{h_j}^2}{2} \;.
\eeq
We hence have
\be
\braket{\hat M^2(p_{\text{mtree}}^2,\xi)}_{ij}  =m_{h_i}^2\delta_{ij} -
\hat\Sigma_{h_i h_j}\braket{\frac{m_{h_i}^2+m_{h_j}^2}{2},\xi} \;,
\label{eq:arithm1}
\ee
and the corresponding mass values denoted by $M_{\text{mtree},H_i}^2$
are obtained through rotation with the matrix $R^{\mtree}$ as
\be
\text{diag}(M_{\mtree,H_1}^2,M_{\mtree,H_2}^2,M_{\mtree,H_3}^2)=
R^\mtree \hat M^2(p_{\text{mtree}}^2,\xi) (R^\mtree)^T \;.
\label{eq:arithm2}
\ee
By this approach we approximately take into account the momentum
dependence of the self-energies (see \cite{Domingo:2017rhb} for a
discussion). \s

The correct OS properties of the loop-corrected mass eigenstates are
obtained by following the procedure described in \cite{Williams:2007dc}, \textit{i.e.}~by multiplying
the tree-level matrix $R^h$ with the finite wave-function
normalization factors given by the $\bfZ$ matrix
\cite{Williams:2007dc} for external OS Higgs bosons at higher orders, 
\be
\bfZ = \begin{pmatrix}
	\sqrt{\hat{Z}_{H_1}}                  & \sqrt{\hat{Z}_{H_1}}
	\hat{Z}_{H_1H_2}                      & \sqrt{\hat{Z}_{H_1}} \hat{Z}_{H_1H_3} \\
	\sqrt{\hat{Z}_{H_2}} \hat{Z}_{H_2H_1} &
	\sqrt{\hat{Z}_{H_2}}                  & \sqrt{\hat{Z}_{H_2}} \hat{Z}_{H_2H_3} \\
	\sqrt{\hat{Z}_{H_3}} \hat{Z}_{H_3H_1} &
	\sqrt{\hat{Z}_{H_3}} \hat{Z}_{H_3H_2} & \sqrt{\hat{Z}_{H_3}}
\end{pmatrix} \;,
\label{eq:Z-resum}
\ee
where
\be
\hat{Z}_i = \frac{1}{\left( \frac{i}{ \Delta_{ii} (p^2)}
\right)^\prime (\tilde M_i^2)} \qquad \mbox{and} \qquad
\hat{Z}_{ij} = \left. \frac{ \Delta_{ij} (p^2)}{ \Delta_{ii} (p^2)}
\right|_{p^2=\tilde M_i^2} \;,
\label{eq:zfactor}
\ee
with the indices $i,j=H_1,H_2,H_3$. The prime denotes the derivative
with respect to $p^2$. The quantity
\be
\Delta = - \left[\hat{\Gamma} (p^2) \right]^{-1}
\ee
is evaluated at the complex poles $\tilde M_i^2$.
In contrast to the rotation matrices $R^0$ and $R^\mtree$, which are
unitary matrices, the $\mathbf{Z}$ matrix is not as it 
contains resummed higher-order contributions.
If we want to compute the decay width at exact
one-loop level, we therefore
have to expand the $\mathbf{Z}$ matrix and take only the one-loop terms
\begin{align}
	\mathbf{Z}^{\text{1l}} \approx \mathbb{I} + \mqty(\frac{-(\hat\Sigma_{h_1 h_1}^{1l})^\prime\qty(m_{h_1}^2,\xi)}{2} &\frac{ - \hat\Sigma_{h_1 h_2}^{1l}\qty(m_{h_1}^2,\xi)}{m_{h_1}^2 - m_{h_2}^2}  &  \frac{-\hat \Sigma_{h_1 h_3}^{1l}\qty(m_{h_1}^2,\xi)}{m_{h_1}^2 - m_{h_3}^2}               \\
	 \frac{-\hat \Sigma_{h_2 h_1}^{1l}\qty(m_{h_2}^2,\xi)}{m_{h_2}^2 - m_{h_1}^2}                            & \frac{-(\hat\Sigma_{h_2 h_2}^{1l})^\prime\qty(m_{h_2}^2,\xi)}{2}              & \frac{- \hat \Sigma_{h_2 h_3}^{1l}\qty(m_{h_2}^2,\xi)}{m_{h_2}^2 - m_{h_3}^2}               \\
	\frac{- \hat \Sigma_{h_3 h_1}^{1l}\qty(m_{h_3}^2,\xi)}{m_{h_3}^2 - m_{h_1}^2}                            & \frac{- \hat \Sigma_{h_3 h_2}^{1l}\qty(m_{h_3}^2,\xi)}{m_{h_3}^2 - m_{h_2}^2} & \frac{-(\hat\Sigma_{h_3 h_3}^{1l})^\prime\qty(m_{h_3}^2,\xi)}{2}) \label{eq:Zmatrixexpand},
\end{align}
where $(\hat \Sigma^{1l}(p^2))^\prime = \partial_{p^2} \hat\Sigma^{1l}(p^2)$. Note
that the $\hat \Sigma^{1l}$ are evaluated at the tree-level mass
values $m_{h_i}^2$, since using loop-corrected masses would introduce
higher-order effects. 

%%%%%%%%%%%%%%%%%%%%%%%%%%%%%%%%%%%%%%%%%%%%%%%%%%%%%%%%%%%%%
%--------------------------------------------------------------------
\section{Electroweak One-Loop Corrections to $H^\pm \to W^\pm h_i$ }
\label{sec:one-loop-decay}
%--------------------------------------------------------------------
%%%%%%%%%%%%%%%%%%%%%%%%%%%%%%%%%%%%%%%%%%%%%%%%%%%%%%%%%%%%%
\subsection{\label{sec:leadingorder} Decay Width at Leading Order}
The decay of the charged Higgs boson $H^\pm$ into a 
$W^\pm$ boson and a CP-even neutral Higgs boson $h_i$ ($i =
	1,2,3$) depends on the coupling
\beq
g_{W^- h_i H^+} = g_2 \qty(R^h_{i 1} \cos{\beta} - R^h_{i 2}
\sin\beta) \;,
\eeq
The leading-order (LO) decay width can be written as
\be
\Gamma_{\process{i}} = \frac{\lambda^{3/2}\left(m_{H^\pm}^2, M_W^2,
	m_{h_i}^2\right)}{64 \pi m_{H^\pm}^3
	M_W^2}\left|\mathcal{M}^{\text{tree}}_{\process{i}} \right|^2\,.
\label{eq:tot-width}
\ee
with $\lambda(x,y,z) = x^2 + y^2 + z^2 - 2xy - 2xz - 2yz$ denoting the usual
K{\"a}ll\'en function and $\mathcal{M}^{\text{tree}}_{\process{i}}$ the reduced 
matrix element, which for the tree-level decay is given by
\begin{align}
  \mathcal{M}^{\text{tree}}_{\process{i}} = i g_{W^- h_i H^+} \;. \label{eq:amp-tree}
\end{align}
We remind the reader that by $m_{h_i}$ we denote the tree-level mass
of the final state Higgs boson. 

%%%%%%%%%%%%%%%%%%%%%%%%%%%%%%%%%%%%%%%%%%%%%%%%%%%%%%%%%%%%%
%SUBSECTION: STRICT ONE LOOP CALCULATION
\subsection{\label{sec:strict-one-loop} Decay Width at Strict One-Loop
  Order}
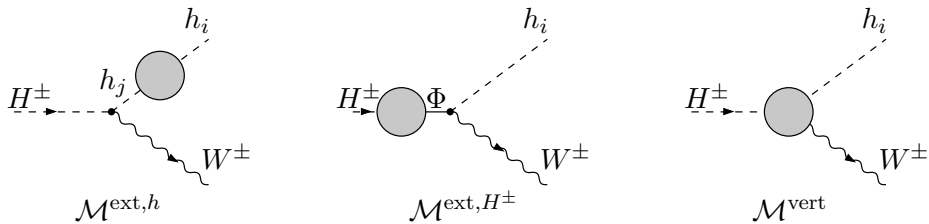
\begin{figure}
\begin{center}
	\begin{overpic}[width = 0.2 \textwidth]{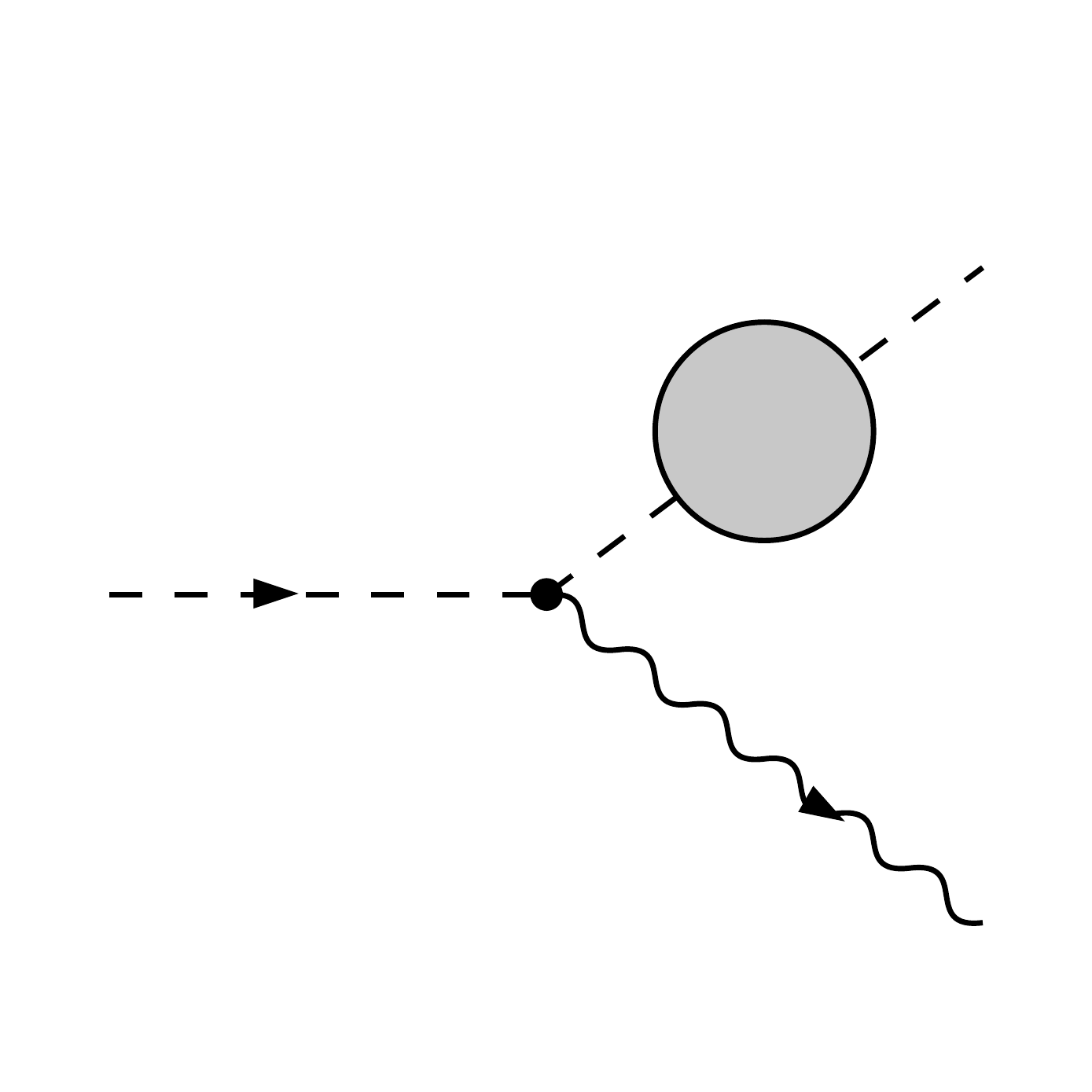}
		\put (7, 47) {$H^\pm$}
		\put (80, 80) {$h_i$}
		\put (87, 22) {$W^\pm$}
		\put (45, 55) {$h_j$}
		\put (35, 1) {$\mathcal{M}^{\text{ext}, h}$}
	\end{overpic}
\hspace*{1cm}
	\begin{overpic}[width = 0.2 \textwidth]{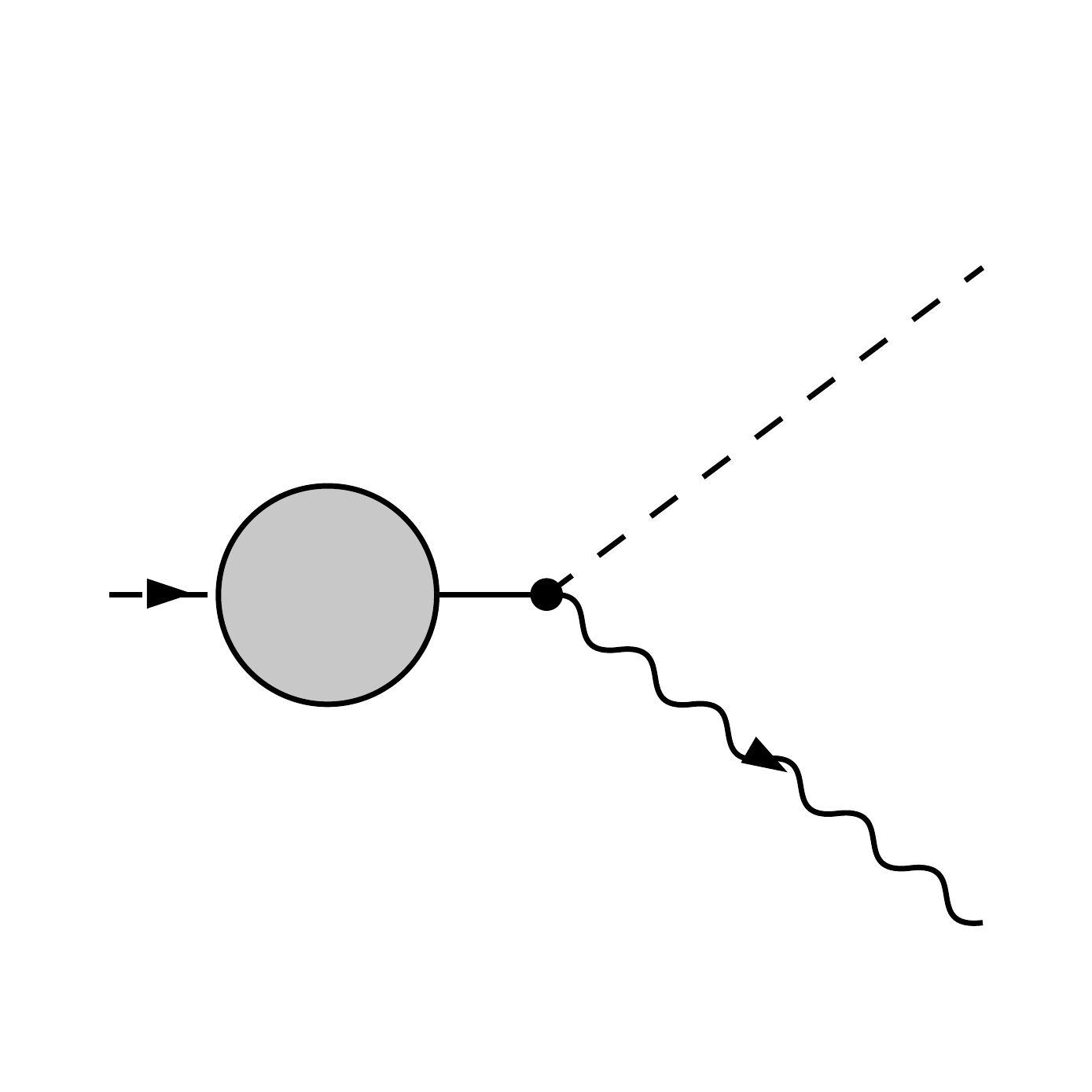}
		\put (3, 47) {$H^\pm$}
		\put (80, 80) {$h_i$}
		\put (87, 22) {$W^\pm$}
		\put (40, 47) {$\Phi$}
		%\put (0,10) {\small $\Phi \in \qty{H^\pm, G^\pm, W^\pm }$}
		\put (33, 1) {$\mathcal{M}^{\text{ext}, H^\pm}$}
	\end{overpic}
\hspace*{1cm}
	\begin{overpic}[width = 0.2 \textwidth]{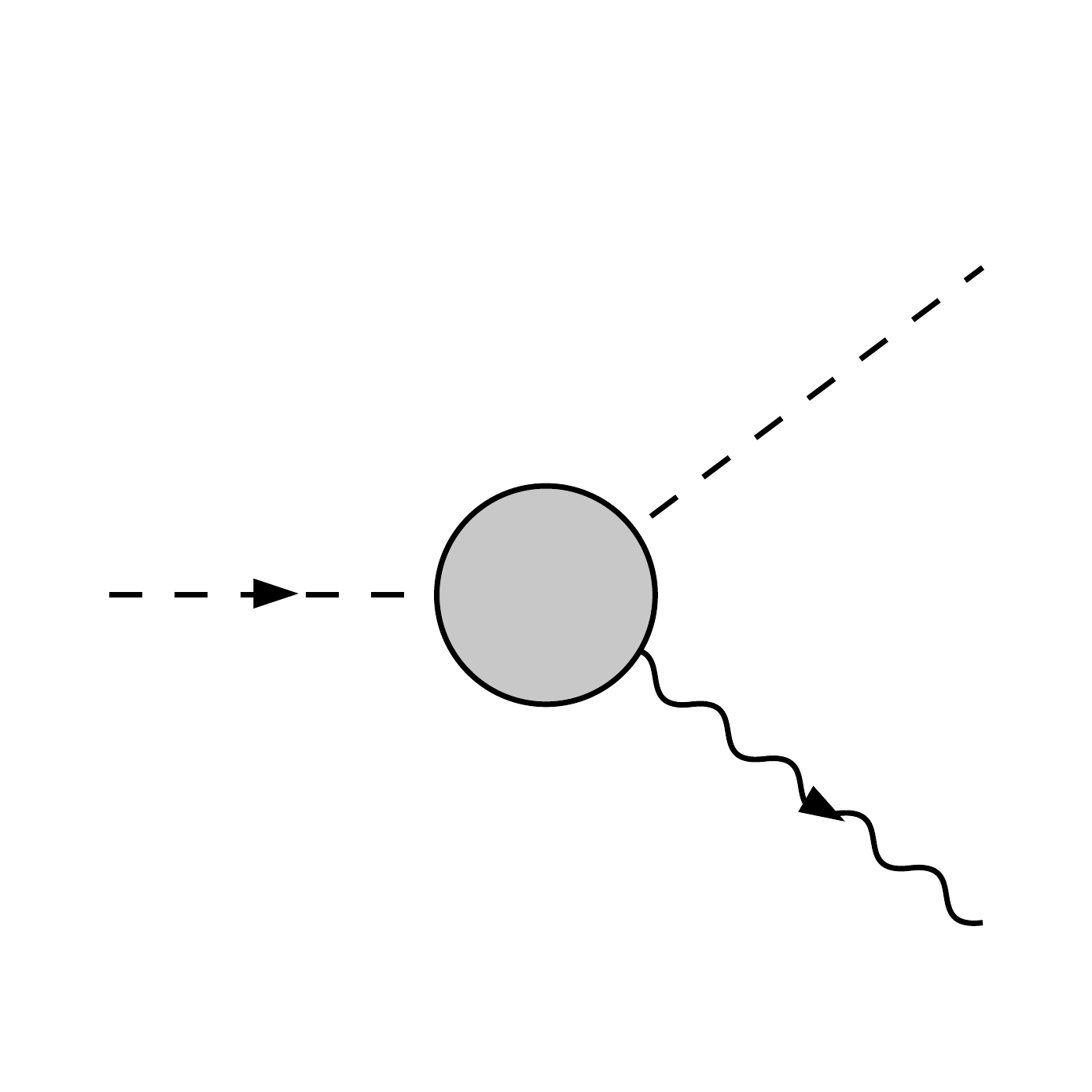}
		\put (7, 47) {$H^\pm$}
		\put (80, 80) {$h_i$}
		\put (87, 22) {$W^\pm$}
		\put (35, 1) {$\mathcal{M}^\text{vert}$}
	\end{overpic}
%	\begin{overpic}[width = 0.205 \textwidth]{diag/HWhCT.pdf}
%				\put (30,50) {$H^\pm$}
%				\put (63,65) {$h_i$}
%				\put (63,18) {$W^\pm$}
%		\put (50, 1) {$\mathcal{M}^\text{CT}$}
%			\end{overpic}
	%	\begin{overpic}[width = 0.245 \textwidth]{diag/extW.pdf}
	%		\put (7, 47) {$H^\pm$}
	%		\put (80, 80) {$h_i$}
	%		\put (80, 25) {$W^\pm$}
	%		\put (47, 33) {$\Phi$}
	%		%\put (0,10) {\small $\Phi \in \qty{H^\pm, G^\pm, W^\pm }$}
	%		\put (50, 1) {$\mathcal{M}^{\text{ext.} W^\pm}$}
	%	\end{overpic}
\end{center}
\caption{Generic diagrams contributing to $\mathcal{M}^\text{virt}$:
  corrections to the external 
  legs of $h_i$ and $H^\pm$,  $\mathcal{M}^{\text{ext}, h}$ and
  $\mathcal{M}^{\text{ext}, H^\pm}$, and genuine vertex corrections
  $\mathcal{M}^\text{vert}$. $\Phi$ stands for the fields $\Phi \in \{
  H^\pm, G^\pm, W^\pm \}$.} 
	\label{fig:one-loop}
\end{figure}

The next-to-leading order (NLO) decay width $\Gamma ^\text{NLO}$ is
given by the sum of the LO width $\Gamma ^\text{LO}$, the
virtual corrections $\Gamma ^\text{virt}$ and the real corrections
$\Gamma ^\text{real}$ as 
\beq
\Gamma^{\text{NLO}} = \Gamma^{\text{LO}} + \Gamma^{\text{virt}} +
\Gamma^{\text{real}} \;.
\eeq
The virtual corrections contain the counterterm contributions that
cancel the UV divergences. The IR divergences in the real
corrections cancel those in the virtual corrections. $\Gamma^{\text{virt}}$ is given by
\beq
\Gamma^\text{virt} = \frac{\lambda^{3/2}\left(m_{H^\pm}^2,
M_W^2,m_{h_i}^2\right)}{64\pi m_{H^\pm}^3 M_W^2} 
 2\,  \text{Re}\qty(\mathcal{M}^{\text{tree}*}_{\process{i}}
 \mathcal{M}^{\text{virt}}) \;.
\label{eq:one-loop-virt}
\eeq
Note that in Eq.~(\ref{eq:one-loop-virt}) we set
$p_{h_i}^2=m_{h_i}^2$, {\it i.e.}~we set the external Higgs boson on
its tree-level mass shell. We do the same in the LO
  decay width and the real corrections. In this way, we ensure that the NLO decay
width $\Gamma^{\text{NLO}}$ remains at strict one-loop order
by avoiding admixtures of higher orders through loop corrections to the
mass. $\mathcal{M}^\text{virt}$ consists of the
one-particle-irreducible (1PI) diagrams depicted in
\fig{fig:one-loop}. They show the external leg corrections
$\mathcal{M}^{\text{ext},h}$ and $\mathcal{M}^{\text{ext},H^\pm}$ to
the neutral Higgs boson and to the charged Higgs boson, respectively, and the genuine
  vertex corrections $\mathcal{M}^\text{vert}$.
They already include the counterterms and are hence finite.  
The corrections to the $W$ boson leg vanish due to the OS
renormalization of the $W$ boson. We hence have 
\be
\mathcal{M}^\text{virt}=\mathcal{M}^\text{vert}+\mathcal{M}^{\text{ext},
  H^\pm}+ \mathcal{M}^{\text{ext}, h} \;. 
\label{eq:mvirt}
\ee 

The amplitudes for the external leg contributions to the neutral and
charged Higgs bosons, $\mathcal{M}^{\text{ext}, h}$ and
$\mathcal{M}^{\text{ext}, H^\pm}$, respectively, can be factorized into the
tree-level amplitude and the propagator corrections to the external
legs. For $\mathcal{M}^{\text{ext}, h}$ we obtain 
\begin{align}
\mathcal{M}^{\text{ext}, h} & = \sum_{j = 1}^3\delta\mathbf{Z}^{\text{1l}}_{ij}
\mathcal{M}^\text{tree}_{\process{j}} \;.
\label{eq:Mexth}
\end{align}
Note that here we apply $\delta\mathbf{Z}^{\text{1l}}_{ij}$ at strict one-loop
order, defined as
\beq
\delta\mathbf{Z}^{\text{1l}}_{ij} =\mathbf{Z}^{\text{1l}}_{ij}- \delta_{ij} \;,
\eeq
with the $\mathbf{Z}$ matrix at strict one-loop order,
$\mathbf{Z}^{\text{1l}}$, given in Eq.~\eqref{eq:Zmatrixexpand}. The 
charged Higgs WFRC is determined in
the $\overline{\mbox{DR}}$ scheme so that there are finite
contributions to the LSZ factor $\sqrt{\hat{Z}_{H^\pm}}$ at one-loop order,
\begin{align}
\sqrt{\hat{Z}_{H^\pm}} & 
\approx 1 - \frac{\delta \hat Z_{H^\pm}}{2}
= 1 - \frac{\text{Re} \Sigma_{H^+ H^-}^\prime(m_{H^\pm}^2)}{2} - \frac{\cos^2\beta \,\delta Z_{H_d} + \sin^2\beta\, \delta Z_{H_u}}{2}\,,
	\label{eq:ZHpm}
\end{align}
where the prime denotes the
derivative with respect to the squared four-momentum. \s

The mixing between $H^\pm$ and $G^\pm$ can be related to the mixing
between $H^\pm$ and $W^\pm$ by using the Slavnov-Taylor identity for
the renormalized self-energies, 
\begin{align}
\hat \Sigma_{H^+ G^-}(m_{H^\pm}^2) = \frac{m_{H^\pm}^2}{M_W}
  \hat \Sigma_{H^+ W^-}(m_{H^\pm}^2) \;,
\end{align}
where $\hat{\Sigma}_{H^+ W^-}(m_{H^\pm}^2)$ denotes the renormalized
truncated self-energy of the transition $H^+\,\to\,W^+$. The
correction to the $H^\pm$ propagator thereby results in 
\beq
\mathcal{M}^{\text{ext},H^\pm} = & \frac{\delta \hat Z_{H^\pm}}{2}
 \mathcal{M}^\text{tree}_{\process{i}}+
 \frac{1}{M_W^2}  \hat 
\Sigma_{H^+ W^-}(m_{H^\pm}^2) \, g_{h_i WW}\,,
\eeq 
where the coupling is given by
\be  
g_{h_i WW} =\frac{g_2^2}{2} v (c_\beta R_{i1} + s_\beta R_{i2}) \;.
\ee

\begin{figure}
	\center
	\includegraphics[width=0.87\textwidth]{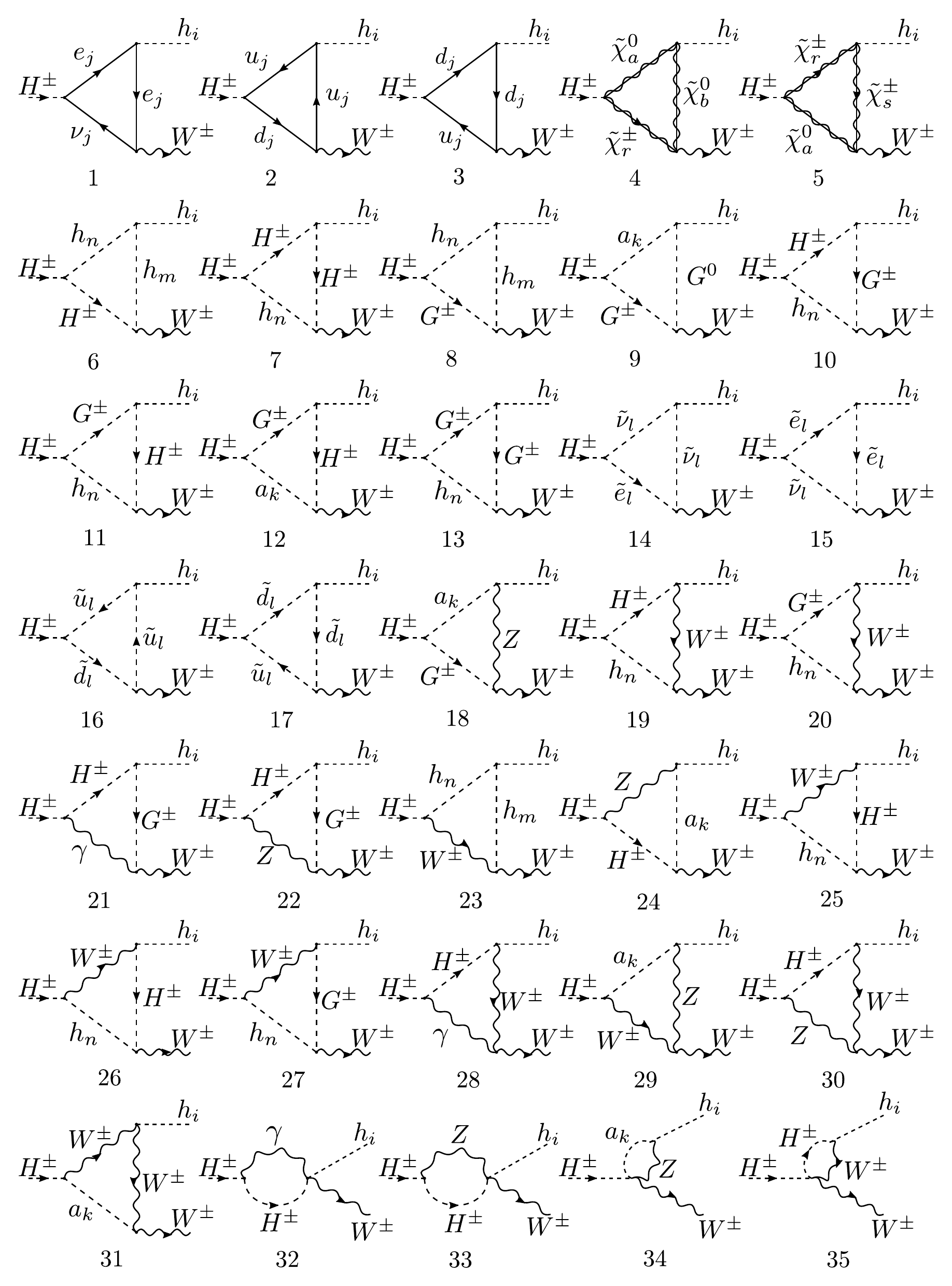}
	\caption{One-loop diagrams contributing to the pure vertex
          corrections at NLO. Here $u_j,d_j$
          denote up- and down-type quarks and $\nu_j,e_j$
          neutrinos and charged leptons for all three generations
          ($j=1,2,3$), $\tilde{u}_l, \tilde{d}_l$ denote up- and
          down-type squarks and $\tilde{e}_l,\tilde{\nu } _l$ denote
          charged sleptons and sneutrinos ($l=1,...,6$), $a_k$
          ($k=1,2$) denote the  pseudoscalar and $h_{i,m,n}$ 
          ($i,m,n=1,2,3$) the scalar Higgs bosons, and
        $\tilde{\chi}_{a,b}^0$ ($a,b=1,...,5$) and
        $\tilde{\chi}_{r,s}^\pm$ ($r,s=1,2$) represent the neutralinos
        and charginos, respectively.} 
	\label{fig:vertex-corrections}
\end{figure}

The genuine vertex corrections $\mathcal{M}^\text{vert}$ are given by
the diagrams depicted in \fig{fig:vertex-corrections} plus the
corresponding counterterm contributions that are not shown here. The vertex
corrections comprise the 1PI diagrams given by the
triangle diagrams with scalars, fermions and gauge bosons in the
loops, and the diagrams involving four-particle vertices. 
The counterterm amplitude is given by 
\beq
{\cal M}^{\text{CT}} &= & i\frac{\delta g_2}{2} \qty(R^h_{i 1}
\cos{\beta} - R^h_{i 2} \sin\beta) +  i\frac{g_2}{2}
\qty(R^h_{i 1} \cos{\beta} \delta Z_{H_d} -  
R^h_{i 2} {\sin\hspace{-2pt}\beta \hspace{2pt}} {\delta Z_{H_u}}) \nonumber \\
&+& i\frac{g_2}{2} \qty(R^h_{i 1} \cos{\beta} - R^h_{i 2}
\sin\beta) \frac{\delta Z_W}{2} \;,
	\label{eq:CTvertex}
\eeq
in terms of the WFRCs $\delta Z_{H_d}$, $\delta Z_{H_u}$ and $\delta Z_W$ and the
counterterm $\delta g_2$ for the $SU(2)_L$ gauge coupling constant
$g_2$, which in terms of the counterterms of our input parameters,
{\it cf}.~Eq.~(\ref{eq:defparset}), reads
\beq
	\delta g_2 = \delta \left(\frac{e}{\sin{\theta_W}} \right) =
        \frac{g_2}{e} \delta e - g_2 \frac{\cos^2\theta_W}{2
          \sin^2\theta_W}\qty(\frac{\delta M_Z^2}{M_Z^2} -
        \frac{\delta M_W^2}{M_W^2}) \;. 
\eeq

\begin{figure}
\begin{center}
	\includegraphics[scale=0.3]{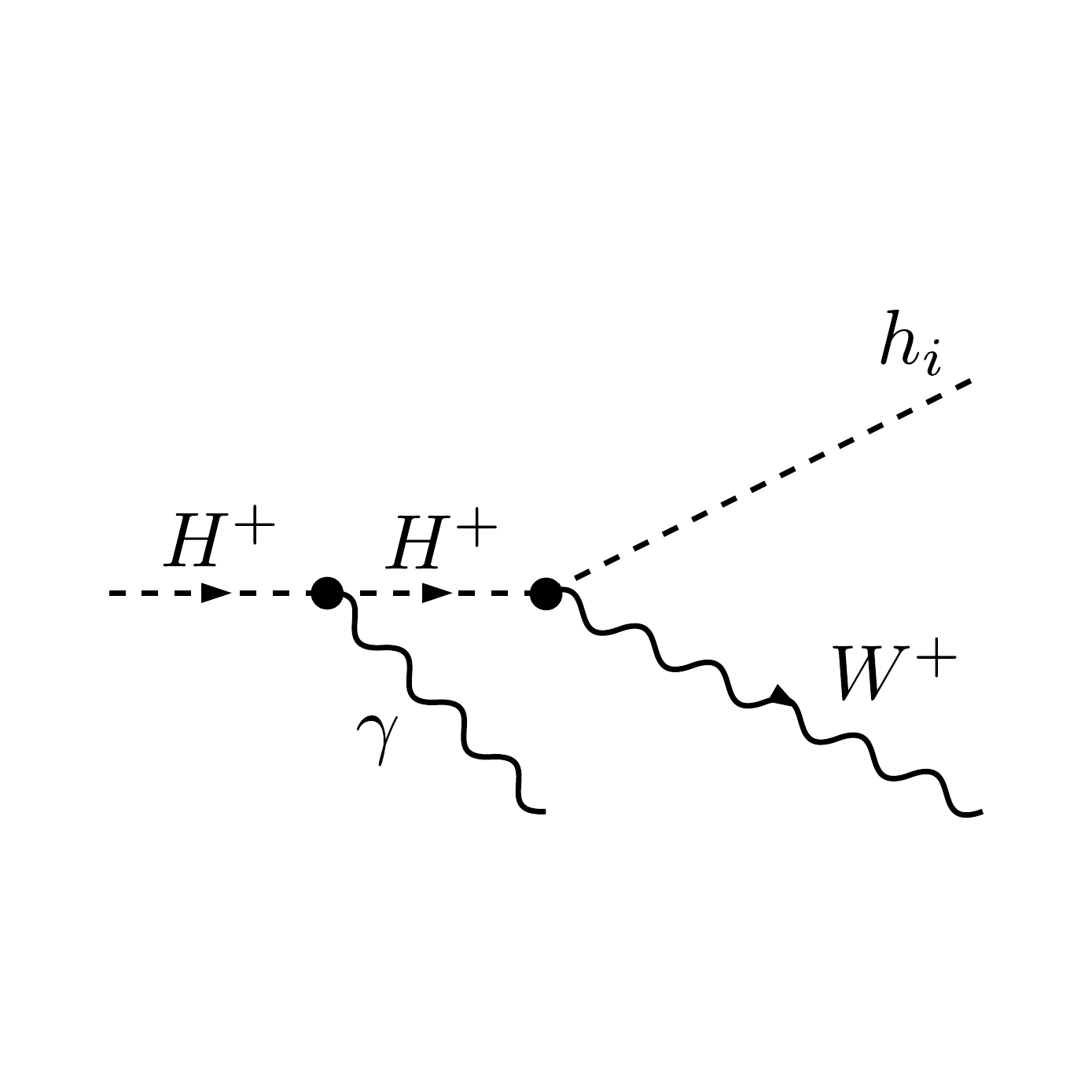}
	\includegraphics[scale=0.3]{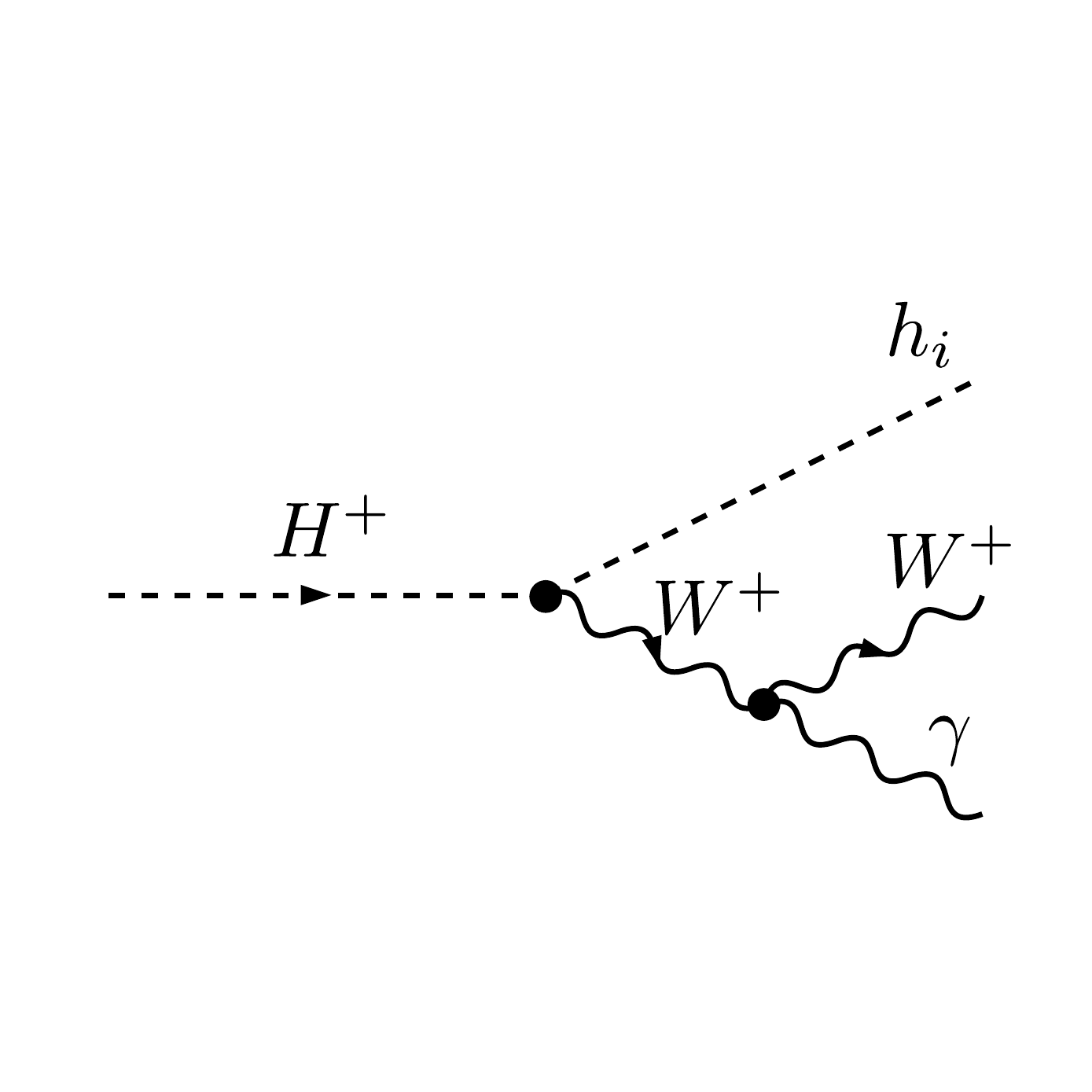}
	\includegraphics[scale=0.3]{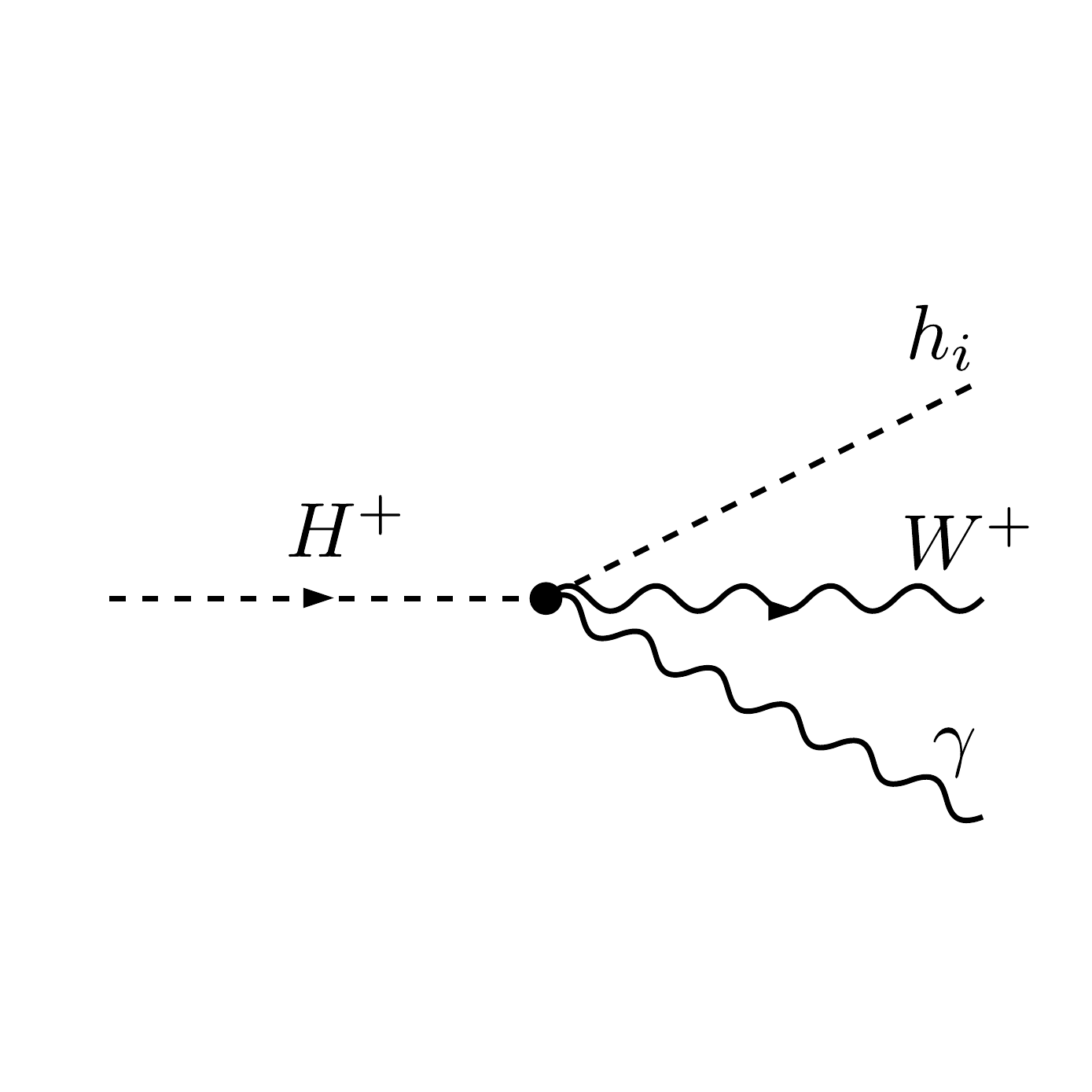}
	\caption{Feynman diagrams contributing to \process{i}$\gamma$
          in the unitary gauge.} 
	\label{fig:realcorr}
\end{center}
\end{figure}

The vertex diagrams also contain IR divergences. These arise from the
exchange of a soft virtual photon between the external legs ({\it
  cf.}~diagrams 21 and 28 of \fig{fig:vertex-corrections}). Also
$\delta M_W^2$, $\delta Z_W$ and $\hat Z_{H^+ H^-}$ are IR-divergent.
These soft singularities in the virtual corrections are canceled by
the IR-divergent contributions from real photon 
radiation \cite{Kinoshita:1975bt, lee1964degenerate} in the process
\process{i}$\gamma$. The process is independently gauge-invariant and
most easily calculated in unitary gauge. The diagrams that contribute
in unitary gauge to the process are shown in \fig{fig:realcorr}. They consist of the
proper bremsstrahlung contributions, where a photon is radiated from
the charged initial and final state particles, and the diagram
involving a four-particle vertex with a photon. 
The decay width for the real emission is given by
\beq
\Gamma^{\text{real}}\equiv\Gamma_{\process{i}\gamma} &=& \frac{\alpha
  g^2_{W^-h_iH^+}}{4 \pi^2 m_{H^\pm}}
\bigg[  -m_{H^\pm}^2 I_{H^\pm H^\pm} + \qty(m_{h_i}^2 - m_{H^\pm}^2
- M_W^2) I_{H^\pm W} - M_W^2 I_{WW} \nn \\ 
&& - I_{H^\pm} - I_{W}+\frac{2
  M_W^2}{\lambda(m_{H^\pm}^2,M_W^2,m_{h_i}^2) }\qty(I_{WW}^{H^\pm 
  H^\pm} + 2 I_W^{H^\pm} + I)\bigg], 
\label{eq:one-loop-real}
\eeq
in terms of the bremsstrahlung integrals \cite{0709.1075v1}
\beq
I_{i_1 \ldots i_n}^{j_1 \ldots j_m} = \frac{1}{\pi^2} \int
\frac{d^3 p_{W}}{2 E_W} \frac{d^3 p_{h_i}}{2
  E_{h_i}} \frac{d^3 p_{\gamma}}{2 
  E_\gamma} \delta\qty(p_{H^\pm} - (p_W + p_{h_i} +
  p_\gamma)) \frac{(\pm 
  2 p_\gamma  p_{j_1}) \ldots (\pm 2 p_\gamma 
  p_{j_m})}{(\pm 2 p_\gamma  p_{i_1}) \ldots (\pm 2 p_\gamma
   p_{i_n})}\,,
\eeq
where $E_W$, $E_{h_i}$ and $E_\gamma$ denote the energies of the
corresponding particles and the plus sign corresponds to the outgoing
momenta $p_W$ and 
$p_{h_i}$ while the minus sign belongs to the incoming momentum
$p_{H^\pm}$. No upper (lower) index $j_k$ ($i_l$) means that the 
corresponding $\pm 2p_\gamma p_{j_k}$ $(\pm 2p_\gamma p_{i_l})$ in the numerator
(denominator) of the fraction is replaced by 1. 
The total \nlo{} width $\Gamma^{\text{\nlo{}}}_{\process{i}}$ is then
both UV- and IR-finite. Furthermore, since $\Gamma^{\text{\nlo{}}}$ has been
calculated at strict one-loop level with $p_h^2 = m_{h_i}^2$, it is
also independent of the gauge parameter as we explicitly checked.
\s

We finish with the remark that for the computation of the loop-corrected
decay width we used a {\tt{FeynArts-3.10}} \cite{HAHN2001418} model
file for the NMSSM generated by
{\tt{SARAH-4.12.3}}\cite{Staub:2009bi,Staub:2010jh,Staub:2013tta,Staub:2012pb}. The
various pieces of the one-loop corrected decay width were obtained
with the help of 
{\tt{FormCalc-9.6}} \cite{Hahn:1998yk}, {\tt{FeynCalc-8.2.0}}
\cite{Mertig:1990an,1604.06709v1} and {\tt{LoopTools-2.14}}
\cite{Hahn:1998yk}. Both for the computation of the loop-corrected
Higgs boson masses and the decay widths two independent calculations
have been performed which are in full agreement.

%%%%%%%%%%%%%%%%%%%%%%%%%%%%%%%%%%%%%%%%%%%%%%%%%%%%%%%%%%%%%%
\subsection{\label{sec:gaugedep} The Issue of Gauge Dependence} 
Phenomenology requires that the loop corrections to the masses of the
external Higgs bosons  should be taken into account. This is
particularly important when the external Higgs boson is the SM-like scalar,
as its upper mass bound at tree level is well below the measured value
of 125.09~GeV. Therefore, in the decay of the charged 
Higgs boson $H^\pm$ into a final state with neutral Higgs bosons, we
should consider the loop-corrected Higgs states $H_i$ ($i=1,2,3$)
with the corresponding loop-corrected masses $M_{H_i}$. For
the decay $H^\pm \to W^\pm h_i$, this means that we should set the external
momentum to $p^2 = M_{H_i}^2$. However, this introduces contributions
beyond the one-loop order in the one-loop decay width, which has two
implications. First, it invalidates the tree-level relation
between the couplings of the charged and neutral Higgs bosons with 
a charged Goldstone boson or a $W^\pm $ boson, {\it
  i.e.}~between the couplings $g_{h_i H^- G^+}$ and $g_{h_i H^-
  W^+}$. This relation needs to be satisfied, however, in order to cancel the
IR divergences occuring in the decay $H^\pm \to W^\pm h_i$ at one-loop
order. Additionally the relation between $g_{h_i a_j G^0}$ and $g_{h_i a_jZ}$ is spoiled, but
this does not influence IR divergence. Second, the 
introduction of loop-corrected neutral 
Higgs masses leads to mixing of different orders of perturbation theory, and the
gauge independence of the matrix element is no longer
guaranteed, and it is indeed violated as will be shown in the
following. The problem of gauge dependences arising from
loop-corrected Higgs masses, and their restoration via the inclusion of partial
two-loop terms in the case of neutral Higgs decays in the NMSSM
with complex parameters was recently discussed in
\cite{Domingo:2018uim}. In this work, the gauge dependence arose from the
mixing of the neutral Higgs bosons with the $Z$ boson. This does not
apply to our work as we are working in the CP-conserving NMSSM
and we are considering  only decays into CP-even neutral Higgs bosons. The gauge
dependence in our case originates from other sources and cannot be
remedied easily, if at all. \s

In this paper, we want to investigate the impact of this gauge dependence on the treatment of the wave-function normalization factors and on the parameters of the model. We
also investigate the issue of the gauge dependence of the
loop-corrected Higgs masses themselves. As long as there is no recipe on
how to  achieve gauge-independent results\footnote{Since the
  gauge dependence
  arises from the admixture of different loop orders it cannot be
  cured at fixed order in perturbation theory and most probably
  requires the resummation of all loop orders which is well beyond the
  scope of this paper.}, the value of the gauge parameter
applied in the computation of a specific Higgs 
observable needs to be specified in order to consistently relate measured
observables with the parameters of the underlying model. 

%%%%%%%%%%%%%%%%%%%%%%%%%%%%%%%%%%%%%%%%%%%%%%%%%%%%%%%%%%%%%%
\subsection{\label{sec:improved-one-loop}Decay Width at Improved One-Loop Order} 
In the following, we look more closely into the relation between the
gauge dependence of the loop-corrected decay $H^\pm \to W^\pm h_i$ and
the  treatment of the external Higgs boson, in particular the
treatment of the {\bf Z} matrix. While curing the IR
divergence beyond strict one-loop order is fairly straightforward, the
intricacies of gauge dependence in our calculation with respect to setting
$p_h^2=M_{H_i}^2$ are much more involved. In order to study this in
more detail, we proceed in two steps:
\begin{enumerate}
\item In the first instance, we modify our result obtained in
  Sec.~\ref{sec:strict-one-loop} by changing $p_h^2$ from the tree-level
  value $m_{h_i}^2$ to the loop-corrected one $M_{H_i}^2$, and
  ensure that all IR divergences cancel by enforcing the correct
  relations between the gauge couplings $g_{h_i H^- G^+}$ and $g_{h_i
    H^- W^+}$ beyond tree level. However, we retain the use of the
  one-loop diagrammatic expansion of the $\bold{Z}$~matrix as applied in
  Eq.~(\ref{eq:Mexth}). It is clear that in order to get correct OS
  properties for the external neutral Higgs boson, we need to make use
  of the resummed $\bold{Z}$~matrix defined in Eq.~(\ref{eq:Z-resum}).
  However, it is instructive to demonstrate the
  breaking of the gauge symmetry that occurs simply by using
  loop-corrected masses $M_{H_i}^2$, before we discuss the full result
  obtained by using the resummed $\bold{Z}$~matrix.

\item For the next step, we set $p_h^2 =M_{H_i}^2$ and apply the
  resummed $\bold{Z}$~matrix in our calculation, treating it as a part
  of the LO amplitude. This means that we no longer need to
  explicitly include external leg corrections
  $\mathcal{M}^{\text{ext},h}$. As a result of this 
  modification we will be required to include $\bold{Z}$~factors also
  for the real corrections, as well as to modify the gauge coupling
  relation between $g_{h_i H^- G^+}$ and $g_{h_i H^- W^+}$, in order
  to obtain an IR-finite result. 
\end{enumerate}

%\vspace{1mm}
\noindent\textbf{Step 1:} The first modification of our strict one-loop
decay width consists of calculating
$\Gamma_{\process{i}}^{\text{virt}}$ in Eq.~(\ref{eq:one-loop-virt})
and $\Gamma^{\text{real}}$ in Eq.~(\ref{eq:one-loop-real}) with $p_h^2
= M_{H_i}^2$\footnote{In order to avoid confusion
    we denote the decay always by $H^\pm \to W^\pm h_i$, {\it i.e.}~we
  use lowercase $h_i$ and not capital $H_i$ also when we include loop
  corrections in the mass of the external neutral Higgs boson. Only the
  notation for its mass is changed from lowercase to capital
  letter. From this and the text, it will always become clear how we
  treat the external neutral Higgs boson.}. The reduced matrix element
$\mathcal{M}^{\text{tree}}_{\process{i}}$ does not depend on $p_h^2$,
so this modification does not affect its gauge independence. Similarly,
the real decay width $\Gamma^{\text{real}}$ is separately gauge independent
even when computed at $p_h^2 = M_{H_i}^2$. The virtual amplitude
$\mathcal{M}^{\text{virt}}$, however, which is gauge independent when
computed at strict one-loop order, acquires a dependence on the
gauge parameters due to higher-order mass effects. Moreover,
the cancelation of the IR divergences in
$\Gamma_{\process{i}}^\text{\nlo{}}$ only takes place when the relation 
\begin{align}
M_W g_{h_i H^- G^+} = \qty(m_{H^\pm}^2 - p_h^2) g_{h_i H^- W^+}\,, \label{eq:enforce}
\end{align}
is satisfied. Due to the gauge structure of the Lagrangian, the relation
\begin{align}
M_W g_{h_i H^- G^+} = \qty(m_{H^\pm}^2 - m_{h_i}^2) g_{h_i H^- W^+}
\end{align}
holds, with $m_{h_i}^2$ being the tree-level CP-even Higgs mass
calculated from Eq.\eqref{eq:tree-level-masses}, {\it
  cf.}~\cite{PAM_doi:10.1140/epjc/s10052-017-5259-x}. We can enforce
the relation Eq.~\eqref{eq:enforce} beyond tree level by modifying the
coupling $g_{h_i H^- G^+}$ where necessary\footnote{We do not change it in
UV-divergent diagrams {\it e.g.} in the external leg corrections to
$h_i$ as that would lead to UV divergences.} such that it is expressed
in terms of the loop-corrected masses, 
\begin{align}
M_W g_{h_i H^- G^+} = \qty(m_{H^\pm}^2 - M_{H_i}^2) g_{h_i H^- W^+}.
	\label{eq:gauge-coup-mod}
\end{align}
This is equivalent to an effective potential approach
\cite{PAM_doi:10.1140/epjc/s10052-017-5259-x}. This modification
ensures an IR-finite \nlo{} width. The decay width obtained with these
modifications will be referred to as ``off-shell'' and denoted as
$\Gamma_{H^\pm \to W^\pm h_i}^{\text{NLO,off-shell}}$. This nomenclature points towards
the fact that the external loop-corrected neutral Higgs boson does not
have the correct OS properties yet. We emphasize that
while $\Gamma_{\process{i}}^{\text{NLO,off-shell}}$ is UV- and IR-finite,
the modification of the coupling constants in
Eq.~(\ref{eq:gauge-coup-mod}) does not restore gauge independence.
The global modification of the Goldstone couplings $g_{G^+G^-h_i}$,
$g_{G^+W^-h_i}$, $g_{G^0G^0h_i}$ and $g_{G^0a_j h_i}$ is not possible
while keeping the result UV-finite, such that gauge independence
cannot be restored by a modification of these couplings. Additionally,
we have to deal with the gauge dependence of the loop-corrected Higgs
masses themselves. \s

\noindent\textbf{Step 2:}
The corrections from $\mathcal{M}^{\text{ext},h}$ to the "off-shell"
$\Gamma_{\process{i}}^{\text{\nlo{}}}$ can be large. In order to
obtain numerically stable \nlo{} corrections\footnote{We use the term
  'numerically stable' in the sense that the NLO corrections to the LO
decay width do not blow up so that the convergence of the
higher-order corrections must be questioned.} and to
ensure that the external neutral Higgs bosons have proper OS
properties, we need to use the full $\bold{Z}$ matrix defined in
Eq.~(\ref{eq:Z-resum}) which resums higher-order contributions to the
external leg corrections, and treat it as part of the LO
amplitude. The second modification to our 
strict one-loop computation therefore consists of not only using
$p_h^2 = M_{H_i}^2$, but also employing the resummed $\bold{Z}$
factors. The LO and the virtual amplitude Eq.~(\ref{eq:mvirt}) are now
computed as 
\beq
\mathcal{M}_{H^\pm \to W^\pm h_i}^{\text{tree,impr}} \approx
\sum_{j=1}^3 \mathbf{Z}_{ij} 
  \mathcal{M}_{\process{j}}^\text{tree}
\label{eq:loamp-improved}
\eeq
and
\beq
\mathcal{M}^{\text{virt,impr}}_{H^\pm \to W^\pm h_i}
\approx 
\sum_{k=1}^3 \mathbf{Z}_{ik} \qty(\mathcal{M}_{\process{k}}^\text{vert}
  + \mathcal{M}_{\process{k}}^{\text{ext.}H^\pm})\,. \label{eq:virtamp-improved}
\eeq
Including the resummed $\bold{Z}$ factors in the \lo{} amplitude means
that $\mathcal{M}^{\text{ext},h}$ does not have to be calculated anymore, and
that the virtual \nlo{} amplitude contains contributions that are
formally of two-loop order and higher. We refer to these amplitudes as
``improved'' amplitudes and to their corresponding widths as
``improved'' widths. They are given by 
\beq
\Gamma^{\text{LO,impr}} &=& \frac{\lambda^{3/2}\left(m_{H^\pm}^2,
M_W^2,M_{H_i}^2\right)}{64\pi m_{H^\pm}^3 M_W^2} 
 \left|\mathcal{M}^{\text{tree,impr}}_{H^\pm \to W^\pm h_i}\right|^2
\label{eq:gamloimpr} 
\\
\Gamma^\text{virt,impr} &=& \frac{\lambda^{3/2}\left(m_{H^\pm}^2,
M_W^2,M_{H_i}^2\right)}{64\pi m_{H^\pm}^3 M_W}^2
 2\,  \text{Re}\qty(\mathcal{M}^{\text{tree,impr}*}_{H^\pm \to W^\pm h_i}
 \mathcal{M}^{\text{virt,impr}}_{H^\pm \to W^\pm h_i})
\eeq
The $\mathbf{Z}$ matrix and the loop-corrected masses are obtained
from the program {\tt{NMSSMCALC}}
\cite{Ender:2011qh,Graf:2012hh,Muhlleitner:2014vsa,Dao:2019qaz,1312.4788v1,King:2015oxa,Djouadi:325078,Djouadi:2018xqq}
with the new implementation of the full one-loop renormalized
self-energies in general $R_\xi$ gauge as discussed in
subsection~\ref{sec:HiggsSec-Loop}.\s 

The inclusion of resummed higher-order corrections to the external
neutral Higgs boson via the full $\bold{Z}$ factor also needs to be
accounted for in the real corrections, so that the IR divergences
cancel properly. This means that we have  
\beq
\Gamma^{\text{real,impr}}_{H^\pm \to W^\pm h_i\gamma} &=& \frac{\alpha
  |\sum_{j=1}^3 \mathbf{Z}_{ij} g_{W^-h_j H^+}|^2}{4
  \pi^2 m_{H^\pm}}   \bigg[  -m_{H^\pm}^2 I_{H^\pm H^\pm} +
\qty(M_{H_i}^2 - m_{H^\pm}^2 - M_W^2) I_{H^\pm W} \nn 
\\  && - M_W^2 I_{WW} - I_{H^\pm} - I_{W}
+\frac{2 M_W^2}{\lambda(m_{H^\pm}^2,M_W^2,M_{H_i}^2)}
\qty(I_{WW}^{H^\pm H^\pm} + 2 I_W^{H^\pm} + I)\bigg] \;.
\label{eq:realcorr}
\eeq
Finally, since we set $p_h^2 = M_{H_i}^2$, we need to use the same
modified gauge couplings that we introduced in Eq.~(\ref{eq:gauge-coup-mod})
in order to cure the breaking of IR finiteness caused by using
loop-corrected masses. We refer to this width as ``improved'' real
corrections. The complete width obtained after these
modifications will henceforth be referred to as the ``improved''
\nlo{} width and is given by 
\beq 
\Gamma_{H^\pm \to W^\pm h_i}^{\text{NLO,impr}} =
\Gamma^{\text{LO,impr}}
+ \Gamma^{\text{1l,impr}} \equiv
\Gamma^{\text{LO,impr}} + \Gamma^{\text{real,impr}} +
\Gamma^{\text{virt,impr}} \;. \label{eq:gamnloimpr} 
\eeq 

%%%%%%%%%%%%%%%%%%%%%%%%%%%%%%%%%%%%%%%%%%%%%%%%%%%%%%%%%%%%%
\section{Numerical Results}
\label{sec:results}
In this section, we will investigate in detail the gauge
dependence of our results. We start by studying the gauge dependence
of the loop-corrected neutral Higgs boson masses and then investigate
the gauge dependence of the virtual corrections to the decay $H^\pm
\to W^\pm h_i$ before studying the gauge dependence of the complete
NLO width, by applying various treatments of the external Higgs
bosons. We do this for parameter points obtained from a scan in the
NMSSM parameter space as described in the following.

%%%%%%%%%%%%%%%%%%%%%%%%%%%%%%%%%%%%%%%%%%%%%%%%%%%%%%%%%%%%%
\newcommand{\DRb}{\overline{\text{DR}}}
\newcommand{\ti}{\tilde}
\newcommand{\msusy}{M_{\text{SUSY}} }
\newcommand{\mg}{m_{\ti{g}}}
\newcommand{\msbl}{m_{\ti{b}_1}}
\newcommand{\msbh}{m_{\ti{b}_2}}
\newcommand{\gahbb}{\Gamma_{hbb}}
\newcommand{\mHi}{m_{H_i}}
\newcommand{\mHj}{m_{H_j}}
\newcommand{\mhi}{m_{h_i}}
\newcommand{\MSb}{\overline{\text{MS}}}
\subsection{The NMSSM Parameter Scan}
\begin{table*}
\begin{center}
%%%%%%%%%%%%%%
{\small \begin{tabular}{l|cccccccccccccc} \toprule
& $M_1$ & $M_2$ & $A_t$ &
$A_b$ & $A_\tau$ & $m_{\tilde{Q}_3}$ & $m_{\tilde{t}_R}$ &  $m_{\tilde{b}_R}$ & $m_{\tilde{L}_3}$ 
& $m_{\tilde{\tau}_R}$ & $M_{H^\pm}$ & $A_\kappa$ & $\mu_{\text{eff}}$ \\ 
& \multicolumn{13}{|c}{in TeV} \TBstrut \\ \hline 
min & 0.4 & 0.4 & -2.0 & -2.0 & -2.0 & 0.4 & 0.4
& 2.0 & 0.4 & 0.4 & 0.2 & -2.0 & 0.2 \Tstrut \\
max & 1.0 & 1.0 & 2.0 & 2.0 & 2.0 & 3.0 & 3.0 & 3.0
& 3.0 & 3.0 & 1.0 & 2.0 & 0.3 \Bstrut \\ \bottomrule
\end{tabular}}
\caption{Scan ranges for the NMSSM scan. All parameters are varied
  independently between the given minimum and maximum
  values, denoted by ``min'' and ``max'', respectively. \label{tab:nmssmscan}} 
\end{center}
\end{table*}
%%%%%%%%%%%%%%
In order to find scenarios that are compatible with the recent
experimental constraints for the purpose of our numerical analysis, we
perform a scan in the NMSSM parameter space. We apply the same procedure as 
in~\cite{Costa:2015llh,King:2014xwa,Azevedo:2018llq}, where also
further details can be found. The parameters $\tan\beta$, $\lambda$ and
$\kappa$ are varied in the ranges
\beq
1.5 \le \tan\beta \le 10 \;, \quad 10^{-4} \le \lambda \le 0.4 \;, \quad 0
\le \kappa \le 0.6 \;,
\eeq
so that we obey the rough perturbativity constraint 
\beq
\lambda^2 + \kappa^2 < 0.7^2 \;.
\eeq
The scan ranges of further parameters are listed in Table~\ref{tab:nmssmscan}\,. We set
\beq 
M_3 = 1.85 \mbox{ TeV}
\eeq
and the mass parameters of the first and second generation sfermions
are chosen to be
\begin{equation}   
m_{\tilde{u}_R,\tilde{c}_R} = 
m_{\tilde{d}_R,\tilde{s}_R} =
m_{\tilde{Q}_{1,2}}= m_{\tilde L_{1,2}} =m_{\tilde e_R,\tilde{\mu}_R}
= 3\;\mbox{TeV} \;. \label{eq:lightsquatmasses}
\end{equation}
The soft SUSY-breaking trilinear couplings of the first two
generations are set equal to the corresponding values of the third
generation. We follow the SUSY Les Houches Accord (SLHA)
format~\cite{Skands:2003cj,Allanach:2008qq}, which means that the 
soft SUSY-breaking masses and trilinear couplings are understood as
$\DRb$ parameters at the scale 
\be 
\mu_R = M_{\text{SUSY}} = \sqrt{m_{\ti Q_3} m_{\ti t_R}} \;. 
\ee
%\mm{
The SM input parameters have been chosen to
be~\cite{PhysRevD.98.030001,Dennerlhcnote}  
\begin{equation}
\begin{tabular}{lcllcl}
\quad $\alpha(M_Z)$ &=& 1/127.955, &\quad $\alpha^{\MSb}_s(M_Z)$ &=&
0.1181 \\
\quad $M_Z$ &=& 91.1876~GeV &\quad $M_W$ &=& 80.379~GeV  \\
\quad $m_t$ &=& 172.74~GeV &\quad $m^{\MSb}_b(m_b^{\MSb})$ &=& 4.18~GeV \\
\quad $m_c$ &=& 1.274~GeV &\quad $m_s$ &=& 95.0~MeV \\
\quad $m_u$ &=& 2.2~MeV &\quad $m_d$ &=& 4.7~MeV \\
\quad $m_\tau$ &=& 1.77682~GeV &\quad $m_\mu$ &=& 105.6584~MeV  \\
\quad $m_e$ &=& 510.9989~keV &\quad $G_F$ &=& $1.16637 \cdot 10^{-5}$~GeV$^{-2}$\,.
%\label{eq:param1} 
\end{tabular}
\end{equation}
We calculate the spectrum of the Higgs particles including corrections
up to two-loop order ${\cal O}(\alpha_t \alpha_s + \alpha_t^2)$
\cite{Dao:2019qaz} with the recently published {\tt NMSSMCALC} version 3.00.
For the scan, $M_{H^\pm}$ has been used as input parameter,
and not $A_\lambda$ which is also provided as an
option by {\tt NMSSMCALC}. We choose OS renormalization for the
top/stop sector (see~\cite{Muhlleitner:2014vsa,Dao:2019qaz} for details).
One of the neutral CP-even Higgs bosons is identified with the SM-like
Higgs boson and its mass is  required to lie in the range  
\beq
122 \mbox{ GeV } \le m_h \le 128 \mbox{ GeV} \;.
\eeq
We use {\tt HiggsBounds 5.3.2}
\cite{Bechtle:2008jh,Bechtle:2011sb,Bechtle:2013wla}
to check for agreement with the Higgs exclusion limits from LEP,
Tevatron and LHC, and {\tt HiggsSignals 2.2.3}~\cite{Bechtle:2013xfa}
to verify agreement with the Higgs rates.
We demand the total $\chi^2$ computed by {\tt HiggsSignals} with our
given effective coupling factors to be compatible 
with the total  $\chi^2$ of the SM within 2$\sigma$.
The input required for {\tt HiggsSignals} is calculated with {\tt NMSSMCALC}. \s

Furthermore, the most relevant LHC exclusion bounds on the SUSY 
masses are taken into account. These constrain the gluino mass and the
lightest squark mass of the second generation to lie above 1.8~TeV,
see~\cite{Aad:2015iea}. The stop and sbottom masses in general have to
be above 800~GeV~\cite{Aaboud:2016lwz,Aad:2015iea}, and the slepton masses above
400~GeV~\cite{Aad:2015iea}.\s  

%%%%%%%%%%%%%%%%%%%%%%%%%%%%%%%%%%%%%%%%%%%%%%%%%%%%%%%%%%%%%
% Nhung started
\newcommand{\non}{\nonumber} 
\newcommand{\gev}{~\text{GeV}}
\newcommand{\mev}{~\text{MeV}}
\newcommand{\tev}{~\text{TeV}}
For the numerical analysis 
 we have chosen two sample parameter points among all parameter sets
 obtained by our scan. For the first scenario, denoted by P1,
 the lightest CP-even Higgs boson is singlet-like and the
 second-lightest CP-even Higgs boson is the SM-like Higgs boson. In
 this scenario the mixing between the singlet- and the SM-like states is quite
 significant. The second point is denoted by P2 and features
 an SM-like Higgs boson that is the lightest CP-even Higgs boson with
 the mixing between the singlet and the SM-like state being
 non-significant. \s

\textbf{Parameter point P1:} Besides the SM values defined above, the
parameter point is given 
by the following soft SUSY-breaking masses and trilinear couplings,
\beq
&&  
m_{\tilde{t}_R}=1036\,\gev \,,\;
 m_{\tilde{Q}_3}=2365\,\gev\,,\; m_{\tilde{b}_R}=2360\,\gev\,, \; \non\\ 
&& 
m_{\tilde{L}_{3}}= 1170\,\gev\,,\; m_{\tilde{\tau}_R}=2872\,\gev \,,\; 
 \non\\ 
&& |A_{u,c,t}| = 2178\,\gev\, ,\; |A_{d,s,b}|=358\,\gev\,,\;
|A_{e,\mu,\tau}| =1401\,\gev\,,\; \label{eq:param4scen1} \\ \non
&& |M_1| = 423\,\gev,\; |M_2|= 669\,\gev\,,\; |M_3|=1850\,\gev \,,\\ \non
&&  \varphi_{A_{e,\mu,\tau}}=0\, ,\; \varphi_{A_{d,s,b}}=\pi\,,\; 
\varphi_{A_{u,c,t}}=\varphi_{M_1}=\varphi_{M_2}=\varphi_{M_3}=0
 \;, 
\eeq
and the remaining input parameters are set to\footnote{The
  imaginary part of $A_\kappa$ is obtained from the minimum conditions
  via the tadpole equations.} 
\beq
&& |\lambda| = 0.367 \;, \quad |\kappa| = 0.584\; , \quad \Re A_\kappa
= -1423\,\gev\;,\quad  
|\mu_{\text{eff}}| = 226.5\,\gev \;, \non \\ 
&&\varphi_{\lambda}=\varphi_{\kappa}=\varphi_u=0\;,
\quad \varphi_{\mu_{\text{eff}}}=0 \;,  
\quad \tan\beta = 3.11 \;,\quad M_{H^\pm} = 624 \,\gev \;.
\label{eq:param4scen1_b}
\eeq

\textbf{Parameter point  P2:} Besides the SM values defined above, the
parameter point is given 
by the following soft SUSY-breaking masses and trilinear couplings,
\beq
&&  
m_{\tilde{t}_R}=872\,\gev \,,\;
 m_{\tilde{Q}_3}=1883\,\gev\,,\; m_{\tilde{b}_R}=2341\,\gev\,, \; \non\\ 
&& 
m_{\tilde{L}_{3}}= 1867\,\gev\,,\; m_{\tilde{\tau}_R}=2394\,\gev \,,\; 
 \non\\ 
&& |A_{u,c,t}| = 2150\,\gev\, ,\; |A_{d,s,b}|=1189\,\gev\,,\;
|A_{e,\mu,\tau}| =159\,\gev\,,\; \label{eq:param4scen2} \\ \non
&& |M_1| = 443\,\gev,\; |M_2|= 748\,\gev\,,\; |M_3|=1850\,\gev \,,\\ \non
&&  \varphi_{A_{e,\mu,\tau}}= \varphi_{A_{d,s,b}}=\pi\,,\; 
\varphi_{A_{u,c,t}}=\varphi_{M_1}=\varphi_{M_2}=\varphi_{M_3}=0
 \;. 
\eeq
The remaining input parameters are set to
\beq
&& |\lambda| = 0.331 \;, \quad |\kappa| = 0.408\; , \quad \Re A_\kappa
= -402\,\gev\;,\quad  
|\mu_{\text{eff}}| = 224\,\gev \;, \non \\ 
&&\varphi_{\lambda}=\varphi_{\kappa}=\varphi_u=0\;,
\quad \varphi_{\mu_{\text{eff}}}=0 \;,  
\quad \tan\beta =  4.46 \;,\quad M_{H^\pm} = 988 \,\gev \;.
\label{eq:param4scen2_b}
\eeq

\subsection{Gauge Dependence of the Neutral Higgs Boson Masses \label{sec:massgauge}}
%%%%%%%%%%%%%%%%%%%%%%%%%%%%%%%%%%%%%%%%%%%%%%%%%%%%%%%%%%%%%
As it has been discussed in Sec.~\ref{sec:HiggsSec-Loop}, the masses of the
neutral Higgs bosons are obtained through an iterative method. While
this method yields precise solutions for the poles of the
propagator matrix, it may introduce intricate gauge dependences due to
the mixing of different orders of perturbation theory. In this section
we investigate how large this gauge dependence can become for the Higgs
boson masses. We consider the first scenario in this subsection and the following
Sec.~\ref{sec:widthgauge}, and the second scenario in Sec.~\ref{sec:p2}.
\s

 The Higgs boson masses and their main compositions in terms of
singlet/doublet and scal\-ar/pseu\-do\-sca\-lar components at tree level,
one-loop level as well as two-loop ${\cal O}(\alpha_t \alpha_s)$ level
and two-loop ${\cal O}(\alpha_t \alpha_s + \alpha_t^2)$ level 
  are given in Table~\ref{tab:massP1OS} for OS and in Table~\ref{tab:massP1DR} for
  $\overline{\mbox{DR}}$ renormalization in the top/stop sector. They
  have been computed with {\tt NMSSMCALC} in the 't\,Hooft--Feynman gauge
  ($\xi=1$). In the Table, lowercase $h_i$ refers to the tree-level and
  capital $H_i$ to the loop-corrected mass eigenstates.
In our chosen parameter point, the
$h_s$-like  and the $h_u$-like Higgs boson masses are light and
receive significant higher-order corrections. We call the Higgs
boson singlet-like in case its dominant contribution to the mass
eigenstate stems from the singlet admixture. The second-lightest Higgs
boson is dominantly $h_u$-like and has a mass of 125~GeV at ${\cal
  O}(\alpha_t \alpha_s  + \alpha_t^2)$ for OS renormalization in the top/stop sector. 
It reproduces the LHC production rates which proceed dominantly through
gluon fusion for small values of $\tan\beta$. Since the LO process is
dominated by top-quark loops the Higgs coupling to the tops must be
substantial, as is the case for a Higgs boson with large $h_u$ admixture.
\s
\begin{table}[t!]
\begin{center}
 \begin{tabular}{|l||c|c|c|c|c|}
\hline
 &$h_1/{H_1}$&$h_2/{H_2}$&$h_3/{H_3}$&$h_4/{H_4}$&$h_5/{H_5}$\\ \hline \hline
tree-level &9.80&91.38&621.27&627.37&1243.31\\ 
main component&$h_s$&$h_u$&$a$&$h_d$&$a_s$\\ \hline  
one-loop &94.96 & 133.12 & 620.97 & 628.1 & 1216.77 \\ 
main component&$h_s$&$h_u$&$a$&$h_d$&$a_s$\\ \hline  
two-loop ${\cal O}(\alpha_t \alpha_s)$ &93.06 & 118.95 & 620.99 &
627.82 & 1216.8 \\  
main component&$h_s$&$h_u$&$a$&$h_d$&$a_s$\\ \hline
two-loop ${\cal O}(\alpha_t \alpha_s+ \alpha_t^2)$
 &93.95 & 125.08 & 620.99 & 627.96 & 1216.8 \\ 
main component&$h_s$&$h_u$&$a$&$h_d$&$a_s$\\ \hline
\end{tabular}
\caption{P1: mass values in GeV and main components of the neutral Higgs
  bosons at tree level, one-loop level, two-loop ${\cal O}(\alpha_t
  \alpha_s)$ level and at two-loop ${\cal O}(\alpha_t \alpha_s +
  \alpha_t^2)$ level obtained by using OS renormalization in the
  top/stop sector.}
\label{tab:massP1OS}
\end{center}
\end{table}
\begin{table}[t!]
\begin{center}
 \begin{tabular}{|l||c|c|c|c|c|}
\hline
 &$h_1/{H_1}$&$h_2/{H_2}$&$h_3/{H_3}$&$h_4/{H_4}$&$h_5/{H_5}$\\ \hline \hline
tree-level &9.80&91.38&621.27&627.37&1243.31\\
main component&$h_s$&$h_u$&$a$&$h_d$&$a_s$\\ \hline
one-loop &92.15 & 114.44 & 620.92 & 627.6 & 1216.81  \\
main component&$h_s$&$h_u$&$a$&$h_d$&$a_s$\\ \hline
two-loop ${\cal O}(\alpha_t \alpha_s)$ &93.21 & 118.84 & 620.91 &
627.67 & 1216.8 \\ 
main component&$h_s$&$h_u$&$a$&$h_d$&$a_s$\\ \hline
two-loop ${\cal O}(\alpha_t \alpha_s+ \alpha_t^2)$
 &93.26 & 119.1 & 620.91 & 627.68 & 1216.8 \\               
main component&$h_s$&$h_u$&$a$&$h_d$&$a_s$\\ \hline
\end{tabular}
\caption{P1: mass values in GeV and main components of the neutral Higgs
  bosons at tree level, one-loop level, two-loop ${\cal O}(\alpha_t
  \alpha_s)$ level and at two-loop ${\cal O}(\alpha_t \alpha_s +
  \alpha_t^2)$ level obtained by using $\DRb$ renormalization in the top/stop sector.}
\label{tab:massP1DR}
\end{center}
\end{table}

\begin{figure}[t!]
\includegraphics[width=0.5\textwidth]{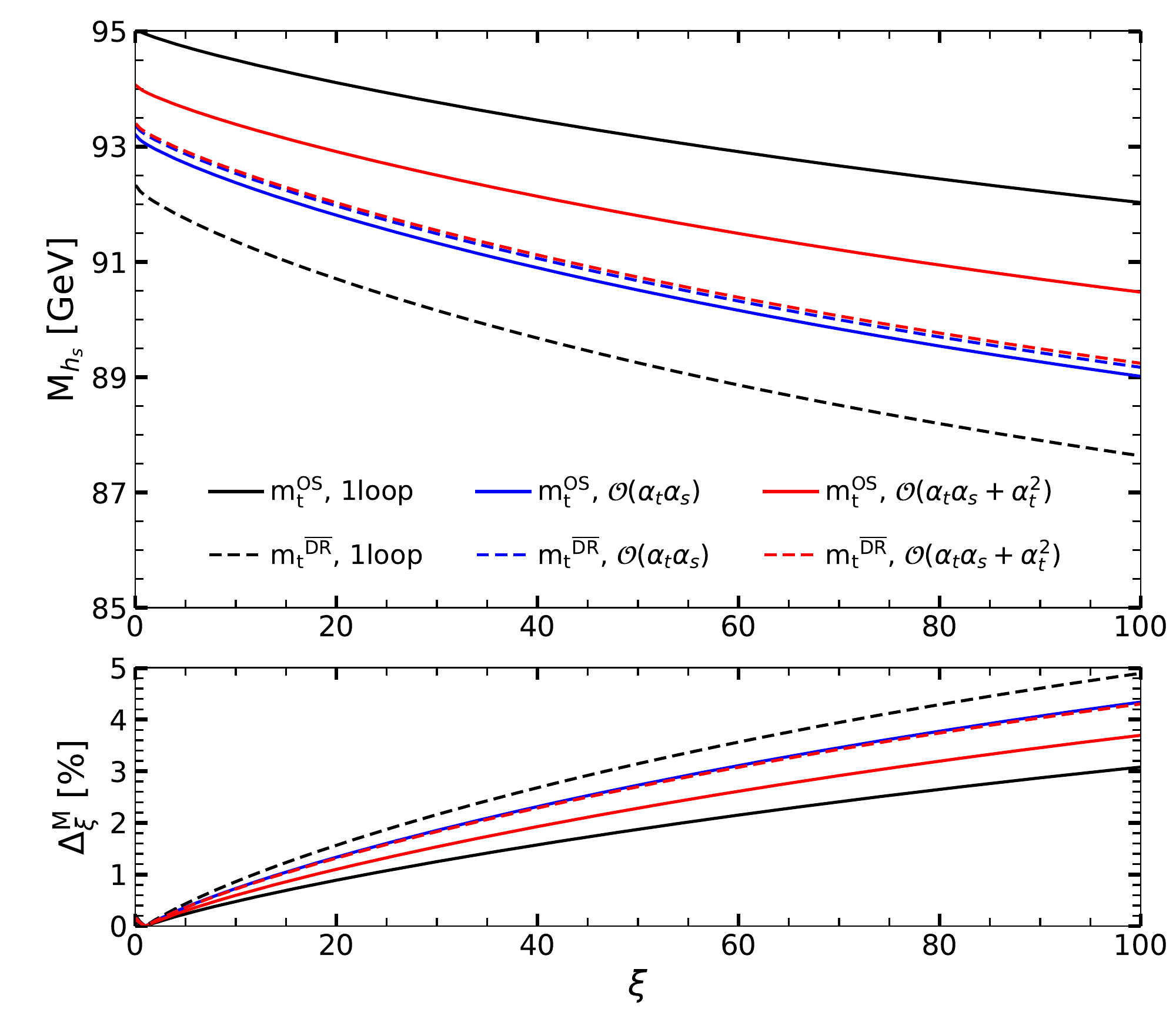}
\includegraphics[width=0.5\textwidth]{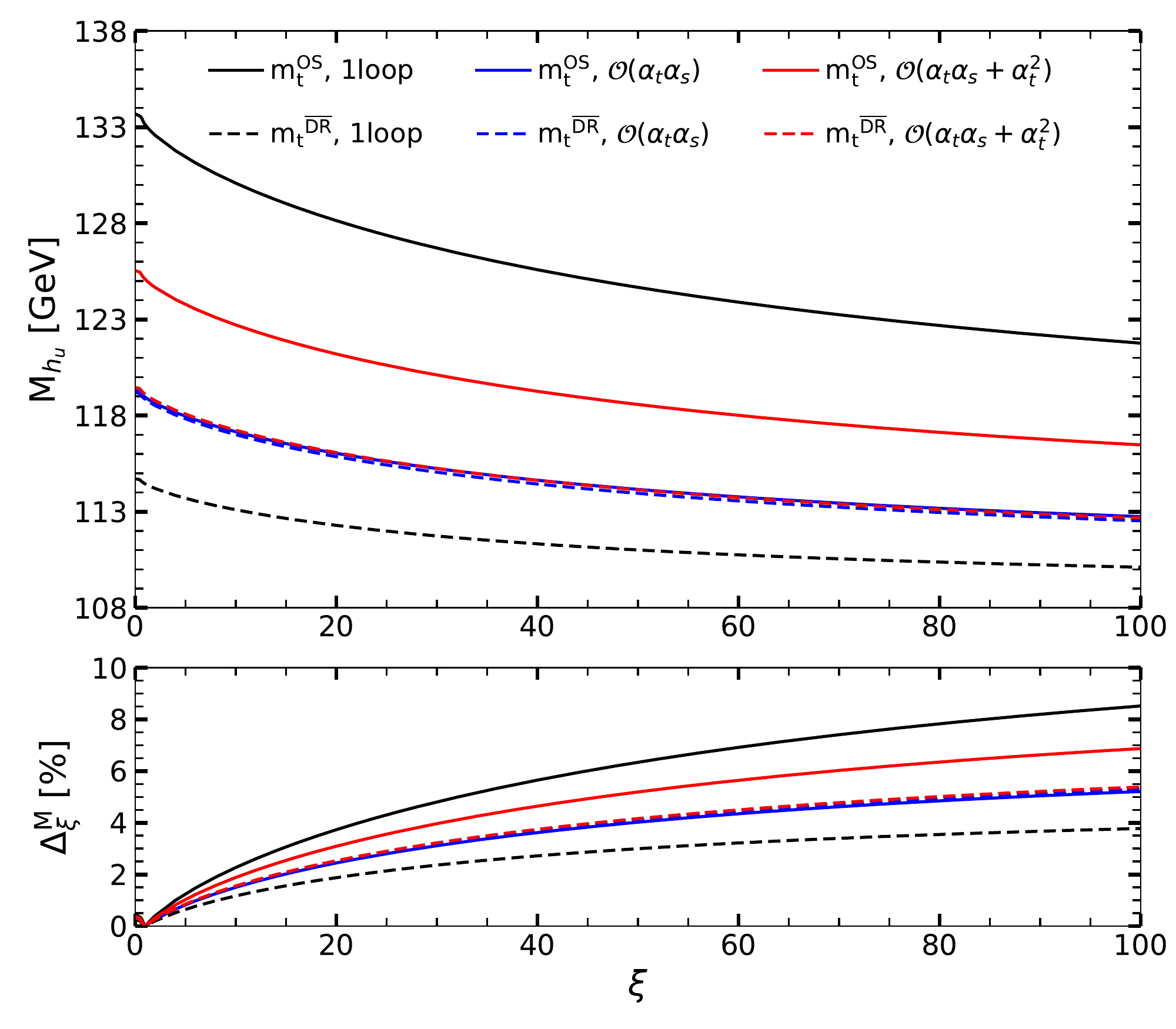}
\caption{Upper Panel: The CP-even singlet-like
  (left) and the SM-like 
  (right) Higgs boson masses  as a function of the gauge parameter $\xi$
at the one-loop (black), two-loop ${\cal
  O}(\alpha_t\alpha_s)$ (blue), two-loop ${\cal O}(\alpha_t \alpha_s +
\alpha_t^2)$ (red) level in the OS (solid lines) and the $\DRb$
(dashed lines) scheme of the top/stop 
sector. Lower Panel: Absolute value of the relative $\xi$ dependence
of the loop-corrected masses, defined as
$\Delta^M_\xi= \abs{M_{x}(\xi) -M_{x}(\xi=1)}/M_{x}(\xi=1)$ ($x=h_s, h_u$),
in percent, as function of $\xi$. The color code in the lower plots is
the same as in the upper plots.}  
\label{fig:Hmass2-xi-lam}
\end{figure}
We first vary the gauge parameter $\xi$ of the general $R_\xi$ gauge,
which we set throughout the section $\xi_W = \xi_Z \equiv \xi$, while all other 
parameters are kept fixed. The masses of the $h_s$- and the $h_u$-like
Higgs bosons depend significantly on $\xi$. 
All remaining Higgs bosons have masses
larger than 600 GeV and show a very small gauge dependence. This is
due to the fact that only the light Higgs bosons receive significant loop
corrections. In \fig{fig:Hmass2-xi-lam}, we show these dependences for the
mass $M_{h_s}$ of the CP-even singlet-like Higgs boson in the 
upper left plot and for the mass $M_{h_u}$ of the SM-like Higgs boson in the upper
right plot including one-loop (black lines), ${\cal O}(\alpha_t\alpha_s)$ two-loop
(blue lines) and ${\cal O}(\alpha_t \alpha_s + \alpha_t^2)$ two-loop
(red lines) corrections. These 
corrections are obtained for the OS (full lines) and $\DRb$  (dashed lines)
renormalization schemes of the top/stop sector. The two lower plots
display the relative difference between the masses in general
$R_\xi$ gauge and in the 't\,Hooft--Feynman gauge $\xi=1$, 
\be 
\Delta_\xi^M = \frac{\abs{M_{x}(\xi) -M_{x}(\xi=1)}}{M_{x}(\xi=1)} ,
\quad \text{with}\quad x=h_s, h_u \;,
\ee  
as functions of $\xi$. Here $M_x$ denotes the loop-corrected
mass value of the Higgs boson $x$ at a fixed loop order, calculated in
general $R_\xi$ gauge ($M_x (\xi)$) and in the 't\,Hooft--Feynman
gauge ($M_x (\xi=1)$). We remind the reader that only the
renormalized one-loop Higgs self-energies are calculated in general
$R_{\xi}$ gauge and therefore depend on $\xi$, while the renormalized
two-loop Higgs self-energies at order  ${\cal O}(\alpha_t\alpha_s)$ 
and ${\cal O}(\alpha_t^2)$ do not depend on $\xi$ as they are computed
in the gaugeless limit.  Note that the tree-level masses for the
$h_s$- and $ h_u$-like Higgs bosons are 9.8\,GeV and 91.38\,GeV, respectively. The
loop corrections to their masses are positive.  From the plots, we can
infer that the loop-corrected masses decrease with increasing 
$\xi$, which we chose to lie between 0 and 100. We
can, however, increase $\xi$ to a larger value and find that for $\xi= 1089$ the
loop-corrected mass of the $h_s$-like Higgs boson becomes negative. In
the lower plots, we see two different behaviors for the $h_s$- and
$h_u$-like Higgs boson. The relative $\xi$ dependence is larger in the
$\DRb$ scheme than in the OS scheme for the loop-corrected $h_s$-like
Higgs boson masses, while the behavior is opposite for the
loop-corrected $h_u$-like Higgs boson masses. The inclusion of the two-loop
corrections of  order  ${\cal O}(\alpha_t\alpha_s)$ and ${\cal
  O}(\alpha_t^2)$ changes $\Delta^M_\xi$ in an intricate way. The relative
differences $\Delta _\xi ^M$ in the OS and $\DRb$ scheme, however,
move closer to each other with the inclusion of the two-loop
corrections. \s 

\begin{figure}[t!] 
\includegraphics[width=0.5\textwidth]{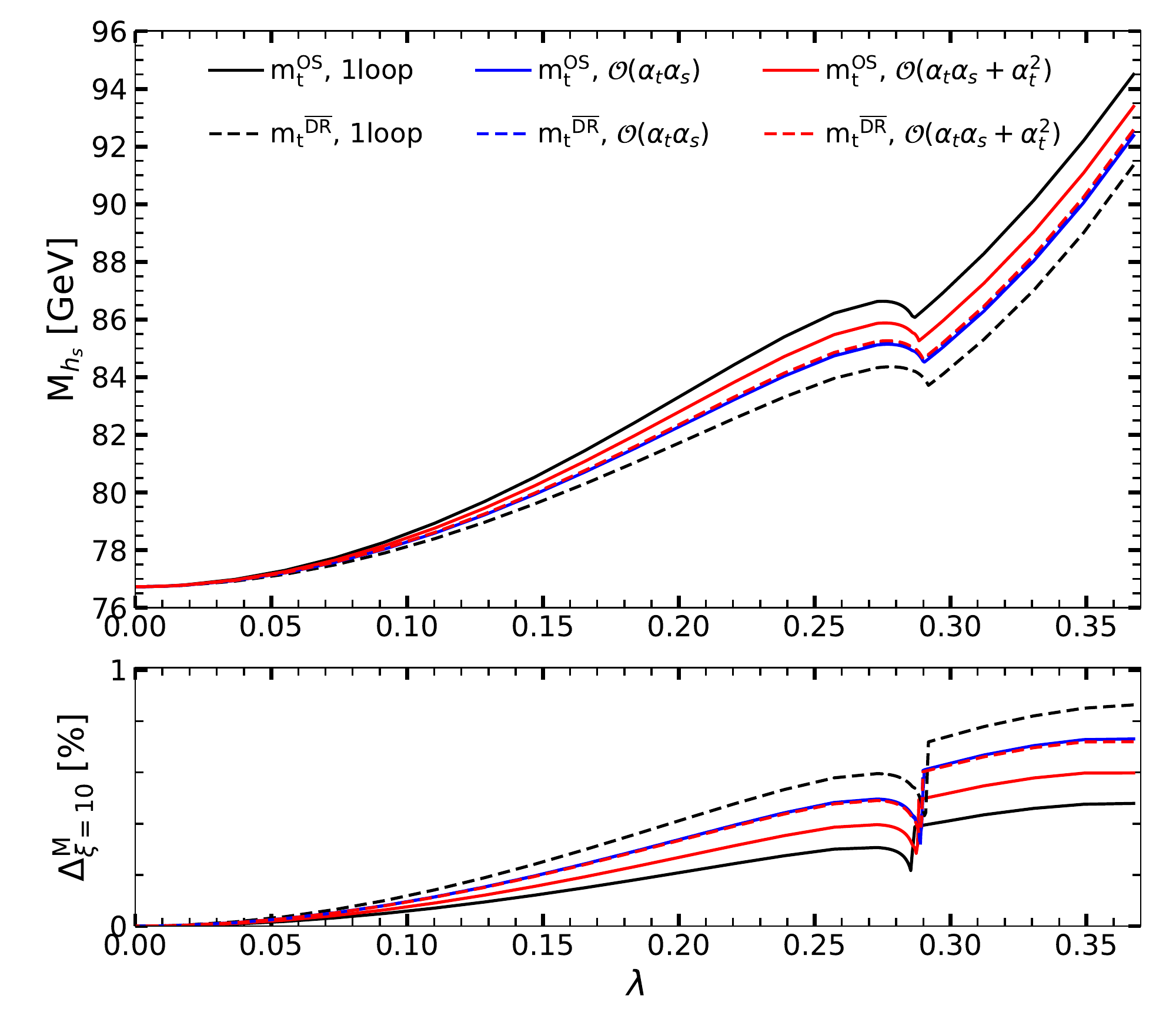}
\includegraphics[width=0.5\textwidth]{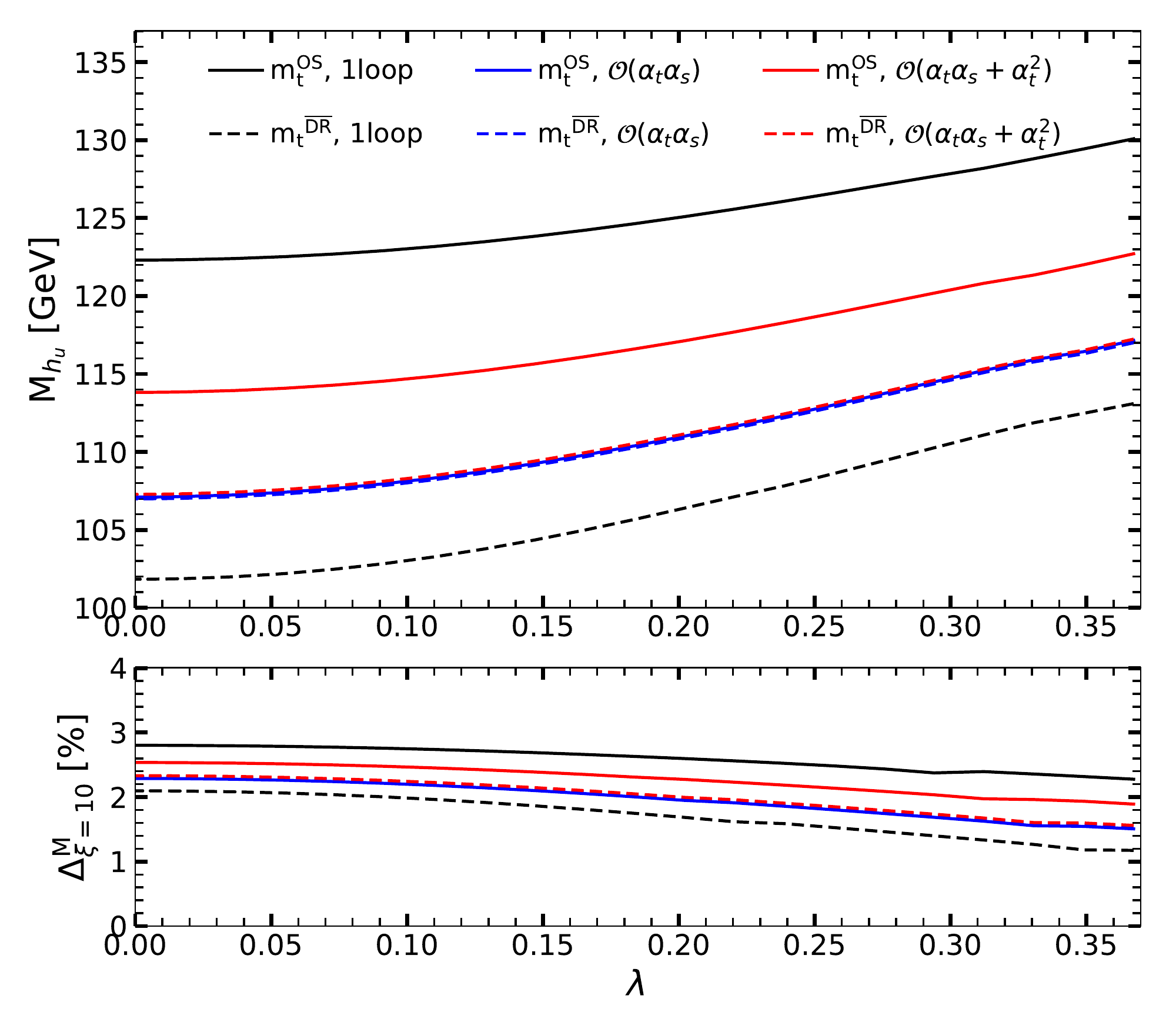}
\caption{Analogous to Fig.~\ref{fig:Hmass2-xi-lam}, but $\lambda$ is varied
  while $\xi$ is kept fixed at $\xi = 10$. 
\label{fig:Hmass2-lam} }
\end{figure}
Next, we fix the value of $\xi$ to 10 and vary $\lambda$ and $\kappa$
at the same time to very small values ($10^{-5}$) while we keep the
ratio $\lambda/\kappa$ constant. In this way we smoothly approach the
MSSM limit, where the singlet and doublet Higgs bosons do not mix. In
Fig.~\ref{fig:Hmass2-lam} we show the thus obtained loop-corrected
masses of the $h_s$-like (left) and $h_u$-like (right) Higgs boson as
function of $\lambda$. The line and color codes are the same as in
Fig.~\ref{fig:Hmass2-xi-lam}. The lower plots show $\Delta^M_{\xi=10}$, {\it
  i.e.}~the deviation of the $h_s$- and $h_u$-like masses,
respectively, calculated for $\xi=10$ from the value obtained in the
't\,Hooft--Feynman gauge. \s

As expected, when $\lambda$ and $\kappa$ are close
to zero, the $h_s$-like Higgs boson decouples and the loop corrections
to the $h_s$-like Higgs boson mass vanish in this limit. Therefore all
lines in the left plots converge at the left endpoint where $\lambda
\approx 0$. As we can see from Fig.~\ref{fig:Hmass2-lam} 
(right), the $\xi$ dependence of the SM-like loop-corrected Higgs boson
masses does not vanish in the MSSM limit. For our chosen parameter point
the relative deviation of the masses for $\xi=10$ and $\xi=1$ even
slightly increases. The kink around $\lambda = 0.29$ appears at the
threshold where the loop-corrected $h_s$ mass is twice the tree-level
$h_s$ mass. \s

\begin{figure}[t!] 
\includegraphics[width=0.5\textwidth]{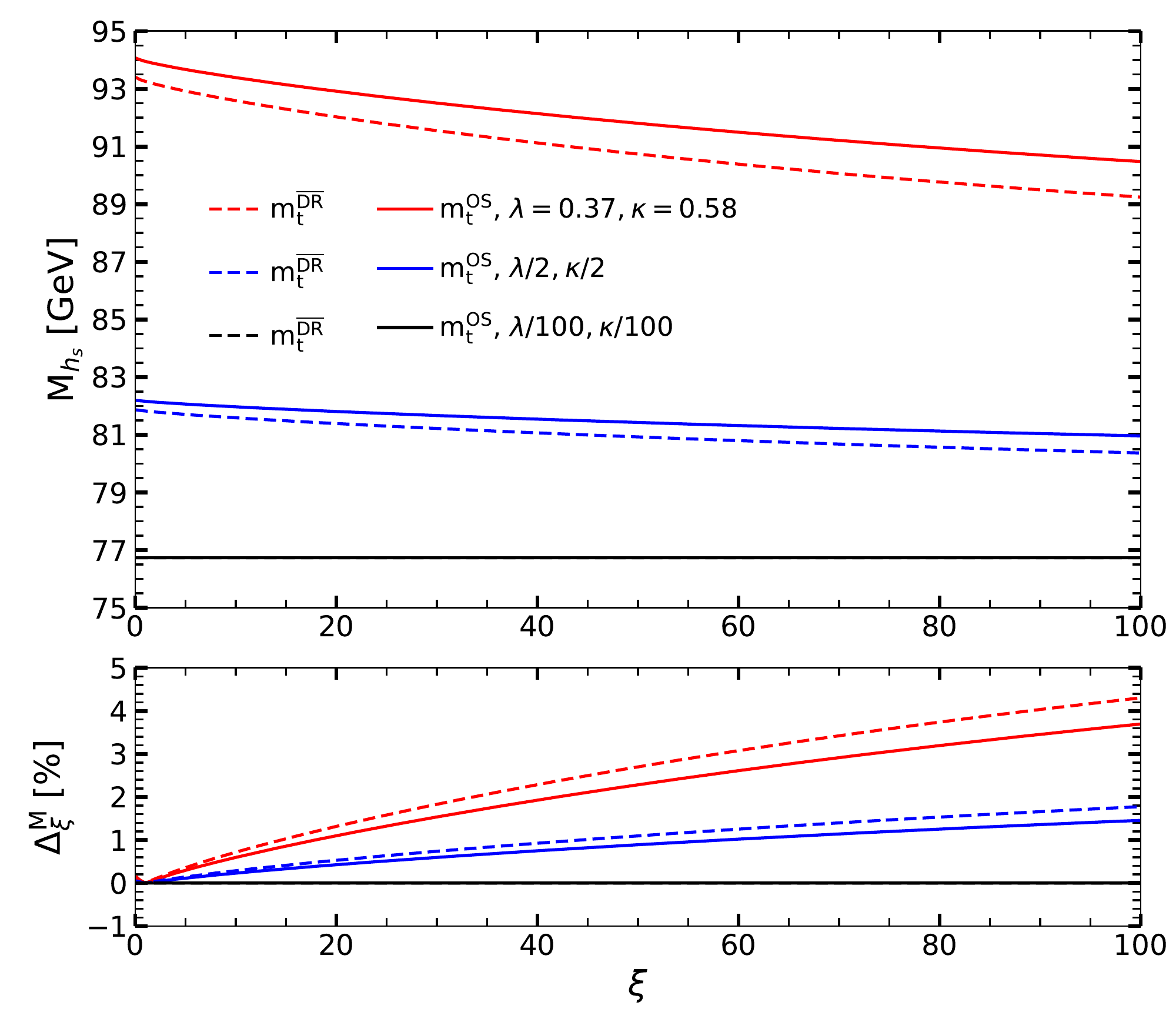}
\includegraphics[width=0.5\textwidth]{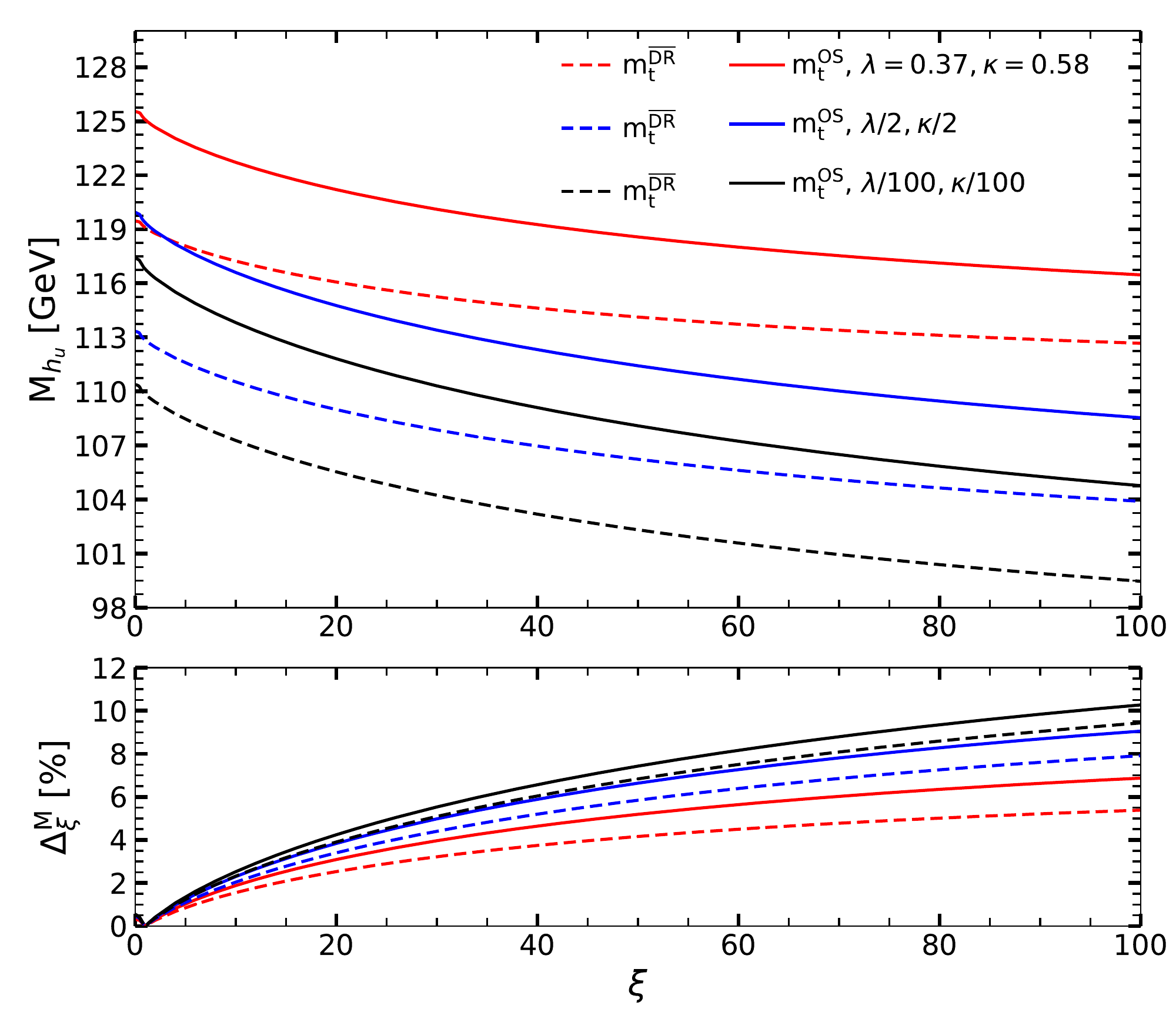}
\caption{Upper panel: The CP-even singlet-like
  (left) and the $h_u$-like 
  (right) Higgs boson masses at two-loop ${\cal O}(\alpha_t \alpha_s +
\alpha_t^2)$ level in the OS (solid lines) and the $\DRb$
(dashed lines) scheme of the top/stop 
sector for $\lambda=0.37, \kappa=0.58$, {\it i.e.}~the P1 values,
(red), for half their values (blue) and for $\lambda/100$,
$\kappa/100$ (black). Lower panel: Corresponding $\Delta^M_\xi$
in percent, as function of $\xi$.} 
\label{fig:hmass-3lam} 
\end{figure}
In order to investigate the influence of the NMSSM-specific
contributions to the mass corrections at ${\cal O}(\alpha_t \alpha_s +
\alpha_t^2)$ and their $\xi$ dependence we simultaneously vary the
couplings $\lambda$ and $\kappa$ and show in 
Fig.~\ref{fig:hmass-3lam} (upper) the two-loop masses of the singlet-like (left)
and the $h_u$-like (right) Higgs bosons for
the original values for $\lambda$ and $\kappa$ of the parameter point
P1 (red lines), half 
their values (blue) and for $\lambda/100$, $\kappa/100$ (black) in the
two renormalization schemes of the top/stop sector. In the left plot
the black dashed and full lines lie on top of each other. The lower plots depict the
corresponding relative $\xi$ dependence. For $h_s$ the
two-loop corrected mass value and the
$\xi$ dependence decrease with smaller values of the NMSSM-like
couplings, as the singlet-like Higgs boson decouples from the
spectrum. The $h_u$-like mass shows the expected behavior
and decreases with smaller singlet admixtures.\footnote{One of the
  virtues of the NMSSM is the increased upper mass bound of the
  SM-like Higgs boson due to the additional NMSSM-like contributions to
  the tree-level mass value.} The relative $\xi$ dependence
increases, however, for the chosen parameter point. The increasing
contribution to the mass corrections for larger $\lambda$, $\kappa$
values from the $h_u-h_s$ admixture mixes with the doublet-like mass
corrections and diminishes their gauge dependence. \s

\begin{figure}[t!] 
\includegraphics[width=0.5\textwidth]{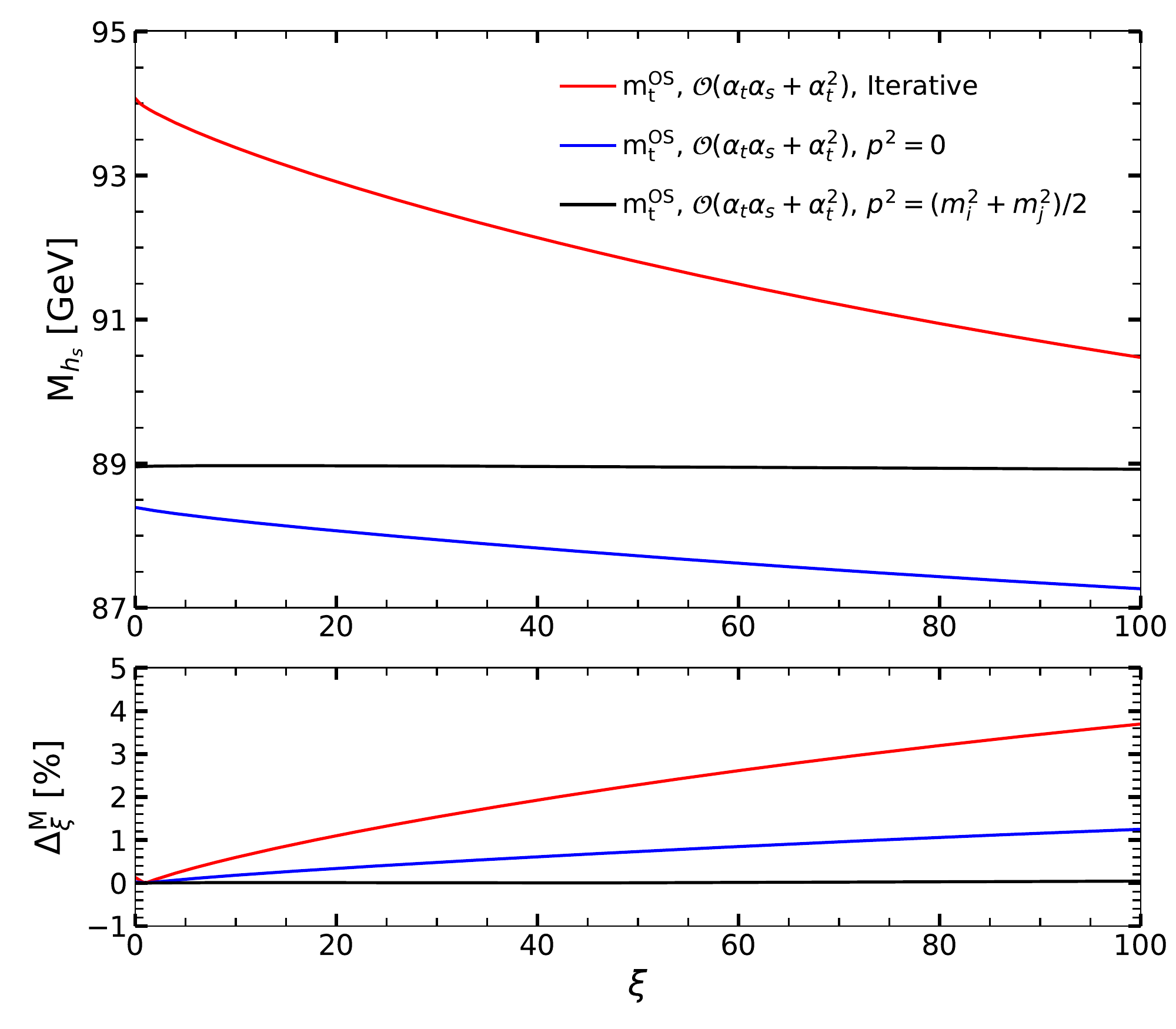}
\includegraphics[width=0.5\textwidth]{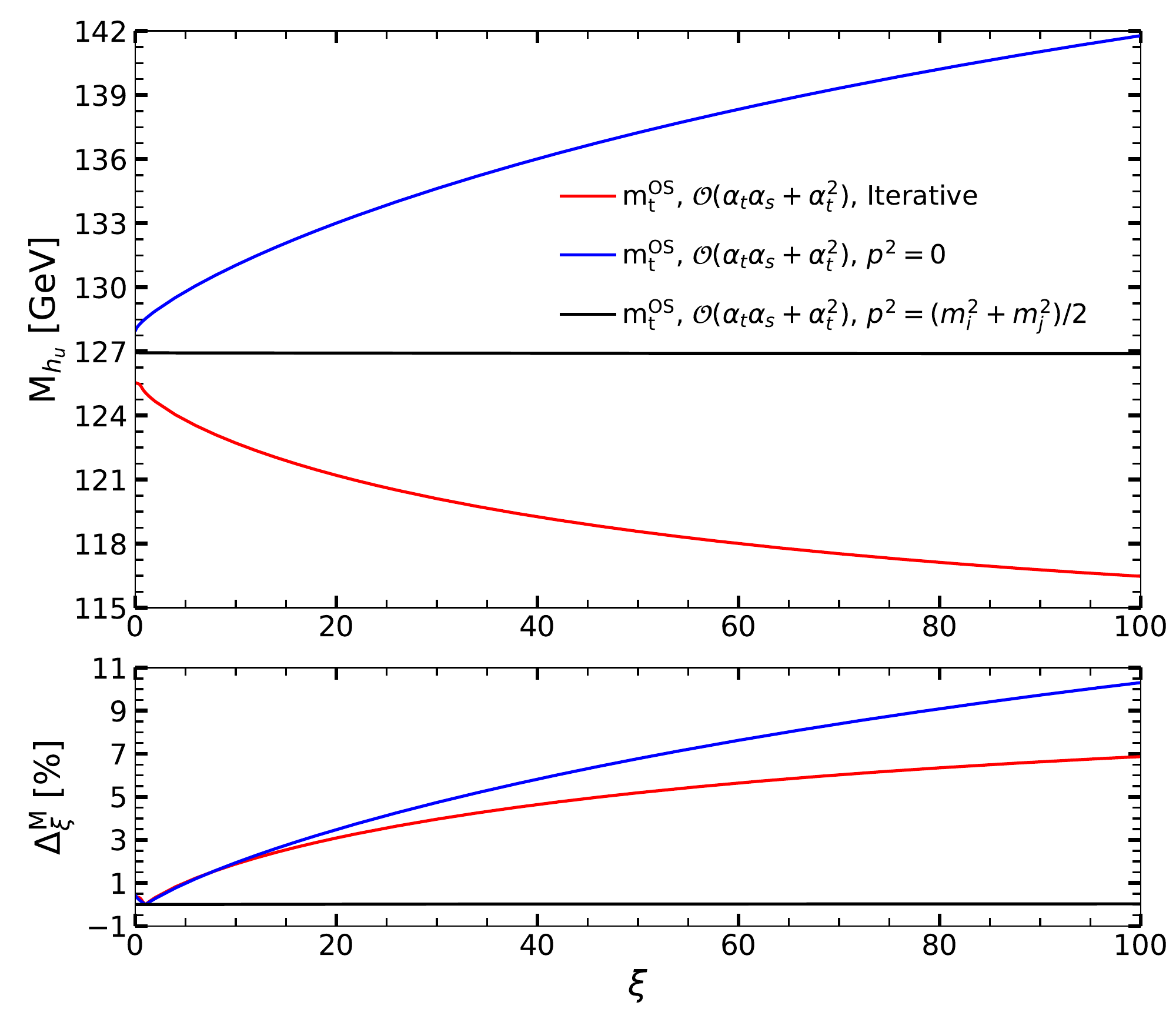}
\caption{Upper Panel: The CP-even singlet-like
  (left) and the $h_u$-like 
  (right) Higgs boson masses at two-loop ${\cal O}(\alpha_t \alpha_s +
\alpha_t^2)$ level in the OS scheme of the top/stop 
sector applying the iterative (red), the $R^0$- (blue) and the
$R^{\text{mtree}}$-method (black). See text, for explanations. Lower
Panel: Corresponding $\Delta^M_\xi$ 
in percent, as function of $\xi$.} 
\label{fig:hmass-mixmat} 
\end{figure} 
The gauge dependence strongly depends on the chosen approach to
determine the loop-corrected masses as we will show
next. Figure~\ref{fig:hmass-mixmat} displays the mass corrections
(upper plots) at ${\cal O}(\alpha_t \alpha_s +\alpha_t^2)$ for OS
renormalization in the top/stop sector and 
their relative $\xi$ dependence (lower plots)
determined through the iterative method to extract the zeros of the
determinant \cite{Ender:2011qh} (red lines) as well as when we apply
the zero momentum approximation ('$R^0$-method' called in the
following, blue lines), {\it 
  cf.}~Eq.~(\ref{eq:R0}), and when  
the mass matrix is diagonalized at the arithmetic squared mass average
('$R^{\text{mtree}}$-method', black lines), {\it cf.}~Eqs.~(\ref{eq:arithm1}) and
(\ref{eq:arithm2}). The gauge dependence
becomes very small for the latter in contrast to the two former
methods. This is because of the fact that the dependence of the
renormalized self-energy $\hat \Sigma_{h_ih_j}$ evaluated at the
arithmetic squared mass average on the gauge parameter is small for $i$
being different from $j$ and  vanishes completely for $i$ being identical to $j$.
Their behavior as a function of $\xi$ depends on the
difference between the tree-level mass and the squared momentum at
which the mass matrix is diagonalized resulting in $\Delta^M_\xi$
values up to about 1\% (4\%) for the $R_0$-method (iterative method)
for the $h_s$-like mass and 10\% (7\%) for the $h_u$-like mass when
$\xi$ is varied up to values of 100.

%%%%%%%%%%%%%%%%%%%%%%%%%%%%%%%%%%%%%%%%%%%%%%%%%%%%%%%%%%%%
\subsection{Gauge Dependence of the Loop-Corrected Decay Width \label{sec:widthgauge}}
In this section, we investigate the gauge dependence of the
loop-corrected decay width computed in Sec.~\ref{sec:one-loop-decay}.  

\subsubsection{Individual Loop Contributions}
We start with the study of the gauge dependence of the various components of
the virtual one-loop correction, namely $\mc{M}^{\text{vert}},
\mc{M}^{\text{ext}, H^\pm}, \mc{M}^{\text{ext},h}$, and finally of the
complete virtual amplitude $\mc{M}^{\text{virt}}$, as defined in
Sec.~\ref{sec:strict-one-loop}. We use the same parameter point as given in
Eqs.~(\ref{eq:param4scen1}) and (\ref{eq:param4scen1_b}).  \s

\begin{figure}[t!]
\includegraphics[width=0.5\textwidth]{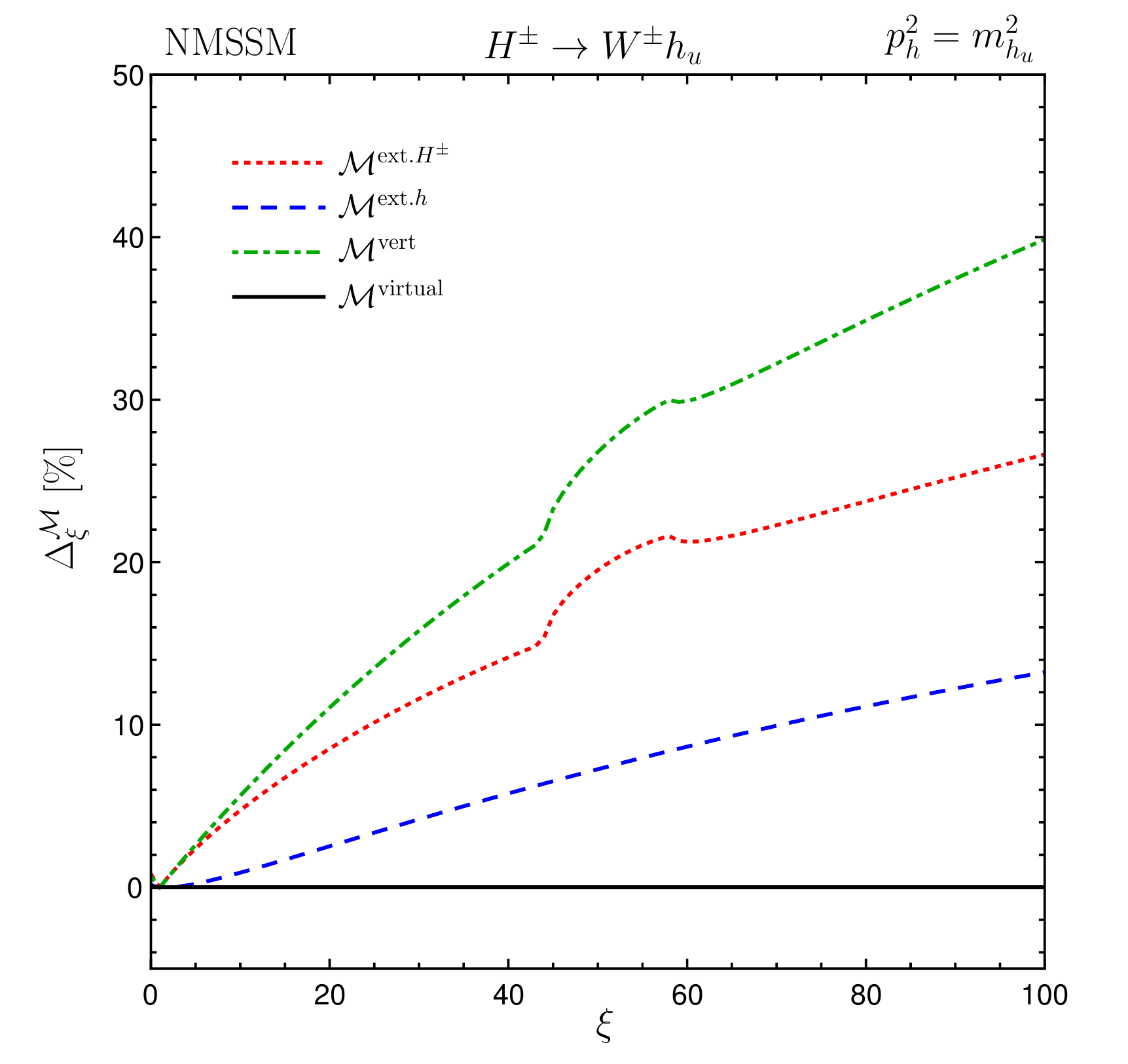}
\includegraphics[width=0.5\textwidth]{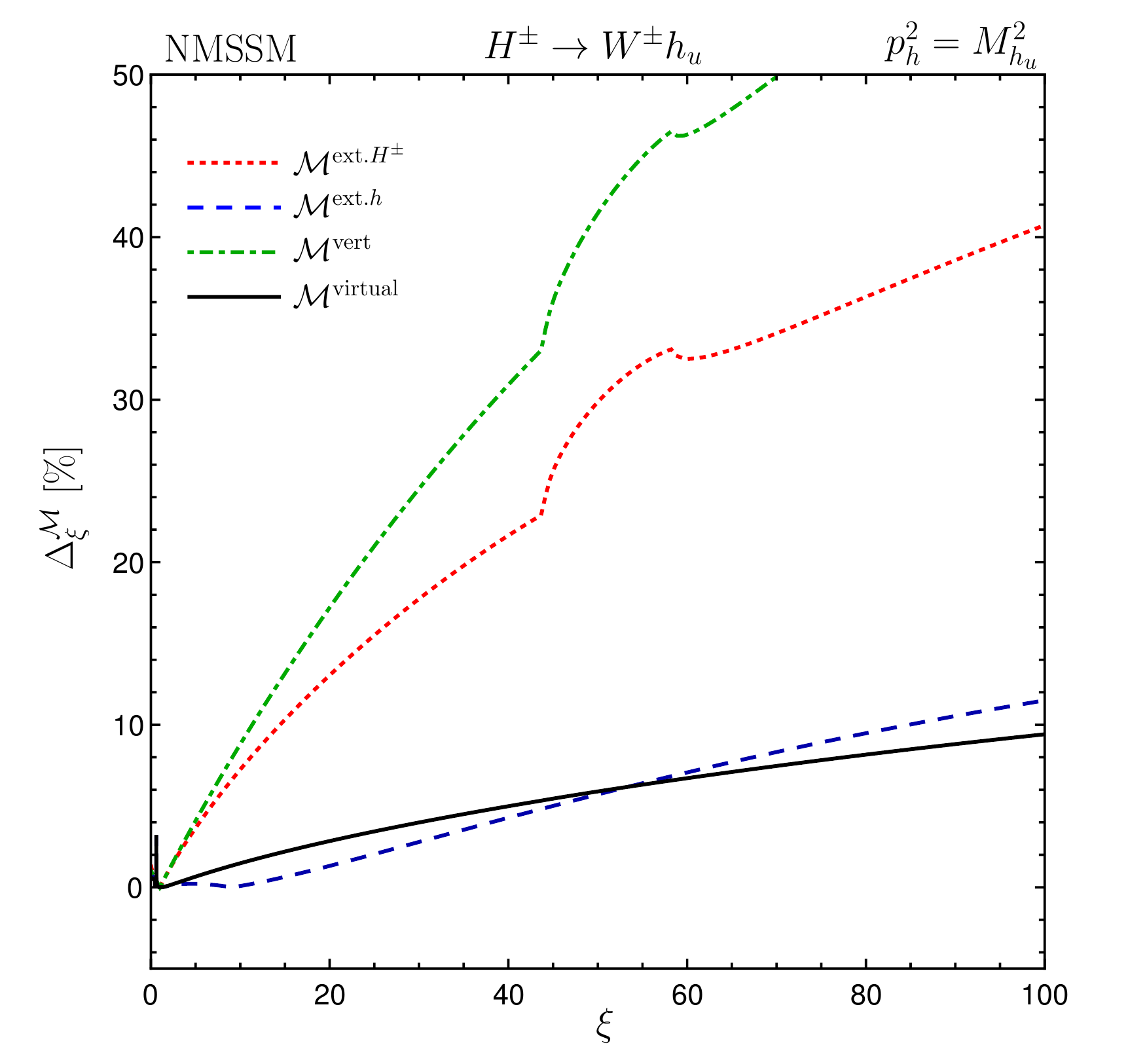}
\caption{Gauge dependence of the virtual
  electroweak corrections to $H^\pm \to W^\pm h_u$, with
  $\mc{M}^{\text{ext}, H^\pm}$ (dotted, red), 
  $\mc{M}^{\text{ext},h}$ (blue, dashed), $\mc{M}^{\text{vert}}$
  (green, dot-dashed) and $\mc{M}^{\text{virt}}$ (black, solid). Left
  plot: virtual corrections computed at strict one-loop order. Right
  plot: virtual corrections computed with the external momentum set to
  the loop-corrected neutral Higgs boson mass $p_h^2 =
  M_{h_u}^2$. For details, we refer to the text.}  
	\label{fig:nmssm-virt}
\end{figure}
In Fig.~\ref{fig:nmssm-virt} we show the relative gauge dependence of the
virtual amplitude and of its individual contributions for the electroweak loop
correction to the decay $H^\pm \to W^\pm h_u$ as a function of the
gauge parameter $\xi$, where the mostly $h_u$-like Higgs boson
corresponds to the SM-like Higgs boson.  We define the
quantity $\Delta^{\cal M}_\xi$ to measure the gauge
  dependence of the amplitude ${\cal M}$, by
\be
\Delta^{\cal M}_\xi = \frac{\abs{\mc{M}_\xi - \mc{M}_{\xi =
    1}}}{\abs{\mc{M}^{\text{virt}}_{\xi = 1}}} \;,
\ee
where ${\cal M}_\xi$ with ${\cal M} \in \lbrace \mc{M}^{\text{vert}}, \mc{M}^{\text{ext}, H^\pm},
\mc{M}^{\text{ext},h}, \mc{M}^{\text{virt}}\rbrace$ denotes the
amplitude in the general $R_\xi$ gauge and ${\cal M}_{\xi=1}$ the
computed in the 't\,Hooft--Feynman gauge $\xi=1$. Note that we normalize to 
${\cal M}^{\text{virt}}$, {\it i.e.}~to the sum of all contributions to the
virtual corrections of the one-loop amplitude, at $\xi=1$. 
We choose to vary $\xi$ between 0 and 100 in order not
  to introduce new scales in the calculation (the Goldstone masses scale with
$\sqrt{\xi} m_{\text{gauge}}$ where  $m_{\text{gauge}}$ denotes the
electroweak gauge boson masses). The individual components of the
virtual corrections include their respective counterterms, such that
the individual parts are UV-finite, but still IR-divergent. The IR
divergence is regulated by a finite photon mass. The red (dotted) curve depicts the
relative gauge dependence of the external leg corrections
$\mc{M}^{\text{ext}, H^\pm}$ to the charged Higgs boson, the blue one
(dashed) is the corresponding curve for the external leg correction
$\mc{M}^{\text{ext},h}$ to the outgoing neutral Higgs $h_u$, and the
green (dot-dashed) curve depicts the relative gauge dependence of the
vertex corrections $\mc{M}^{\text{vert}}$. Finally, the solid black
curve shows the result for the total virtual correction
$\mc{M}^{\text{virt}}$. In the left plot of Fig.~\ref{fig:nmssm-virt}, we
show these curves for the strict one-loop calculation as described in
Sec.~\ref{sec:strict-one-loop}. This means that the external leg
corrections to $h_u$ are accounted for diagrammatically using
Eq.~(\ref{eq:Zmatrixexpand}), as opposed to using the resummed
$\mbf{Z}$ matrix. Moreover, we use the tree-level mass $m_{h_u}$ for
the external momentum such that $p_h^2 = m_{h_u}^2$. Here and in the
following plots we use the  mixing matrix that
diagonalizes the tree-level mass matrices in the computation of the
couplings as otherwise the result will not be
UV-finite. In this strict
one-loop computation, the virtual corrections are 
gauge independent, as can be checked
  explicitly numerically, {\it cf.}~the solid black 
curve of \fig{fig:nmssm-virt} (left): while each individual component
of the virtual correction is gauge dependent, their sum, resulting in
$\mc{M}^{\text{virt}}$, is gauge independent.
Actually, the relative gauge dependences of the external leg contributions to the
charged and the neutral Higgs mass, ${\cal M}^{\text{ext},H^\pm}$ and
${\cal M}^{\text{ext},h}$, and the ones of the
vertex corrections, ${\cal M}^{\text{vert}}$, come with opposite sign
(not visible from the plot as we show the absolute values). 
The reason for the kinks in the red (dotted) and green (dot-dashed)
curves is the following. The masses of the
Goldstone bosons depend on the gauge parameter $\xi$, and these kinks
occur when a production threshold for the Goldstone boson is reached,
{\it i.e.}~at the points 
\be
m_{H^{\pm}} = \sqrt{\xi} M_W + m_{h_j} \quad \text{and} \quad
m_{H^{\pm}} = \sqrt{\xi} M_Z + m_{a_j}\,. 
\ee 
In \fig{fig:nmssm-virt} (right), we investigate how this
gauge independence of the strict one-loop computation
changes when 
we apply the improved one-loop computation as defined in
Sec.~\ref{sec:improved-one-loop} 'Step 1' that we denoted 'off-shell'. The
means, we use loop-corrected masses for the external leg corrections
to the neutral Higgs boson $h_u$, {\it i.e.}~we set $p_h^2 =
M_{h_u}^2$. Note, however, that the $\mbf{Z}$ matrix is evaluated at
pure one-loop order, as defined in Eq.~(\ref{eq:Zmatrixexpand}).
The masses are calculated at ${\cal O}(\alpha_t \alpha_s +
\alpha_t^2)$ for OS renormalization in the
top/stop sector  by  {\tt NMSSMCALC} in general $R_\xi$ gauge as
described in subsection~\ref{sec:HiggsSec-Loop}. Going from the
strict one-loop calculation to the 'off-shell' one, we see that the
dependence of the individual components of the virtual corrections on
the gauge parameter $\xi$ changes, such that their sum
$\mc{M}^{\text{virt}}$ (solid, black curve) is no longer
gauge independent. The overall gauge dependence does not
cancel any more, as we move away from the strict
fixed-order calculation 
and include partial higher-order effects coming from the
loop-corrected Higgs mass $M_{h_u}^2$, such that with increasing $\xi$
values, the \nlo{} amplitude becomes arbitrarily large. The relative
change of the total virtual amplitude is of up to ${\cal O}(10\%)$ for
$\xi$ values up to 100. \s

\begin{figure}[t!]
\begin{center}
\hspace*{-0.2cm}
\includegraphics[width=0.5\textwidth]{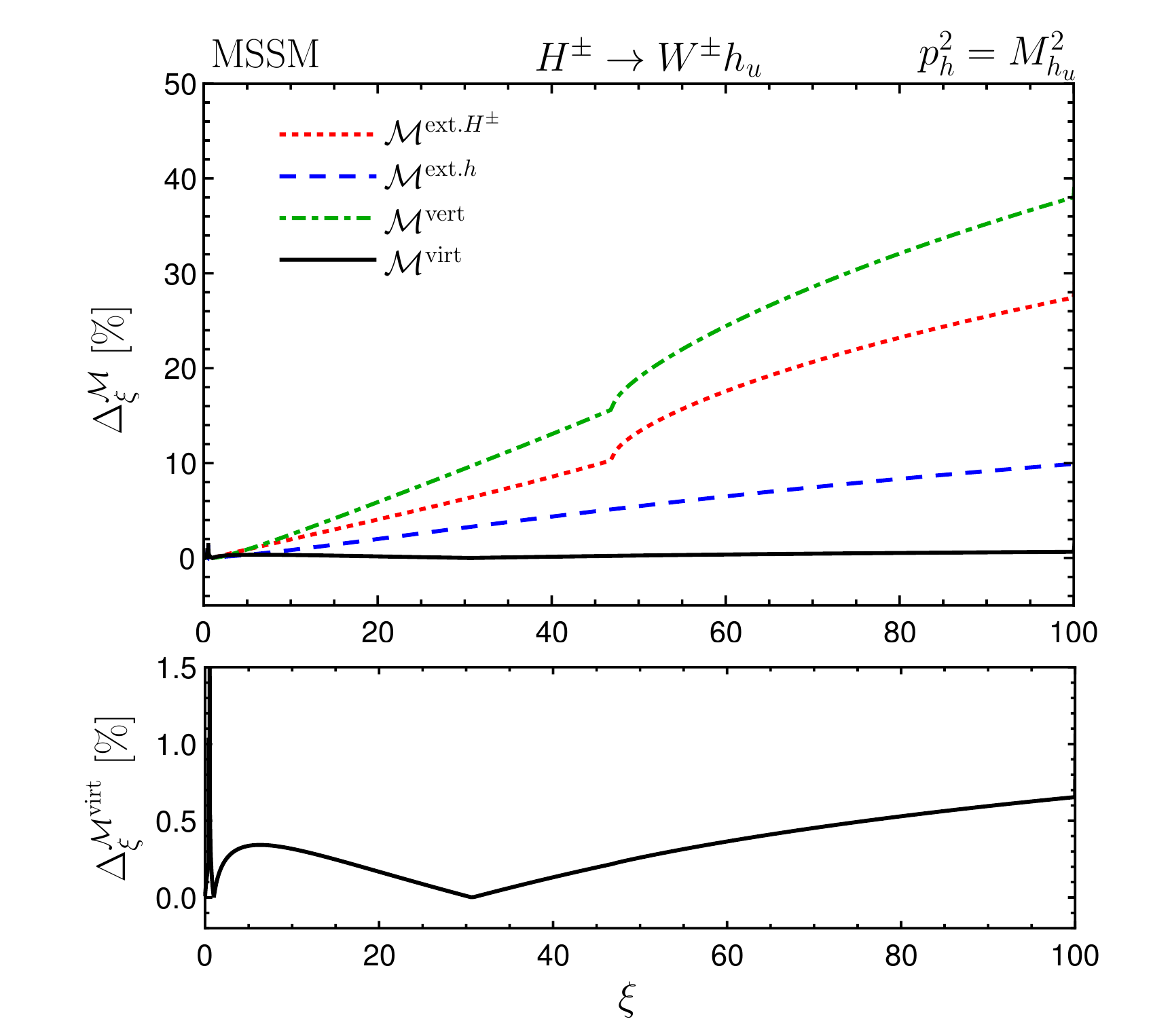}
\hspace*{-0.2cm}
\includegraphics[width=0.5\textwidth]{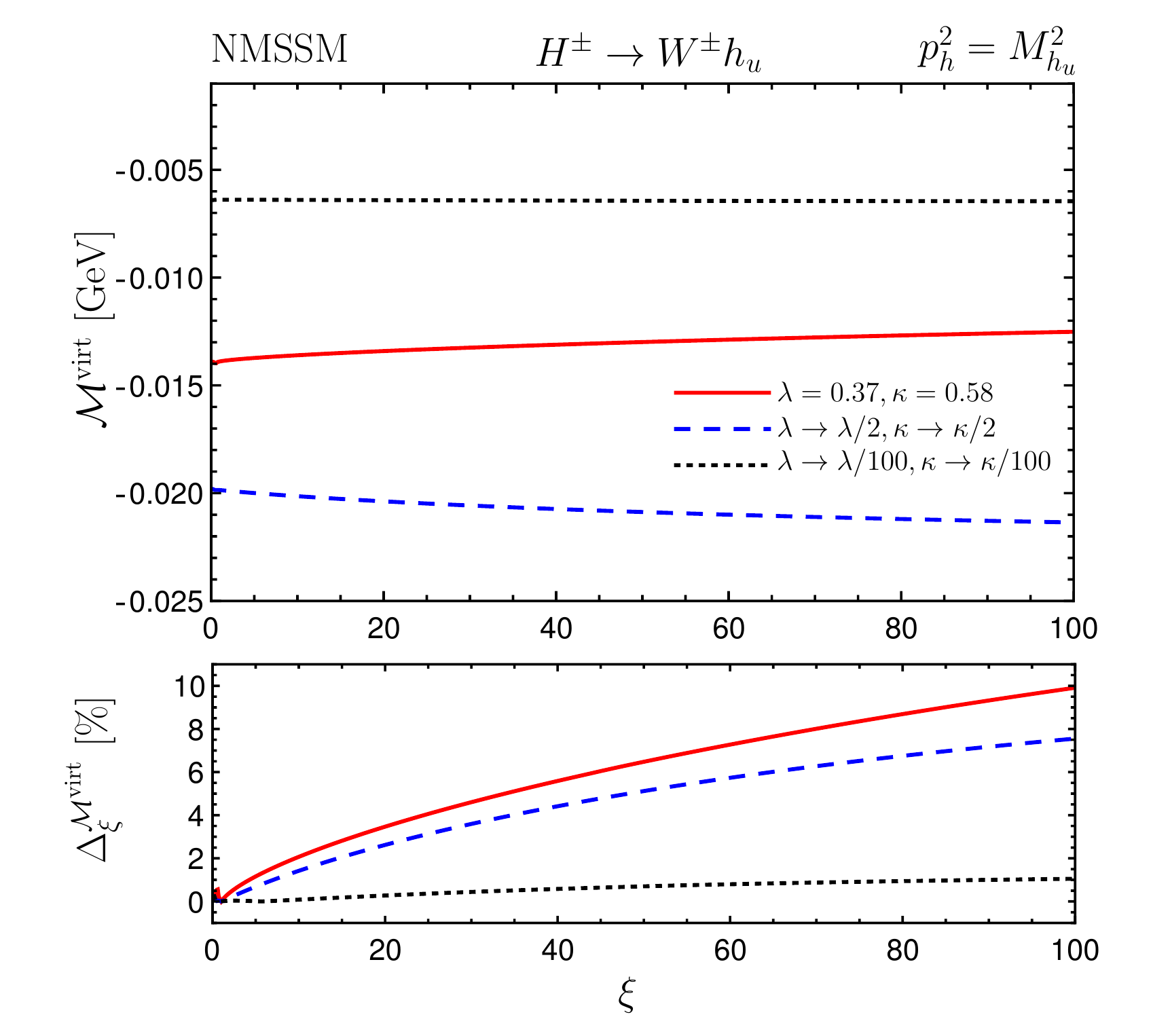}	
\end{center}
\caption{Left: Gauge dependence of the virtual electroweak
  corrections to $H^\pm \to W^\pm h_u$ in the MSSM-limit, using loop-corrected masses for the
  external neutral Higgs boson $p_h^2 = M_{h_u}^2$ calculated in
  general $R_\xi$ gauge. The color and line code is the
  same as in \fig{fig:nmssm-virt}. Lower panel: Gauge dependence of
  $\mc{M}^{\text{virt}}$ with the vertical axis zoomed-in Right:
  ${\cal M}^{\text{virt}}$ as a function of $\xi$ (upper) and its
  relative gauge dependence (lower) for $\lambda=0.37$, $\kappa=0.58$
  (parameter point P1, red/full lines) and their variation to half their
  values (blue/dashed lines) and to $\lambda/100$, $\kappa/100$
  (black/dotted lines).}
\label{fig:mssm-virt}
\end{figure}
In Fig.~\ref{fig:mssm-virt} (left), the curves corresponding to the right plot of
Fig.~\ref{fig:nmssm-virt} are plotted in the MSSM limit of the chosen
parameter point. This limit is taken by setting $\lambda,\kappa \to 0$
and keeping the ratio $\lambda/\kappa$ constant (actually
$\lambda,\kappa= {\cal O}(10^{-8})$). From Fig.~\ref{fig:mssm-virt} we
see that in the MSSM limit the resulting gauge
dependence of $\mc{M}^{\text{virt}}$ has a numerically small
effect. It varies up to 0.7\%\footnote{Per
definition, the line crosses zero at $\xi=1$.}, although we 
are using gauge-dependent loop-corrected masses for the
external momentum $p_h^2 =M_{h_u}^2$. This is illustrated once again
in the right plot of the figure, where we show ${\cal
  M}^{\text{virt}}$ and its relative gauge dependence alone for the
chosen parameter point (red) and after varying $\lambda$ and $\kappa$
to half their values (blue) and to $\lambda/100$ and $\kappa/100$
(black). The relative gauge dependence decreases successively from
10\% to 1\% at $\xi=100$. While the gauge dependence of
the masses increases in the MSSM limit for our chosen parameter point
the opposite is hence the case for the loop corrections to the decay. This
again shows that the singlet admixtures play an important role for the
gauge dependence of the parameters and observables and do not follow a
simple law. 

%%%%%%%%%%%%%%%%%%%%%%%%%%%%%%%%%%%%%%%%%%%%%%%%%%%%%%%%%%%%
\subsubsection{The Complete Loop-Corrected Decay Width}
%%%%%%%%%%%%%%%%%%%%%%%%%%%%%%%%%%%%%%%%%%%%%%%%%%%%%%%%%%%%

\begin{figure}[t!]
\includegraphics[width=0.5\textwidth]{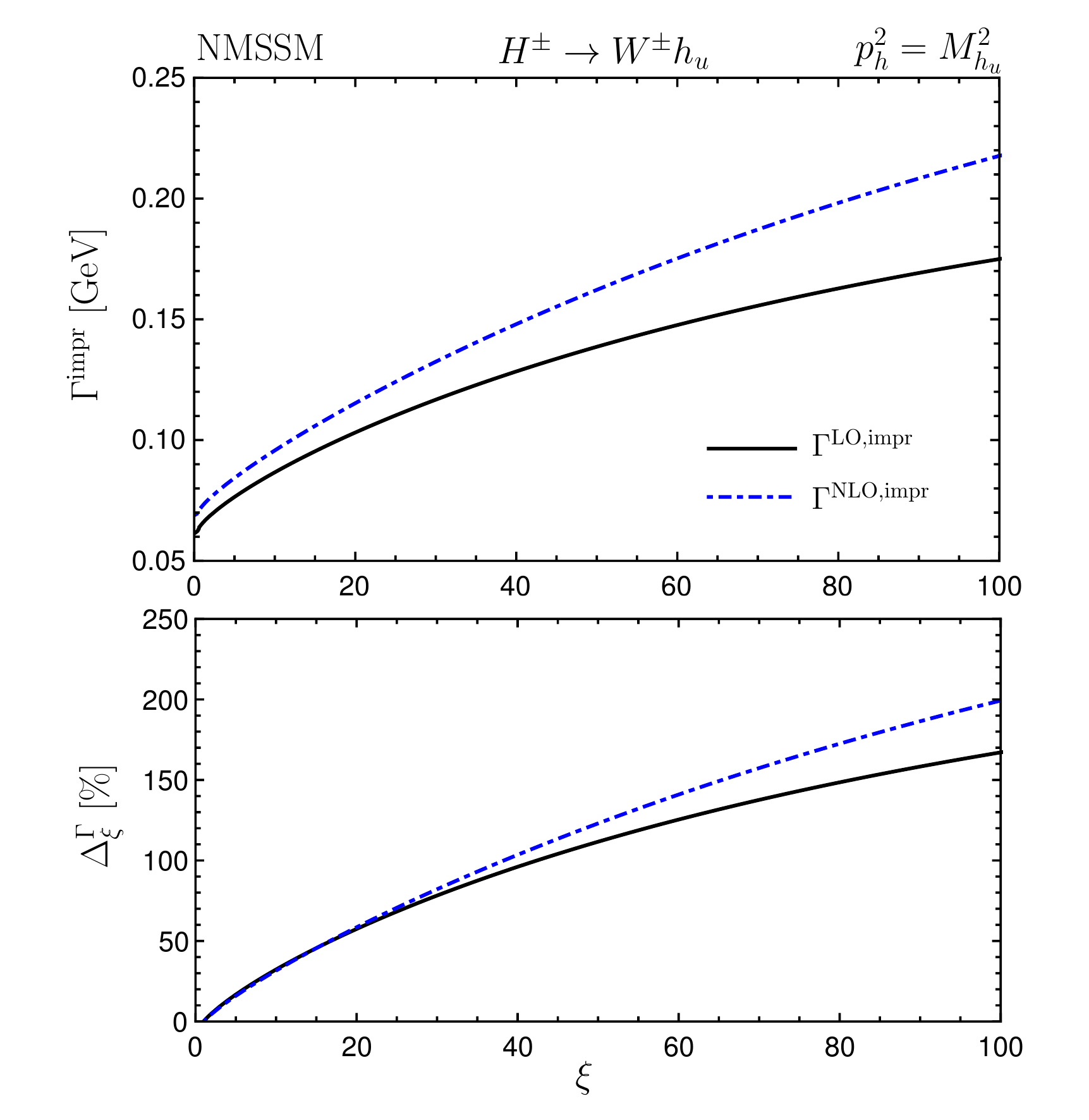}
\includegraphics[width=0.5\textwidth]{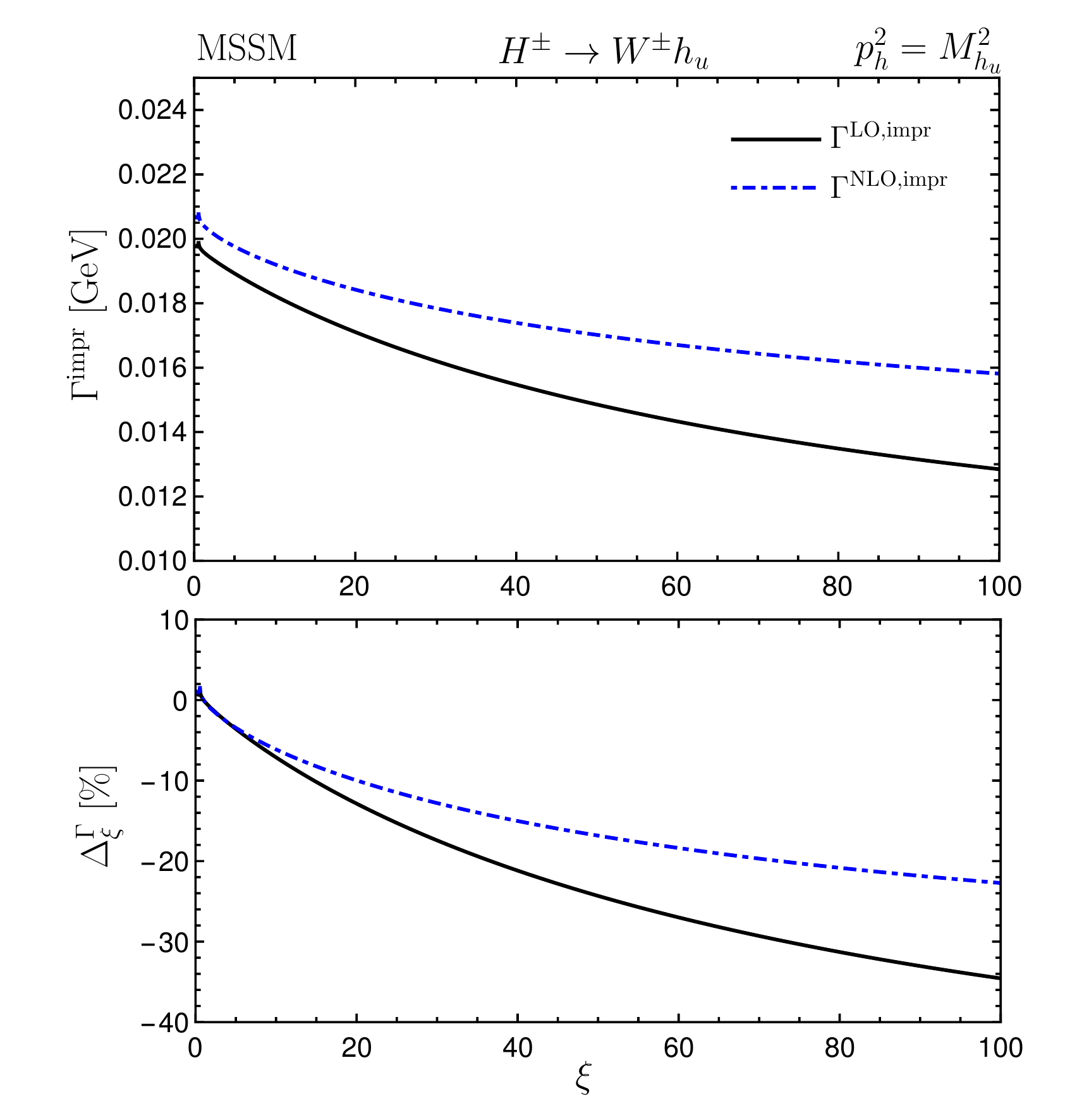}
\caption{Parameter point P1: gauge dependence of the LO and
  NLO decay widths in the NMSSM (left) and its MSSM limit
  (right). $\Gamma^{\text{impr}}$ depicts widths calculated using the
  resummed $\mbf{Z}$ matrix defined in Eq.~(\ref{eq:Z-resum}). The
  upper panels show the gauge dependence of the LO (black, solid) and
  NLO (blue, dot-dashed) widths. In the lower panels we display the
  relative gauge dependence of the LO and NLO widths.} 
	\label{fig:width-Z}
\end{figure}

In the following, we study for the parameter point P1 the
gauge dependence of the complete loop-corrected partial  
decay width of the decay $H^\pm \to 
W^\pm h_u$. In the upper plots of Fig.~\ref{fig:width-Z}, the black (solid)
curve shows, as a function of $\xi$, 
the improved \lo{} decay width for $H^{\pm} \to W^{\pm} h_u$,
{\it i.e.} we apply Eq.~(\ref{eq:gamloimpr}) as denoted with 'Step 2' in
Sec.~\ref{sec:improved-one-loop}. This means that
  we set the external momentum to the loop-corrected Higgs boson mass
$M_{h_u}$,  $p_h^2 = M_{h_u}^2$, which is calculated
at ${\cal O}(\alpha_t \alpha_s + \alpha_t^2)$ with OS renormalization
in the top/stop sector. Additionally, we include in the external-leg
corrections to the neutral Higgs boson the resummed
$\mbf{Z}$ matrix defined in Eq.~(\ref{eq:Z-resum}) in order to ensure the
correct OS properties. The blue (dot-dashed) curve
displays the corresponding improved NLO width, given by
Eq.~(\ref{eq:gamnloimpr}). The left plots are for the NMSSM
parameter point P1, whereas the right plots are for the MSSM limit of
the same benchmark point. The NMSSM widths show a stronger dependence 
on $\xi$ than the ones in the MSSM limit (note that the scales of the
two plots are different).
%%%%%%%%%%%%%%%%%%%%%%%%%%%%%%%%%%%%%%%%%%%%%%%%%%%%%%%%%
In the lower plots we show the relative gauge dependence of the LO and
NLO widths, respectively, as a function of $\xi$, as defined by 
\be
\Delta^\Gamma_\xi = \frac{\Gamma_\xi - \Gamma_{\xi = 1}}{ \Gamma_{\xi =
    1}} \;.
\ee
Here $\Gamma_\xi$ denotes the decay width calculated in
general $R_\xi$ gauge at fixed loop order, {\it i.e.}
$\Gamma^{\text{LO, impr}}$ or $\Gamma^{\text{NLO, impr}}$, and
$\Gamma_{\xi = 1}$ the width calculated in the 't\,Hooft--Feynman gauge. 
%%%%%%%%%%%%%%%%%%%%%%%%%%%%%%%%%%%%%%%%%%%%%%%%%%%%%%%%%
In the NMSSM the relative gauge dependence is larger for the NLO width
than for the LO one while in the MSSM, where the singlet admixture
vanishes, the opposite is the case. For the complete NLO width of
scenario P1 the relative corrections
  $\Delta^{\Gamma}_\xi$ can become as large as 200\% for $\xi=100$.
Such a strong gauge dependence is clearly unacceptable for making
meaningful predictions for the decay widths.  
For phenomenological investigations, on the other
  hand, the interesting quantity is the branching ratio. In order to
  make meaningful predictions, this requires the inclusion of the
  electroweak corrections to all other charged Higgs boson decays, so
  that the total width entering the branching ratio is computed at
  higher order in the electroweak corrections. This is beyond the
  scope of the present paper and left for future work.
Even if one argues not to introduce new mass scales in the process and
to remain below $\xi$ values of 100 the $\xi$ dependence is large, in
particular it is far beyond the relative size of the loop corrections which is about
11\% for $\xi=1$. In the MSSM limit, depicted in the right plot
  of Fig.~\ref{fig:width-Z}, the relative gauge dependence is smaller
  with values of up to about 20\% for $\xi=100$. \s

\begin{figure}[t!]
\begin{center}
\hspace*{-0.2cm}
\includegraphics[width=0.5\textwidth]{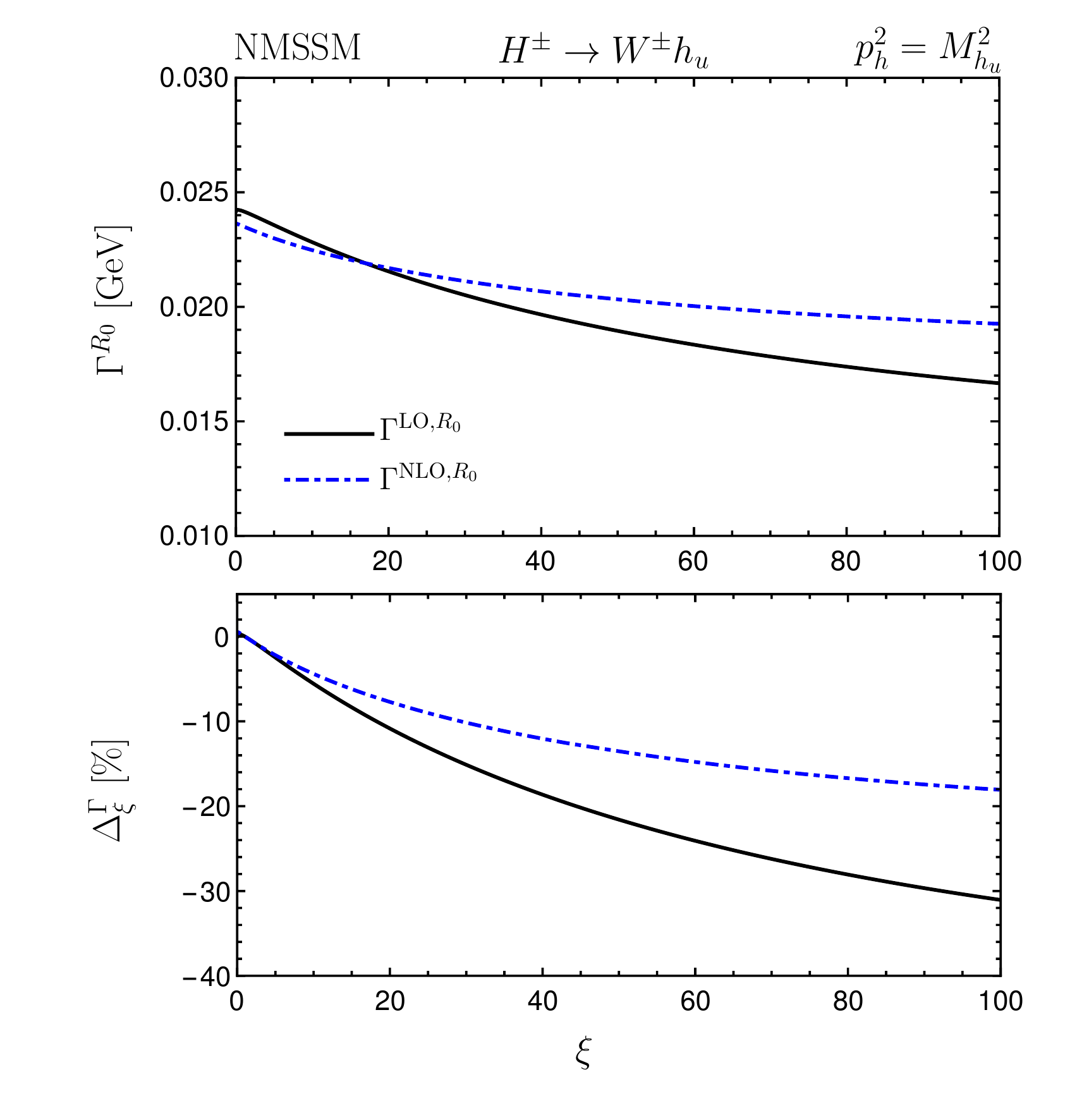}
\hspace*{-0.2cm}
\includegraphics[width=0.5\textwidth]{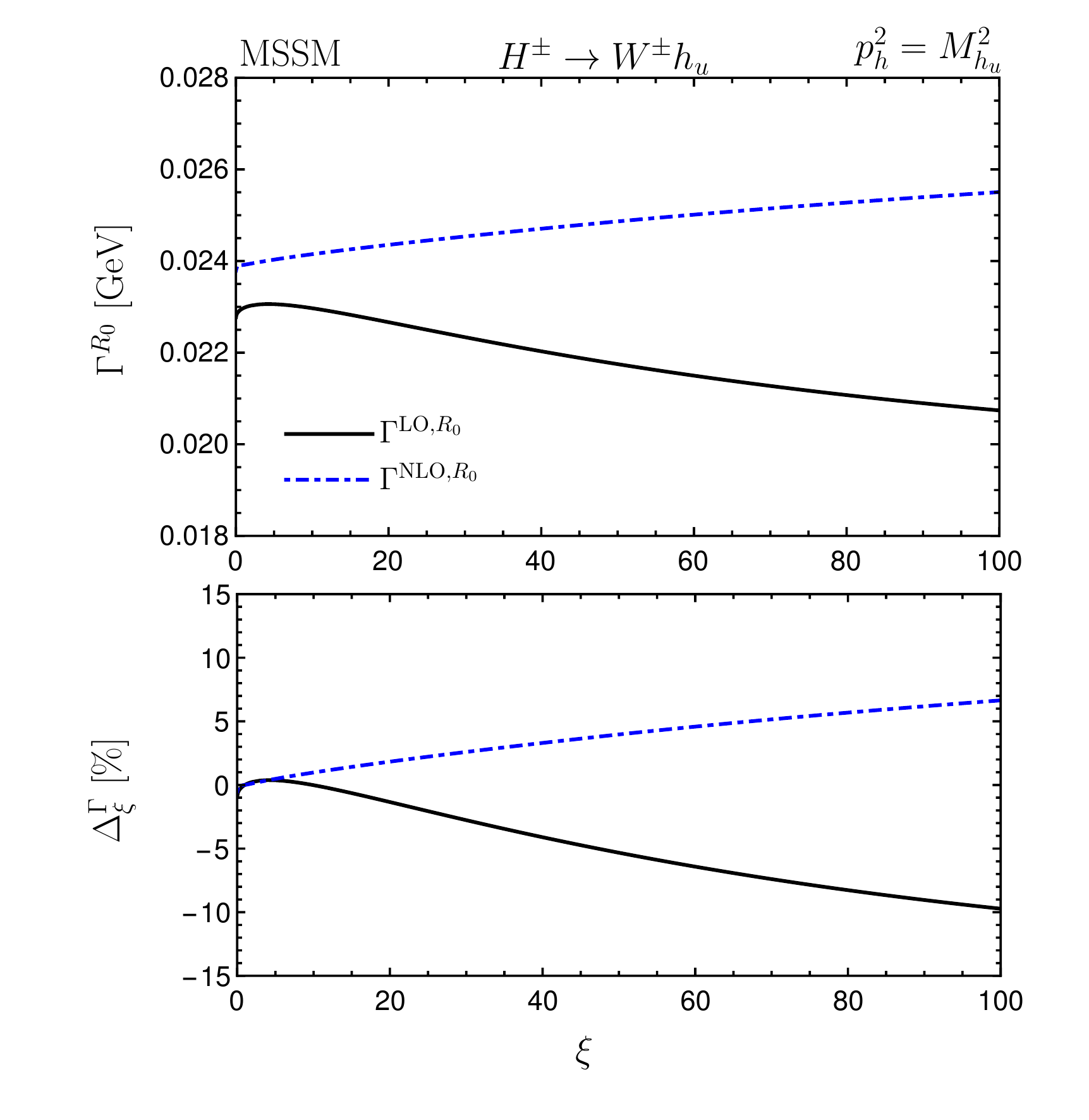}
\vspace*{-0.6cm}
\caption{Analogous to Fig.~\ref{fig:width-Z}, but
  using the $R^0$-method 
  instead of the iterative approach to extract the loop-corrected
  masses and mixing matrix.
	\label{fig:width-R}}
\end{center}
\end{figure}
\begin{figure}[t!]
\begin{center}
\hspace*{-0.2cm}
\includegraphics[width=0.5\textwidth]{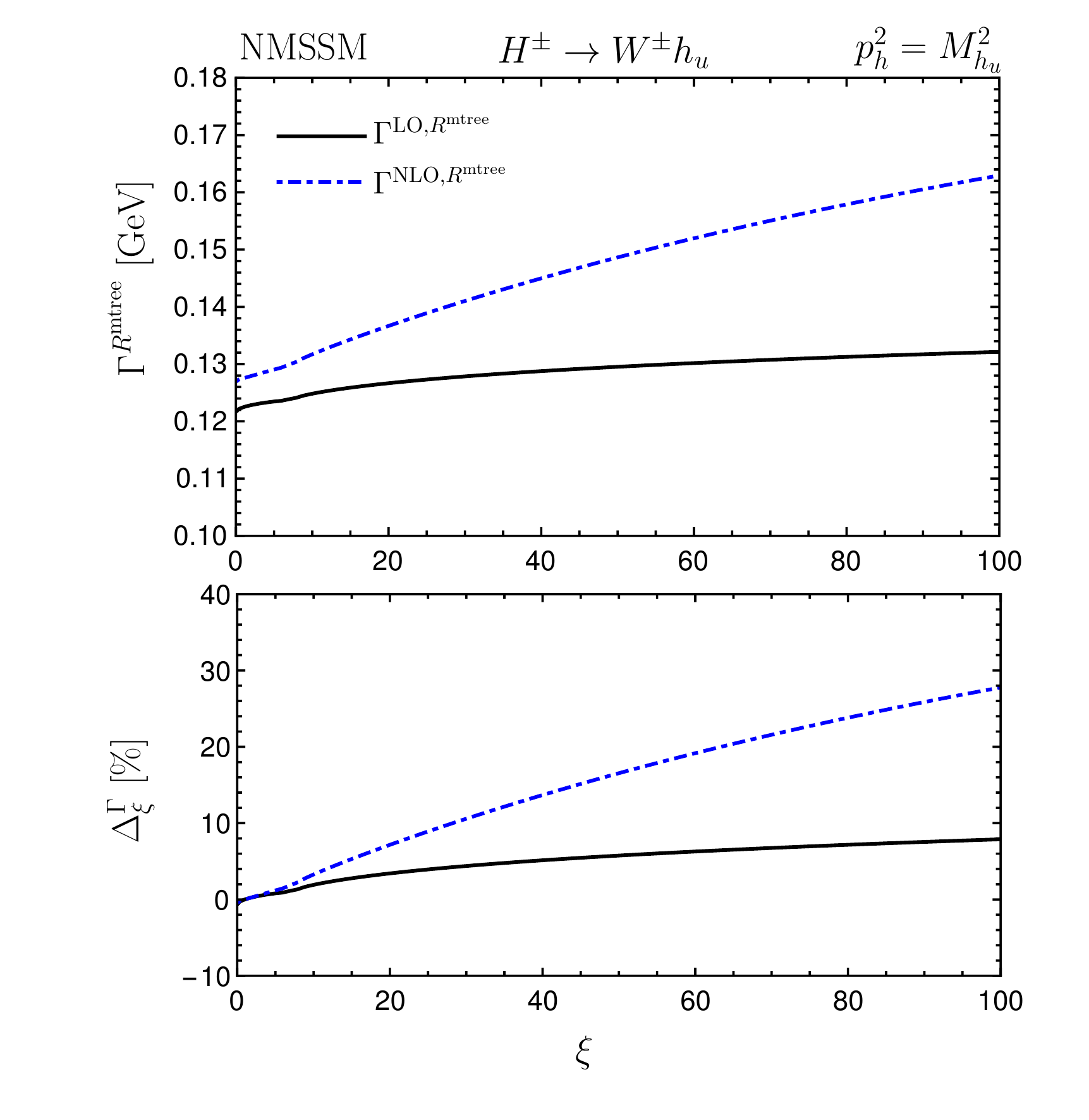}
\hspace*{-0.2cm}
\includegraphics[width=0.5\textwidth]{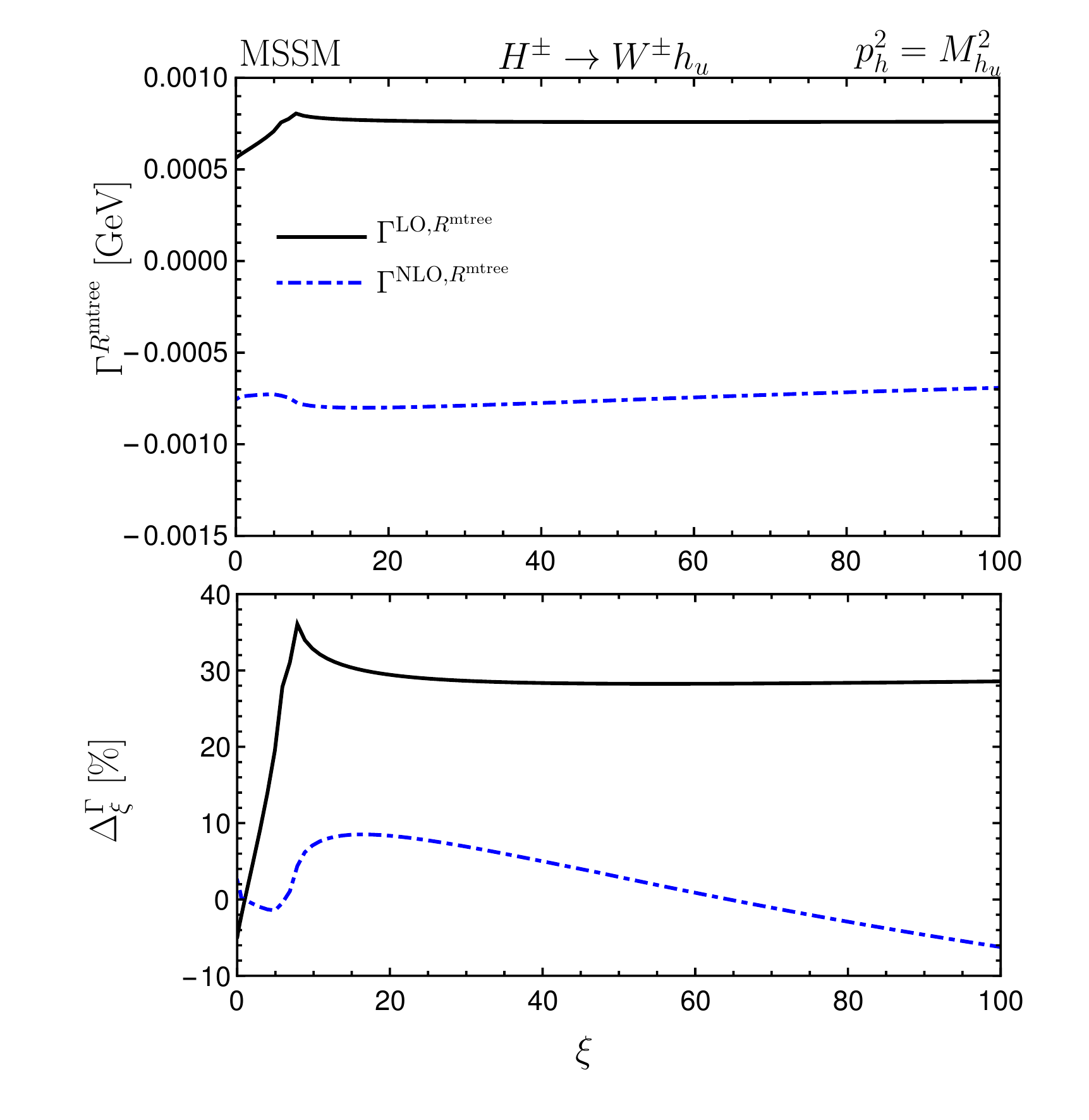}
\vspace*{-0.6cm}
\caption{Same as Fig.~\ref{fig:width-Z}, but using the $R^{\text{mtree}}$-method
  instead of the iterative approach to extract the loop-corrected
  masses and mixing matrix.} 
\label{fig:width-Rmtree}
\end{center}
\end{figure}

In Fig.~\ref{fig:width-R}, we show the corresponding curves
analogous to Fig.~\ref{fig:width-Z}, however now using the $R^0$
method to extract the loop-corrected mass values and mixing
matrix\footnote{As remarked above, in the couplings we 
  always use the tree-level mixing matrix elements, however.}. The LO
and NLO widths are then calculated by applying 
Eqs.~(\ref{eq:loamp-improved}) to (\ref{eq:realcorr}), but with the
${\bf Z}$ matrix replaced by the $R^0$ matrix, defined in
Eq.~(\ref{eq:R0}). We denote the corresponding widths with the superscript $R^0$ as
$\Gamma^{R^0}$. Figure~\ref{fig:width-Rmtree} shows the corresponding
results if the masses are extracted at the arithmetic squared mass
average such that the ${\bf Z}$ matrix is 
replaced by the $R^{\text{mtree}}$ matrix, defined in Eqs.~(\ref{eq:arithm1},\ref{eq:arithm2}). The
corresponding widths are denoted by the superscript
$R^{\text{mtree}}$. \s

The comparison of Figs.~\ref{fig:width-R} and \ref{fig:width-Rmtree}
with Fig.~\ref{fig:width-Z} shows that for this parameter point the
gauge dependence is smallest in the $R^0$ approximation. The relative
change of the complete NLO width with $\xi$ compared to its value for
$\xi=1$, {\it i.e.}~$\Delta_\xi^\Gamma$, is about -18\% for $\xi=100$,
while in the $R^{\text{mtree}}$ approximation it is about +28\%, which
is still smaller than if the ${\bf Z}$ matrix is applied. The corresponding
values in the MSSM limit are 6.5\% ($R^0$) and -6\%
($R^{\text{mtree}}$). The method of extracting the mixing matrix
elements has a strong influence on the $\xi$ dependence of the NLO
width and also on the sign of this $\xi$ dependence. For the parameter
point P1 the $h_u$-like Higgs boson has a strong
singlet admixture. From previous analyses
\cite{Staub:2015aea,Drechsel:2016htw}, we know already that the 
mixing matrix elements are then very sensitive to changes in the
approximation of the loop calculation. Since the mixing matrix
elements enter the Higgs couplings, the computed observables, in this case the
decay width, become very sensitive to the applied approximation. This is confirmed by our results on the $\xi$ dependence but also by the
values of the widths themselves for the various
approximations.\footnote{The NLO width in the MSSM-like scenario is
  negative for the $R^{\text{mtree}}$ approximation which is clearly
  non-physical and which is due to the tiny tree-level width. Here
  higher-order corrections beyond one-loop order would need to be
  included for a meaningful prediction.} \s

\begin{figure}[t!]
\begin{center}
\includegraphics[width=0.49\textwidth]{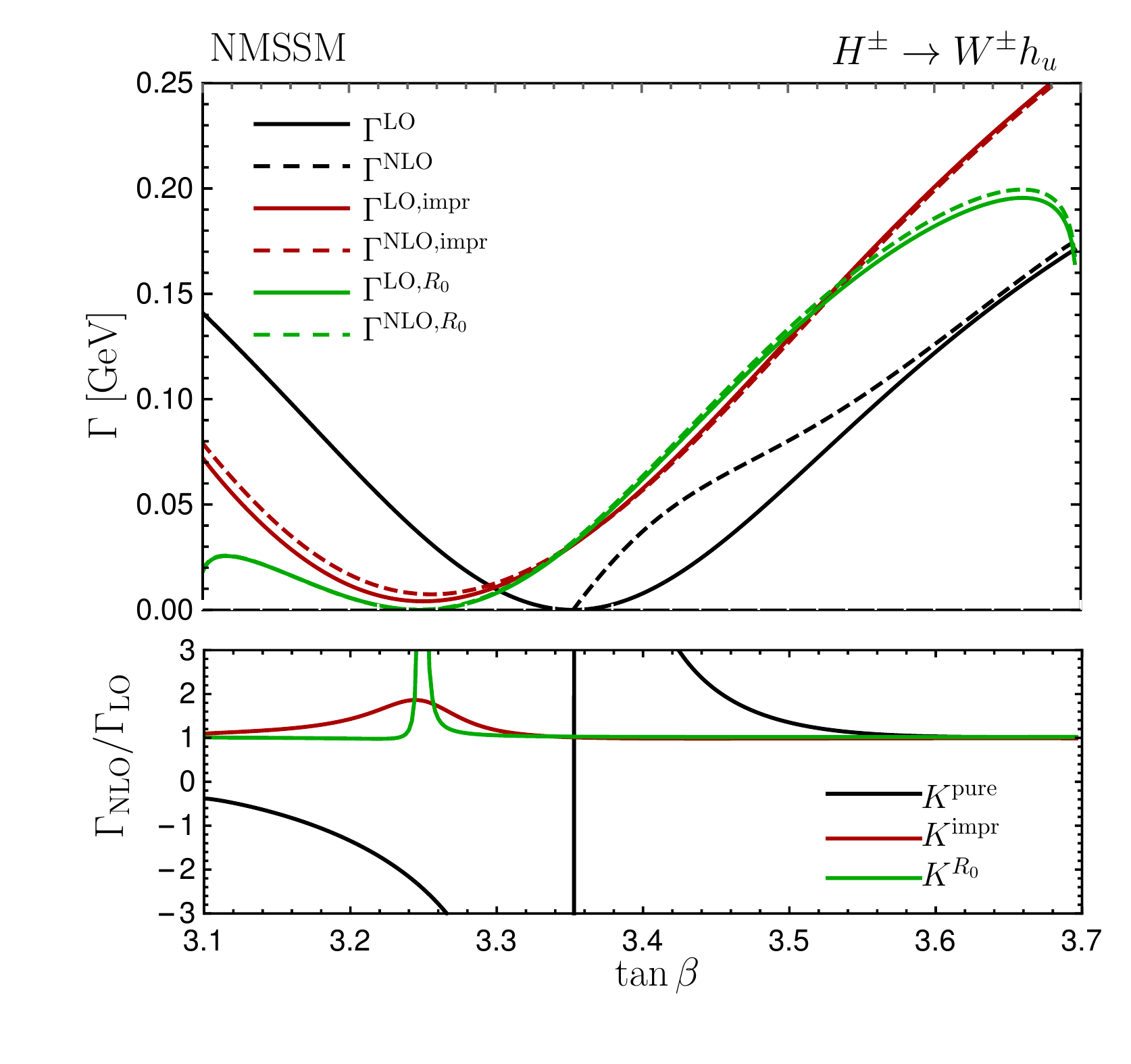}
\hspace*{-0.2cm}
\includegraphics[width=0.49\textwidth]{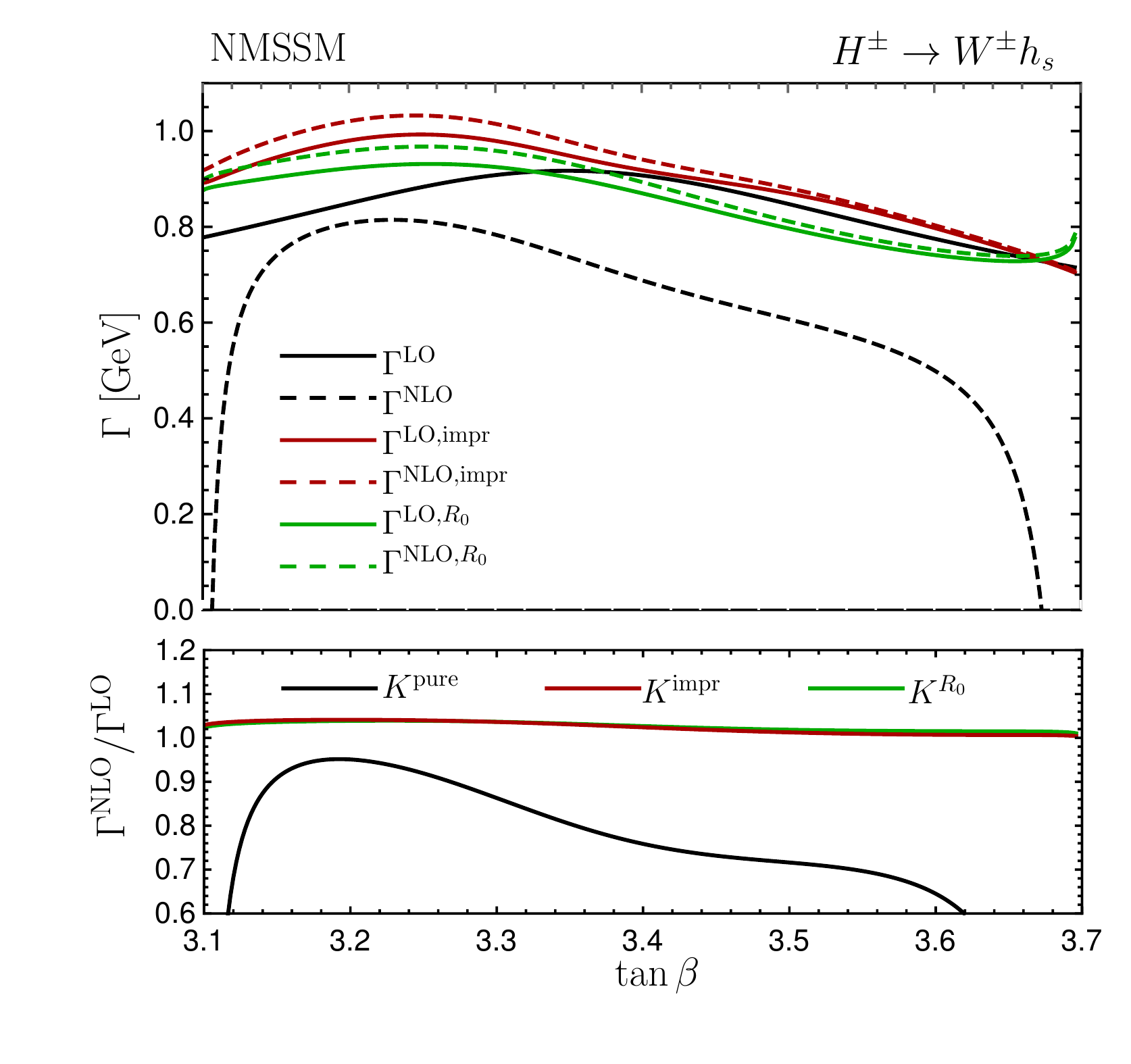}
\caption{Parameter point P1: partial decay width of the decay $H^\pm
  \to W^\pm h_u$ (left) and $H^\pm \to W^\pm h_s$ (right) at LO (full)
and NLO (dashed) at strict LO and one-loop order (black), in the
improved approach applying the ${\bf Z}$ matrix (red) and the $R_0$
matrix (green), as a function of $\tan\beta$. Lower panels:
corresponding $K$ factors.}
	\label{fig:tanbetascan}
\end{center}
\end{figure}

Overall, we found that the gauge dependence of
the loop-corrected mass of the external neutral Higgs boson has a much
smaller influence on the gauge dependence of the NLO width than the
matrix that is used to set the external Higgs boson OS. The
strength of this effect sensitively depends on the chosen parameter
set, as can be inferred from Fig.~\ref{fig:tanbetascan}. The figure
displays the partial decay widths of the decay $H^\pm
  \to W^\pm h_u$ (left plot) and $H^\pm \to W^\pm h_s$ (right plot)
  both at LO and NLO as a function of $\tan\beta$. All other parameter
  values are fixed to those of scenario P1. Shown are the
  results for the pure LO and strict one-loop width (black lines) and
  the ones when we calculate the improved LO and NLO widths applying
  the ${\bf Z}$ matrix (red) and the $R_0$ matrix (green). The lower
  plots show the corresponding $K$ factors, defined as the ratio of
  the NLO width and its corresponding LO width 
\beq
K = \frac{\Gamma^{\text{NLO}}}{\Gamma^{\text{LO}}} \;.
\eeq
As can be inferred from the left plot, the value of the decay width
$\Gamma (H^\pm \to W^\pm h_u)$ strongly depends on the applied
approximation for our chosen parameter point, {\it i.e.}~for
$\tan\beta=3.11$, while the $K$ factor is approximately the same for
the improved widths, with a value around 1.1. The $K$ factor for the pure one-loop result
largely differs from the improved ones as it does not take into
account any resummation of higher orders. For values of $\tan\beta$ between
about 3.26 and 3.52 the improved NLO results are rather close, but
differ otherwise. In the singlet-like final state shown in
Fig.~\ref{fig:tanbetascan} (right), {\it i.e.}~$\Gamma (H^\pm \to W^\pm
  h_s )$, the $K$ factors for the improved widths differ by less than
  5\% over the whole scanned range.  

%%%%%%%%%%%%%%%%%%%%%%%%%%%%%%%%%%%%%%%%%%%%%%%%%%%%%%%%%%%%%
\subsection{Analysis for Scenario P2 \label{sec:p2}}
\begin{figure}[t!] 
\includegraphics[width=0.5\textwidth]{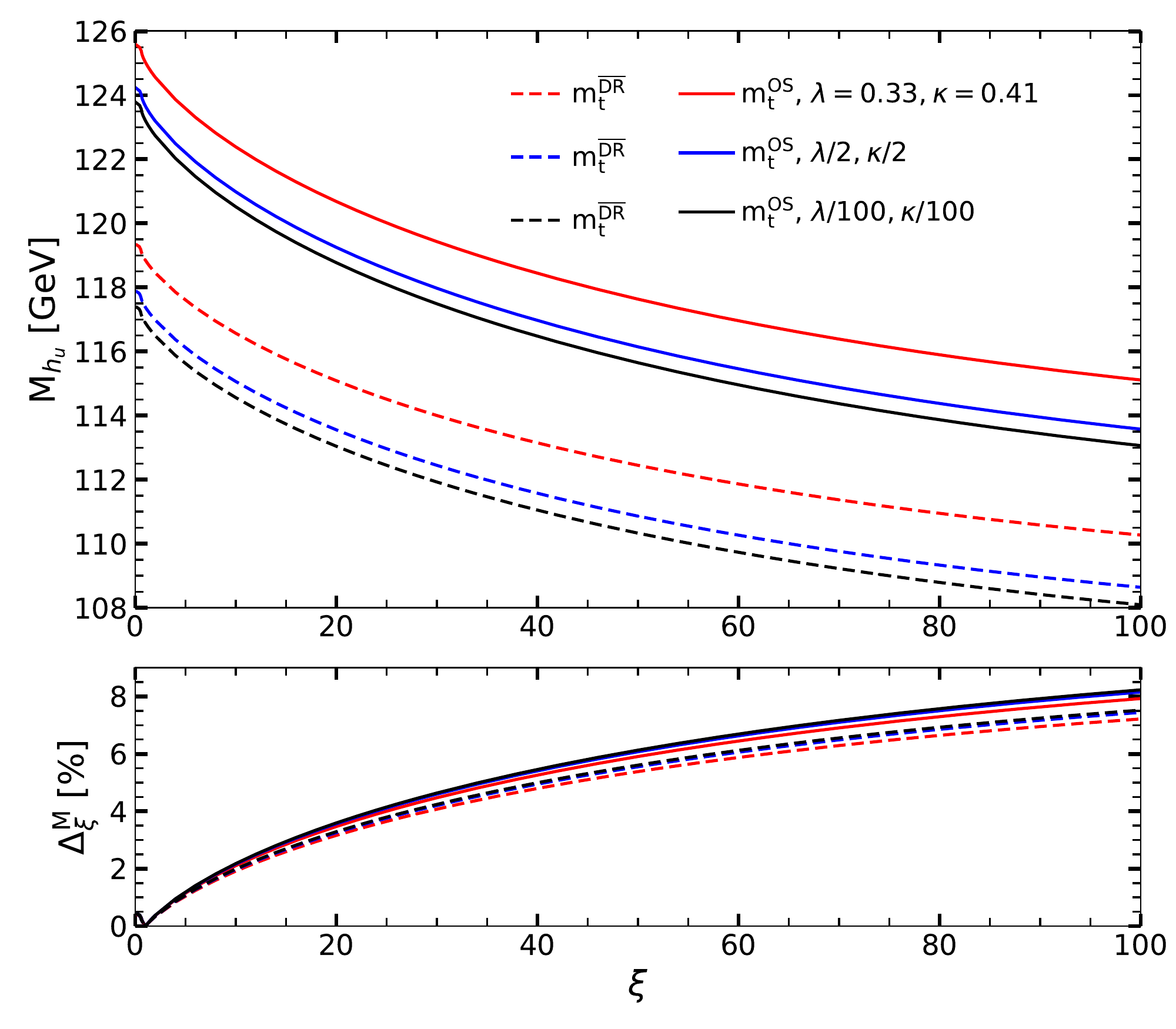}
\includegraphics[width=0.5\textwidth]{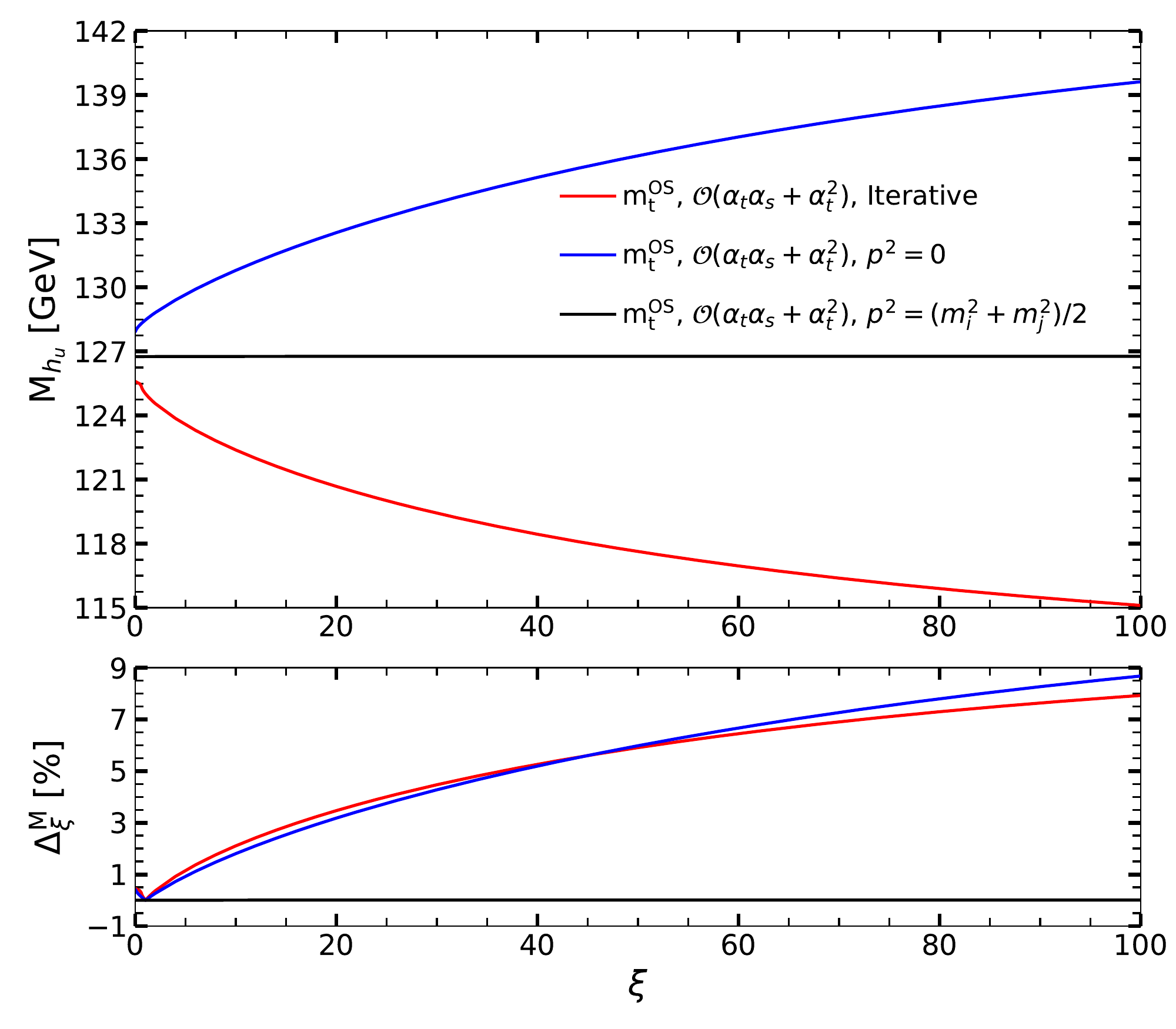}
\caption{Left: The  $h_u$-like 
  Higgs boson masses as a function of $\xi$ at two-loop ${\cal O}(\alpha_t \alpha_s +
\alpha_t^2)$ level in the OS (solid lines) and the $\DRb$
(dashed lines) scheme of the top/stop 
sector for scenario P2, {\it i.e.}~$\lambda=0.33, \kappa=0.41$
(red line), for half their values (blue) and for $\lambda/100$,
$\kappa/100$ (black). Right: The  $h_u$-like Higgs boson mass of
scenario P2 as a function of $\xi$ by applying iterative (red), the
$R^0$ (blue) and the $R^{\text{mtree}}$-method (black).  Lower panels:
The corresponding $\Delta^M_\xi$ values 
in percent, as a function of $\xi$.} 
\label{fig:hmass-p49097} 
\end{figure}

%%%%%%%%%%%%%%%%%%%%%%%%%%%%%%%%%%%%%%%%%%%%%%%%%%%%%%%%%%%%%
In order to further investigate the impact of the gauge dependence, we
analyze the gauge dependence of the Higgs boson masses and the charged
Higgs decay widths for a second parameter point, P2, defined in
Eqs.~(\ref{eq:param4scen2}) and (\ref{eq:param4scen2_b}). We summarize
the Higgs boson masses obtained in the OS renormalization scheme 
of the top/stop sector in Table~\ref{tab:massP2OS} and in the $\DRb$
scheme in Table~\ref{tab:massP2DR}, at tree level, at one-loop level
and at two-loop level including the ${\cal O}(\alpha_t\alpha_s)$ and the
${\cal O}(\alpha_t \alpha_s + \alpha_t^2)$ corrections, respectively. We have
deliberately chosen this scenario in which the $h_u$-like Higgs boson is the  
lightest state with mass around 125 GeV at ${\cal O}(\alpha_t \alpha_s + \alpha_t^2)$
  in the OS renormalization scheme of the top/stop sector while the
  CP-even singlet-like Higgs boson is the second-lightest state with mass
  around 433 GeV. In this scenario we analyze only the mass of the $h_u$-like
  Higgs boson since only its mass is affected substantially by the
  change of the gauge parameter $\xi$. We present in
  Fig.~\ref{fig:hmass-p49097} the  $h_u$-like Higgs boson mass as
  function of $\xi$. The left plot shows its two-loop mass at ${\cal
    O}(\alpha_t \alpha_s + \alpha_t^2)$ for OS (full) and
  $\overline{\mbox{DR}}$ (dashed) renormalization in the top/stop sector for
  three different singlet admixtures. This means we start with the
  $\lambda$ and $\kappa$ values of our original scenario P2 (red
  lines) and compare with the results when we take half (blue lines) and
  1/100 their values (black lines), where the latter means that we
  are close to the MSSM limit. As can be inferred from the plot, the
  gauge dependence $\Delta^M_\xi$ shown in the lower plot is
  only mildly dependent on the renormalization scheme and on the singlet
  admixture, and amounts up to values of about 7 to 8 \% at
  $\xi=100$. \s

\begin{table}[t!]
\begin{center}
 \begin{tabular}{|l||c|c|c|c|c|}
\hline
 &$h_1/{H_1}$&$h_2/{H_2}$&$h_3/{H_3}$&$h_4/{H_4}$&$h_5/{H_5}$\\ \hline \hline
tree-level &85.43&437.27&581.25&986.77&989.68\\ 
main component&$h_u$&$h_s$&$a$&$a_s$&$h_d$\\ \hline  
one-loop &133.59 & 433.79 & 577.39 & 989.71 & 986.69  \\ 
main component&$h_u$&$h_s$&$a$&$a_s$&$h_d$\\ \hline  
two-loop ${\cal O}(\alpha_t \alpha_s)$ &118.38 & 433.76 & 577.42 & 989.61 & 986.7  \\  
main component&$h_u$&$h_s$&$a$&$a_s$&$h_d$\\ \hline  
two-loop ${\cal O}(\alpha_t \alpha_s+ \alpha_t^2)$
 &125.03 & 433.76 & 577.42 & 989.66 & 986.7\\ 
main component&$h_u$&$h_s$&$a$&$a_s$&$h_d$\\ \hline  
\end{tabular}
\caption{P2: mass values in GeV and main components of the neutral Higgs
  bosons at tree level, one-loop level, two-loop ${\cal O}(\alpha_t
  \alpha_s)$ level and at two-loop ${\cal O}(\alpha_t \alpha_s +
  \alpha_t^2)$ level obtained by using OS renormalization in the top/stop sector.}
\label{tab:massP2OS}
\end{center}
\end{table}
\begin{table}[t!]
\begin{center}
 \begin{tabular}{|l||c|c|c|c|c|}
\hline
 &$h_1/{H_1}$&$h_2/{H_2}$&$h_3/{H_3}$&$h_4/{H_4}$&$h_5/{H_5}$\\ \hline \hline
tree-level &85.43&437.27&581.25&986.77&989.68\\
main component&$h_u$&$h_s$&$a$&$a_s$&$h_d$\\ \hline  
one-loop &113.9 & 433.75 & 577.43 & 989.55 & 986.65  \\
main component&$h_u$&$h_s$&$a$&$a_s$&$h_d$\\ \hline  
two-loop ${\cal O}(\alpha_t \alpha_s)$ &118.4 & 433.76 & 577.42 & 989.56 & 986.64 \\ 
main component&$h_u$&$h_s$&$a$&$a_s$&$h_d$\\ \hline  
two-loop ${\cal O}(\alpha_t \alpha_s+ \alpha_t^2)$
 &118.86 & 433.76 & 577.42 & 989.57 & 986.64 \\               
main component&$h_u$&$h_s$&$a$&$a_s$&$h_d$\\ \hline  
\end{tabular}
\caption{P2: mass values in GeV and main components of the neutral Higgs
  bosons at tree level, one-loop level, two-loop ${\cal O}(\alpha_t
  \alpha_s)$ level and at two-loop ${\cal O}(\alpha_t \alpha_s +
  \alpha_t^2)$ level obtained by using $\DRb$ renormalization in the top/stop sector.}
\label{tab:massP2DR}
\end{center}
\end{table}

In the right plot of Fig.~\ref{fig:hmass-p49097} we show the
$\xi$ dependence of $M_{h_u}$ when we apply different approximations
to determine the loop-corrected Higgs mass eigenstates with OS
renormalization of the top/stop sector, namely through the
iterative method (red line), by applying the rotation matrix $R^0$ to
the mass matrix in  the zero momentum approximation (blue), or finally by
applying $R^{\text{mtree}}$ to the mass matrix evaluated at its
arithmetic squared mass average (black). The $\xi$ dependence of the iterative
and the zero momentum procedure is about the same, with $\Delta^M_\xi$
amounting to 8 and
9\%, respectively, at $\xi=100$. For the arithmetic squared mass average
method, however, we again find that the $\xi$ dependence is very small. 
Overall, the gauge dependence of the $h_u$-like mass in scenario P1 is
larger than in P2. \s

\begin{figure}[t!]
\begin{center}
\includegraphics[width=0.45\textwidth]{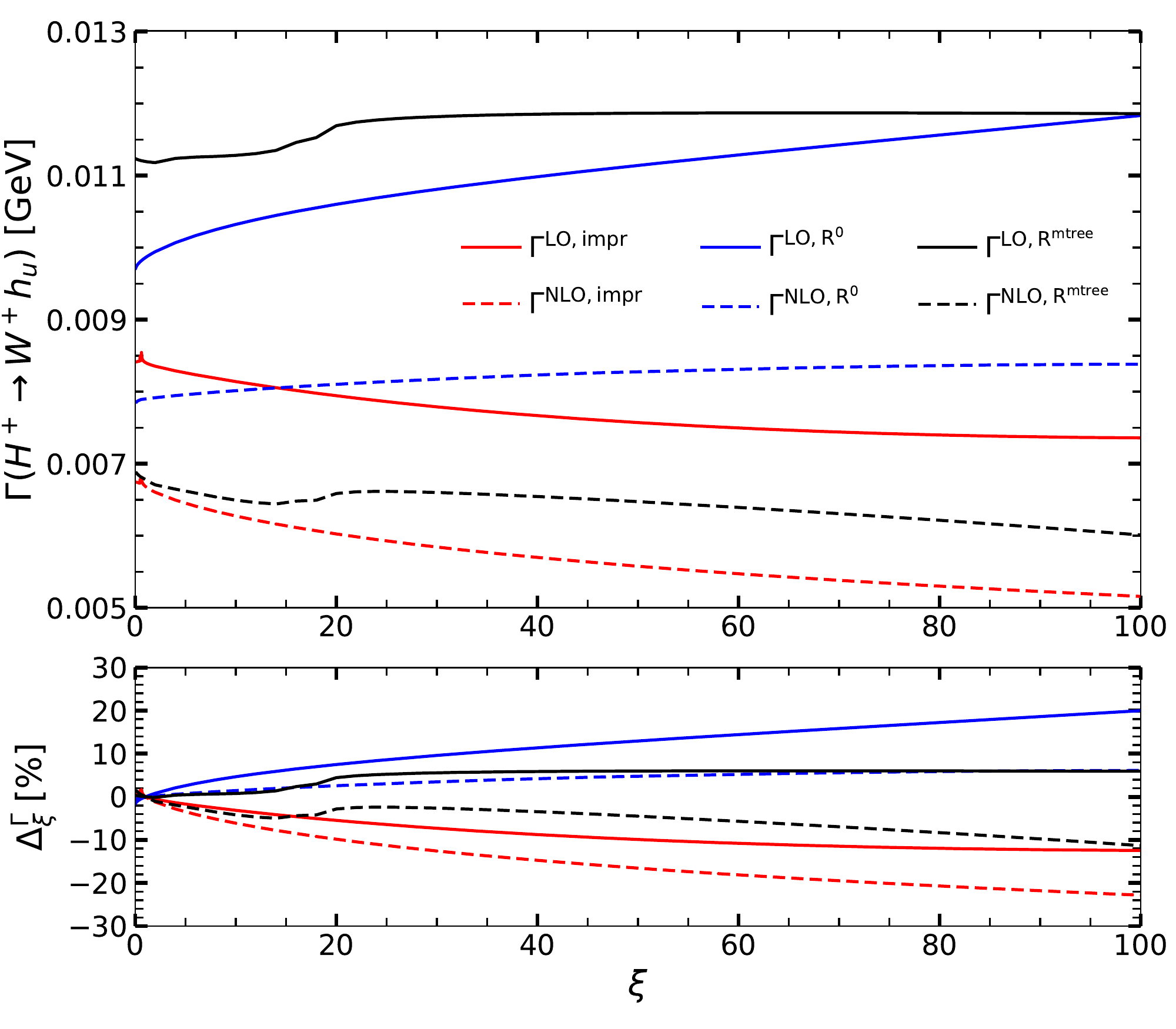}
\includegraphics[width=0.45\textwidth]{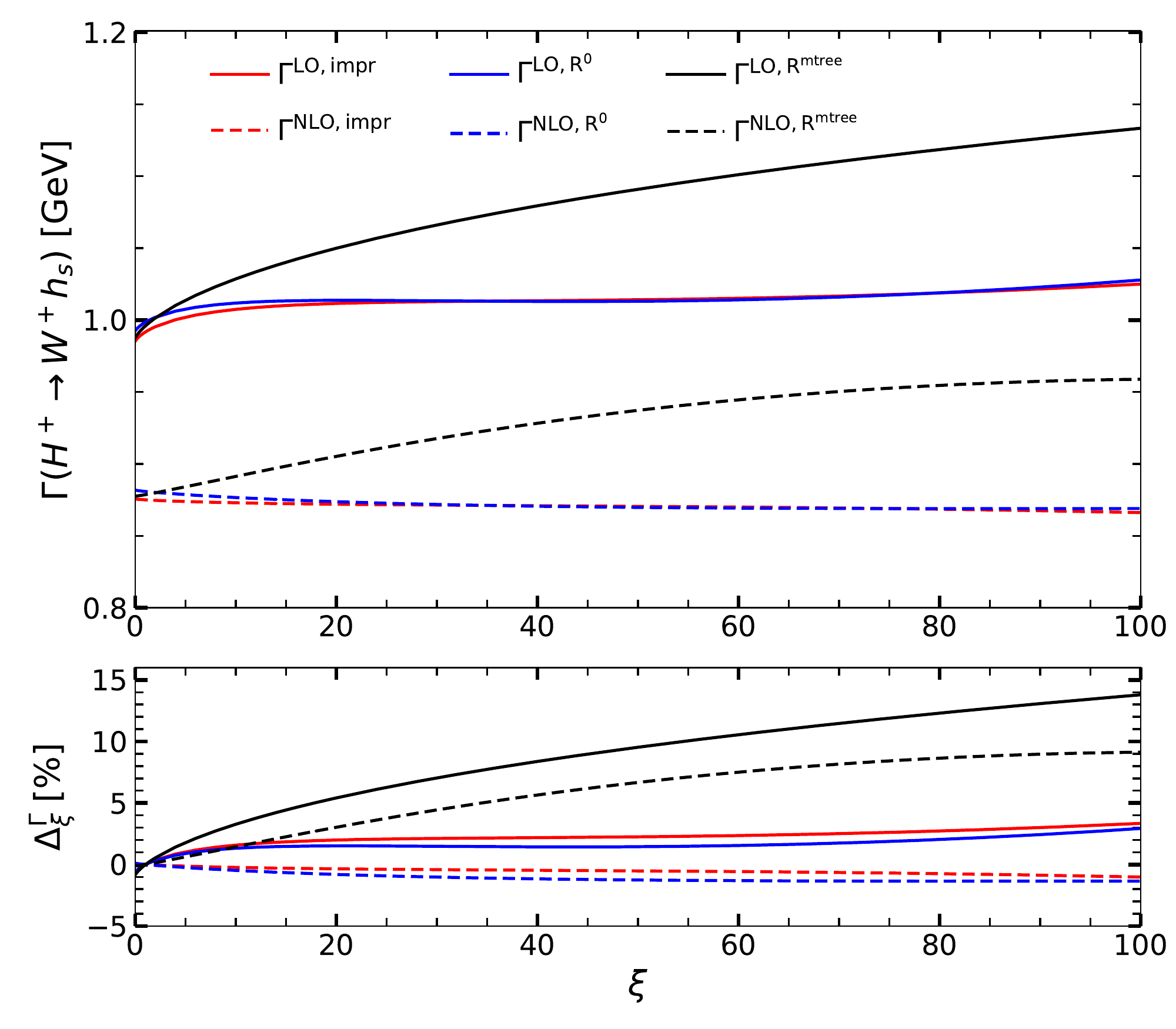}
\caption{P2: Decay widths for  $H^\pm\to W^\pm h_u$ (left
  uppper plot) and  $H^\pm\to W^\pm h_s$ (right uppper plot)  as
  functions of $\xi$ using the iterative (red), the $R^0$ (blue) and the
$R^{\text{mtree}}$ method (black). Lower panels: Corresponding $\Delta^\Gamma_\xi$ 
in percent, as function of $\xi$.}
	\label{fig:width-p49097}
\end{center}
\end{figure}

Figure~\ref{fig:width-p49097} depicts the gauge dependence
  of the partial widths of the decays $H^\pm \to W^\pm h_u$ (left) and
  $H^\pm \to W^\pm h_s$ (right) at LO (full lines) and NLO (dashed
  lines) by applying in Eqs.~(\ref{eq:gamloimpr}) and 
  (\ref{eq:gamnloimpr}), respectively, the three different 
  approximations for the matrices that diagonalize the corresponding loop-corrected
  mass matrices, namely the ${\bf Z}$ matrix (red), $R^0$ (blue) and
  $R^{\text{mtree}}$ (black). The corresponding decay widths are
  denoted by the superscripts 'impr', '$R^0$' and '$R^{\text{mtree}}$'. The lower
  panels display the corresponding $\Delta_\xi^\Gamma$ values. The
  inspection of the plots shows that the $\xi$ dependence of the NLO
  widths not always decreases compared to the LO one. Moreover, there
  is no pattern for the two decays that allows to decide which of
  the three approximations induces the smallest gauge dependence in
  the NLO widths. A closer investigation reveals that the mixing
  between $h_u$ and $h_d$ is responsible for the gauge dependence of
  $H^\pm \to W^\pm h_u$ and the mixing between $h_s$ and $h_d$ for the one in
$H^\pm \to W^\pm h_s$. Overall, however, the gauge dependence of the
partial widths is much smaller than for the parameter point P1, with
maximum values of $\Delta_\xi^\Gamma$ around $- 22\%$ for $H^\pm \to W^\pm h_u$ (for $\Gamma^{\text{NLO},\text{impr}}$) and 14\% for $H^\pm \to W^\pm
h_s$ (for $\Gamma^{\text{LO},R^{\text{mtree}}}$). The relative NLO corrections at $\xi=1$
amount to -20\% (for $\Gamma^{\text{impr}}$) for the 
former and to  -12\% for the latter decay (for
$\Gamma^{R^{\text{mtree}}}$), however, so that the gauge 
dependence is of the order of the loop correction. 

%%%%%%%%%%%%%%%%%%%%%%%%%%%%%%%%%%%%%%%%%%%%%%%%%%%%%%%%%%%%%
\section{Conclusions \label{sec:concl}}
%%%%%%%%%%%%%%%%%%%%%%%%%%%%%%%%%%%%%%%%%%%%%%%%%%%%%%%%%%%%%
In this paper we investigated the influence of the gauge parameter
both on the higher-order corrections to the NMSSM Higgs 
boson masses and the partial decay width of $H^\pm \to W^\pm h_i$,
calculated in general $R_\xi$ gauge. The gauge
dependence enters through the mixing of loop orders: for the masses,
this happens due to the application of an iterative method to determine the
loop-corrected mass values. This is transferred to the decay width as phenomenology
requires the inclusion of the mass corrections to the
external Higgs bosons in order to match the
  experimentally measured values. These are calculated up to two-loop
order including higher-order terms through the application
of the iterative procedure. Gauge dependence then enters the
process through different mechanisms. On the one hand there is a
mismatch between the use of tree-level masses in the propagators of
the internal lines and in the Higgs-Goldstone boson couplings
appearing in the computation of the loops, and the use of the higher-order-corrected
Higgs mass for the external Higgs bosons. The latter prevents the
cancelation of IR divergences when adding up the virtual and real
corrections. While this can be cured by an appropriate adaption of
the involved couplings, the second source of the gauge dependences
persists: it stems from the resummation of higher
  orders that enters both through the 
external Higgs boson mass and the mixing matrix applied to set the
external Higgs boson OS. The latter has a particularly
large impact as we found by applying different approximations to
determine the higher-order masses and mixing matrix elements. The relative
gauge dependence can then largely exceed the relative size of the loop
correction itself, so that for the interpretation of the results the
specification of the used gauge is crucial. By analyzing different
parameter sets, we found that the impact of the gauge
dependence depends on the chosen parameter point and can vary
substantially depending on the applied parameter set.

%%%%%%%%%%%%%%%%%%%%%%%%%%%%%%%%%%%%%%%%%%%%%%%%%%%%%%%%%%%%%%
\section*{Acknowledgments}
We thank Stefan Liebler and Michael Spira for useful discussions. We
thank Philipp Basler for help with the scan for valid parameter points. 
MK and MM acknowledge financial support from the DFG project ``Precision
Calculations in the Higgs Sector - Paving the Way to the New
Physics''. TND thanks for the financial support for her visit at KIT
from the DFG project ``Precision 
Calculations in the Higgs Sector - Paving the Way to the New Physics''.
TND's work is funded by the Vietnam National Foundation for Science
and Technology Development (NAFOSTED) under grant number 103.01-2017.78. 

%\bibliography{refs}
%\bibliographystyle{JHEP}

%%%%%%%%%%%%%%%%%%%%%%%%%%%%%%%%%%%%%%%%%%%%%%%%%%%%%%%%%%

\end{document}